\documentclass[prb,aps,english,twocolumn,notitlepage,superscriptaddress]{revtex4-2}

\usepackage{graphicx}
\usepackage{dcolumn}
\usepackage{rotating}
\usepackage{lipsum}
\usepackage{xcolor}
\usepackage{bm}
\usepackage[T1]{fontenc}  
\usepackage{multirow,amsmath,amssymb}
 \usepackage{babel} 
 \usepackage{txfonts}
 \usepackage{epsfig,epsf,psfrag} 
\usepackage{graphicx} 
\usepackage{pslatex}
\usepackage{epic,eepic} 
\usepackage{color,pstcol}
\usepackage{pstricks} 
\usepackage{fancyhdr} 
\usepackage{tabularx}
\usepackage{physics}
\usepackage{url}
\usepackage{listings}
\usepackage{color} 
\usepackage{caption}
\usepackage{ulem}
\usepackage{balance}
\usepackage{subcaption}
\usepackage[utf8]{inputenc}

\usepackage{hyperref}
\hypersetup{
 colorlinks=true,
 linkcolor=blue,
 anchorcolor = blue,
 citecolor = blue,
 filecolor = blue,
 urlcolor = blue
}
\usepackage{dsfont}

\graphicspath{{Images/}}

\def \be {\begin{equation}} 
\def \ee {\end{equation}}

\begin{document}

\preprint{APS/123-QED}

\title{Reaching Van den Broeck limit in linear response and Whitney limit in nonlinear response in edge mode
quantum thermoelectrics and refrigeration}

\title{Reaching Van den Broeck limit in linear response and Whitney limit in nonlinear response in edge mode
quantum thermoelectrics and refrigeration}
\author{Sachiraj Mishra}%
\email{sachiraj29mishra@gmail.com}

\author{Colin Benjamin}%
\email{colin.nano@gmail.com}
\affiliation{School of Physical Sciences, National Institute of Science Education and Research, HBNI, Jatni-752050, India}
\affiliation{Homi Bhabha National Institute, Training School Complex, AnushaktiNagar, Mumbai, 400094, India }

\begin{abstract}
Quantum heat engines and quantum refrigerators are proposed in three-terminal quantum Hall (QH) and quantum spin Hall (QSH) setups with a voltage-temperature probe in both the linear and nonlinear transport regimes. In the linear response regime, we find that efficiency at maximum power approaches the Van den Broeck limit in both QH and QSH setups. Similarly, in nonlinear response, we find that efficiency at maximum power reaches the Whitney bounds. This is for the first time, we see that in the same setup and using quantum point contacts, the thermoelectric efficiency limits in linear and nonlinear response being achieved.
\end{abstract}

\maketitle


\textit{\underline{Introduction:}}
The quantum Hall resistance \cite{PhysRevLett.45.494} in two-dimensional electron gas in the presence of a normal uniform magnetic field at low temperature arises as a consequence of chiral edge modes \cite{PhysRevB.25.2185}, which is famously known as quantum Hall effect. These edge modes are robust against disorder and are immune to backscattering \cite{PhysRevB.38.9375} being topologically protected. Complimentary to chiral edge modes, helical edge modes are seen in quantum spin Hall experiments with Mercury Telluride/Cadmium Telluride heterostructures \cite{zhang, Konig, Roth_2009}, which are also topologically protected. The study of thermoelectric properties in quantum Hall (QH) and quantum spin Hall (QSH) setups has received considerable attention, particularly through the application of the Landauer-Büttiker formalism \cite{datta_1995, Shen_2017, PhysRevLett.114.146801, Brandner_2013, Haack_2021, PhysRevB.100.235442, PhysRevLett.112.130601, PhysRevB.91.115425, PhysRevE.97.022114, PhysRevB.108.195435, PhysRevB.84.201307, PhysRevLett.123.186801, PhysRevLett.112.057001, PhysRevLett.114.067001, mani2019designing, PhysRevE.96.032118, PhysRevB.91.155407, Sanchez_2015}. This formalism forms the basis for understanding the behavior of quantum heat engines and quantum refrigerators in time-independent steady-state quantum thermoelectric studies, which is the main focus of our investigation. Similarly, in the realm of quantum thermodynamics, heat engines and refrigerators are also broadly categorized into finite-time cyclic systems \cite{chen1989effect, PhysRevLett.117.190601, holubec2016maximum, PhysRevE.98.042112} and autonomous (steady-state) systems \cite{kosloff2014quantum, wineland1975proposed, hansch1975cooling}. Quantum transport can be described in two regimes. First is linear, wherein the applied voltage and temperature biases are quite small, and thus leads to a linear relation between current and the temperature or voltage biases applied~\cite{BENENTI20171}. The second is nonlinear, wherein the applied voltage and temperature biases are not small. In the linear response regime, the efficiency at maximum power ($\eta|_{P_{max}}$) \cite{curzon1975efficiency} and the maximum efficiency ($\eta_{max}$) are closely related to the figure of merit $ZT \hspace{0.1cm}  \left( = \frac{G S^2 T}{K}\right)$, where $G$ = conductance, $S$ = Seebeck coefficient, $K$ = thermal conductance and $T$ is the reference temperature of the setup. In this regime, there exists a bound to $\eta|_{P_{max}}$, which is given by Van den Broeck (VDB) \cite{PhysRevLett.95.190602} limit with the efficiency ($\eta|_{P_{max}}$)= $\eta_c/2$, where $\eta_c = 1 - T_2/T_1$ is the Carnot efficiency with $T_1$ and $T_2$ being the temperatures of the hot and cold reservoirs respectively \cite{BENENTI20171}. To achieve this, it is necessary to have a large $ZT$. This can be accomplished by ensuring a high Seebeck coefficient ($S$) and low thermal conductance ($K$). Similarly, in the non-linear transport regime, there also exists an upper bound to $\eta|_{P_{max}}$, the Whitney limit \cite{PhysRevLett.112.130601, PhysRevB.91.115425}, i.e., efficiency at maximum power output ($\eta|_{P_{max}^{Wh}}$) = $\eta|_{P_{max}^{Wh}}$ = $\eta_c / (1 + 0.936 (1 + T_2/T_1))$. This is the nonlinear equivalent of the VDB efficiency in linear response and has been derived utilizing BoxCar-type transmission. The study of two-terminal setups in nonlinear response was extended to a three-terminal setup with a voltage probe, and it was shown that the Whitney limits are universal~\cite{Whitney_2016}. Similarly, in \cite{PhysRevE.92.042165}, it has been proved that a generic mesoscopic thermoelectric heat engine can be highly efficient only when the transmission is Box-car type. Boxcar-type scattering can be implemented via a chain of quantum dots~\cite{PhysRevLett.112.130601, PhysRevB.91.115425, Whitney_2016}. However, electron-electron interaction between quantum dots, which can lead to coulomb blockade effects, is neglected in the calculation in Refs.~\cite{PhysRevLett.112.130601, PhysRevB.91.115425, Whitney_2016}. This assumption is quite extreme, as in real systems, having a chain of quantum dots without the coulomb blockade effect is almost impossible to implement~\cite{BENENTI20171}. In contrast, this work considers quantum point contact (QPC) type scattering, which is easier to implement experimentally \cite{PhysRevLett.60.848, wharam1988one, van1992thermo, PhysRevLett.68.3765, PhysRevB.57.1838} than the boxcar-type transmission. In this work, we reach the Whitney bound in the nonlinear transport regime using QPC-type scattering. Previously, in Refs.~\cite{PhysRevLett.112.130601, PhysRevB.91.115425, Whitney_2016}  a QH setup was only considered; however, in this letter, we consider both QH and QSH setups, albeit with the easier-to-implement QPC type contacts for both linear as well as non-linear response regimes.
 
\begin{figure}
     \centering
     \begin{subfigure}[b]{0.235\textwidth}
         \centering
         \includegraphics[width=\textwidth]{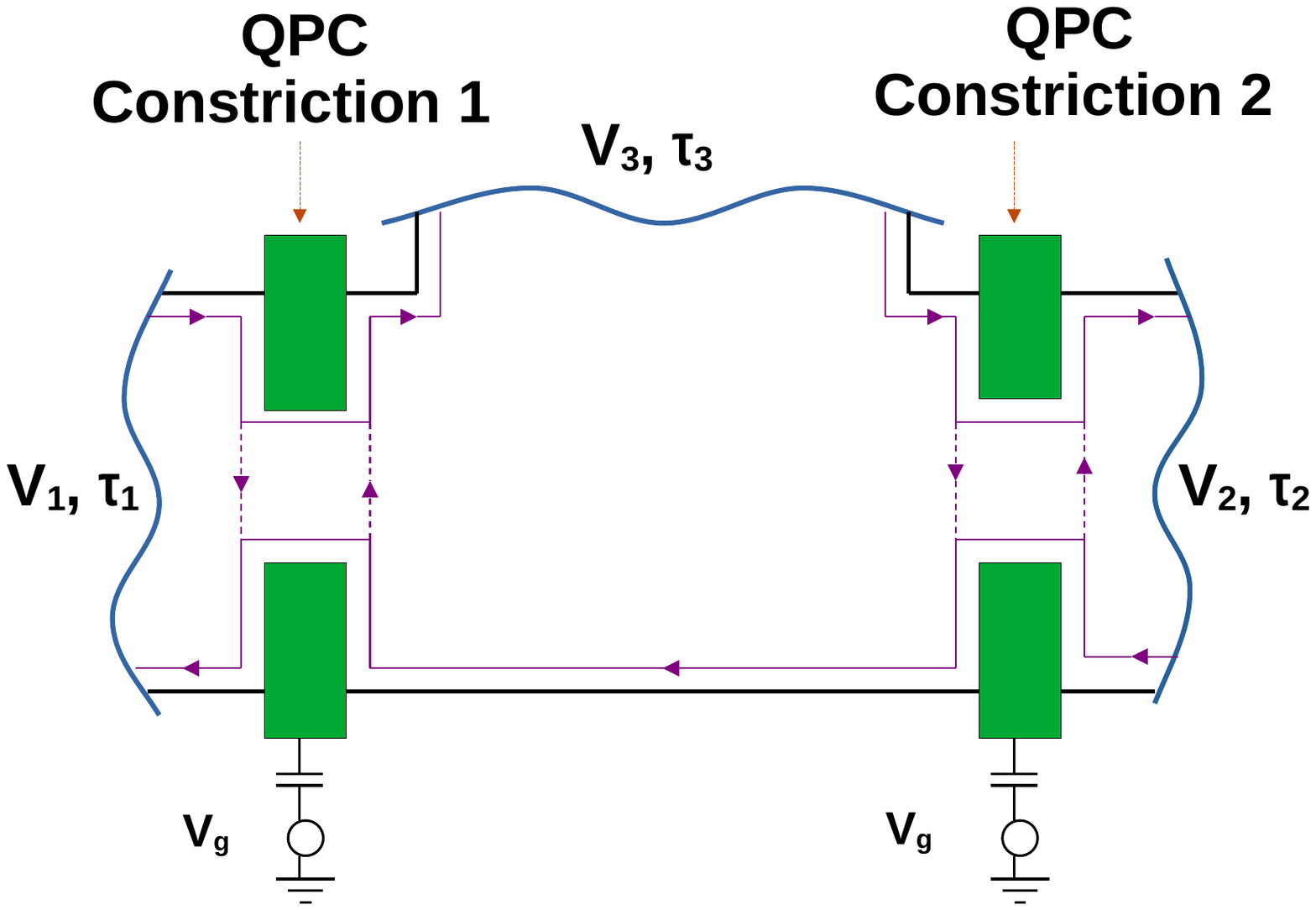}
         \label{fig:3(a)}
         \caption{}
     \end{subfigure}
     \hspace{0.07cm}
     \begin{subfigure}[b]{0.235\textwidth}
         \centering
         \includegraphics[width=\textwidth]{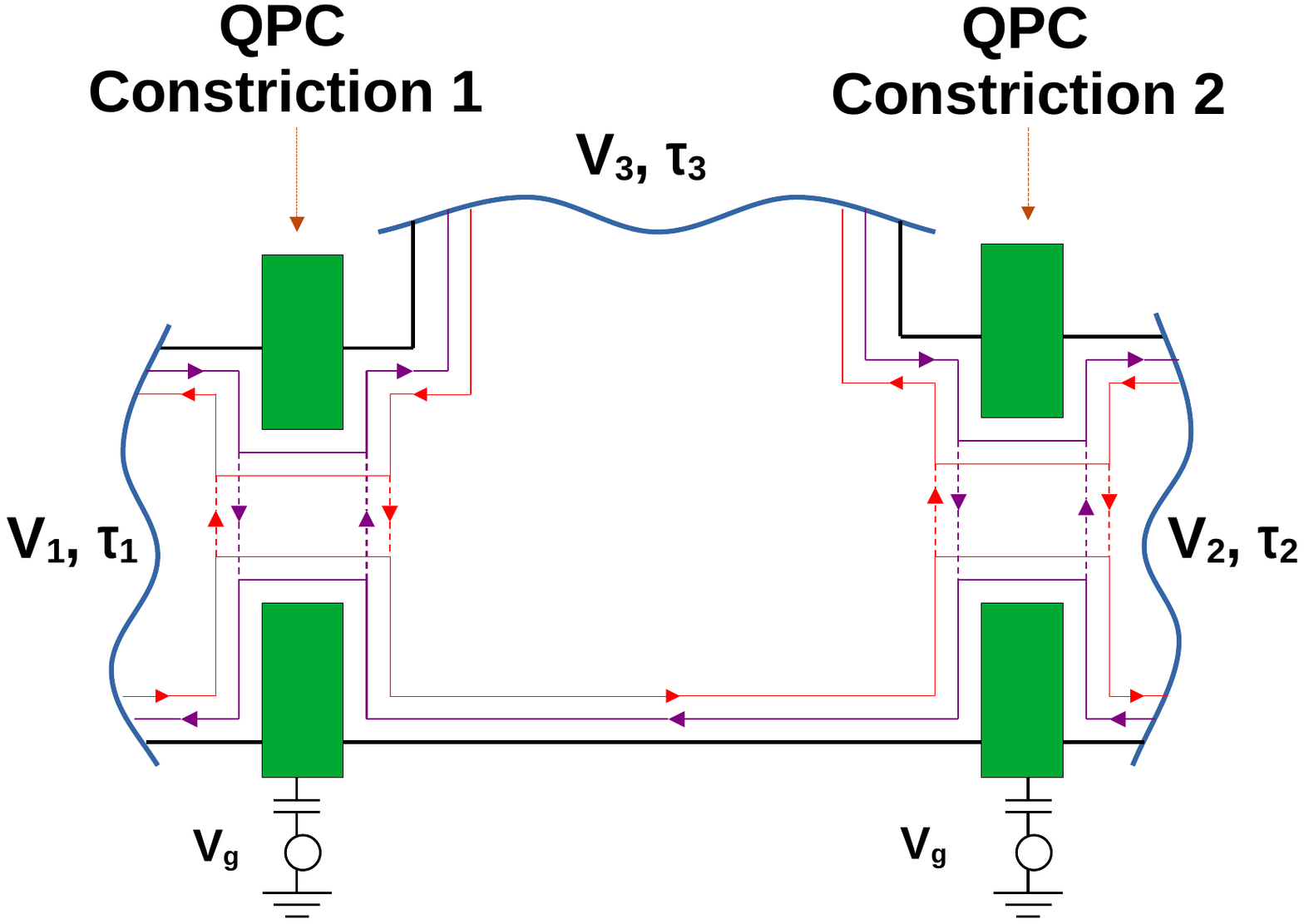}
         \label{fig:3(d)}
         \caption{}
     \end{subfigure}
        \caption{(a) 3T QH sample, (b) 3T QSH sample with a VT probe and two QPC-type constrictions connected to an external gate voltage ($V_g$) in either sample. Purple lines depict electron edge modes in QH samples, while those in QSH samples are shown by purple and red lines for spin-up and spin-down electrons. }
         \label{Fig2}
       \end{figure}

{This letter examines the thermoelectric performance of three-terminal QH and QSH setups using a voltage-temperature (VT) probe in both linear and nonlinear regimes (see Fig. \ref{Fig2}). The VT probe incorporates inelastic scattering processes from electron-electron or electron-phonon interactions, ensuring no net charge ($I_3 = 0$) and heat current ($J_3 = 0$) through terminal 3. This is more accurate than a voltage probe, which insists on a vanishing net charge current \cite{PhysRevE.97.022114, PhysRevLett.114.146801}.
A VT probe ~\cite{das2023majorana} has been used to compare the Lorentz ratio scaling with the Luttinger-liquid model~\cite{PhysRevB.105.L081403}. This letter proposes three-terminal QH and QSH setups with the third terminal as a VT probe, functioning as a highly efficient quantum heat engine (QHE) or quantum refrigerator (QR) in both linear and nonlinear response regimes.}

This letter is divided into two parts. In the first part, we investigate the thermoelectric transport of the QH and QSH setups within the linear response regime and show that efficiency at maximum power approaches Van den Broeck for both QH and QSH setups. In the second part, we focus on non-linear transport, wherein we see that the efficiency at maximum power approaches the Whitney limit for both QH and QSH setups. The technical details of the calculation are presented in the Supplementary Material (SM).

\textit{\underline{Thermoelectricity in QH and QSH setups:}}\label{sec:3}
We start by discussing the general theory for thermoelectric transport in QH as well as QSH setups within the Landaeur-Buttiker scattering theory in both linear and non-linear regimes \cite{BENENTI20171, datta_1995}. The net charge and heat currents in a multiterminal QH sample, see Sec. \ref{Sec:II} of SM,  are given as,

\begin{align} \label{eq1}
\begin{split}
    I_{\alpha} &= \frac{2e}{h}\int_{-\infty}^{\infty}dE \sum_{\beta} \mathcal{T}_{\alpha \beta} (f_{\alpha}(E) - f_{\beta}(E)),\\
    J_{\alpha} &= \frac{2}{h}\int_{-\infty}^{\infty}dE \sum_{\beta} (E - \mu_{\alpha}) \mathcal{T}_{\alpha \beta} (f_{\alpha}(E) - f_{\beta}(E)),
    \end{split}
\end{align}
 where the transmission probability for an electron to scatter from terminal $\beta$ to terminal $\alpha$ is $\mathcal{T}_{\alpha \beta}$. Here, $f_{\alpha (\beta)} = \left(1+e^{(E - \mu_{\alpha (\beta)})/k_B T_{\alpha (\beta)}}\right)^{-1}$, $\mu_{\alpha (\beta)}$ and $T_{\alpha (\beta)}$ are the Fermi function, chemical potential and temperature of terminal $\alpha (\beta)$ respectively. 

Similarly, for QSH setup, with helical edge modes, the currents carry an extra spin degree of freedom and are written as $I_{\alpha} = \sum_{\sigma \in \{\uparrow, \downarrow\}}I_{\alpha}^{\sigma}$ and $J_{\alpha} = \sum_{\sigma \in \{\uparrow, \downarrow\}} J_{\alpha}^{\sigma}$, where \cite{PhysRevE.97.022114, PhysRevB.91.155407}, 
\begin{align} \label{eq2}
\begin{split}
    I_{\alpha}^{\sigma} &= \frac{e}{h}\int_{-\infty}^{\infty}dE \sum_{\beta} \mathcal{T}_{\alpha \beta}^{\sigma} (f_{\alpha}(E) - f_{\beta}(E))],\\
    J_{\alpha}^{\sigma} &= \frac{1}{h}\int_{-\infty}^{\infty}dE (E - \mu_{\alpha}) \sum_{\beta} \mathcal{T}_{\alpha \beta}^{\sigma} (f_{\alpha}(E) - f_{\beta}(E))],
    \end{split}
\end{align}
 with the transmission probability for an electron to transmit from terminal $\beta$ with spin $\rho$ to terminal $\alpha$ with spin $\sigma$ being $\mathcal{T}_{\alpha \beta}^{\sigma} = \sum_{\rho \in \{\uparrow, \downarrow\}} \mathcal{T}_{\alpha \beta}^{\sigma \rho}$. 
In this three-terminal setup (both QH and QSH) as shown in Fig. \ref{Fig2}, $T_1 = T + \tau_1, T_2 = T + \tau_2$ and $T_3 = T + \tau_3$, where $T$ is the reference temperature and $\tau_1, \tau_2$ and $\tau_3$ are the temperature biases applied in terminals 1, 2 and 3 respectively. For our setups, we apply temperature bias only at terminal 1, i.e., $\tau_1 = \tau$ and $\tau_2 = 0$. We study the thermoelectric properties for a QHE, considering terminal 3 to be a VT probe. We also assume both terminals 1 and 2 to be current probes with $V_1 = -V$ and $V_2 = 0$.

\textit{\underline{Quantum point contact vs. boxcar-type transmission:}}  In this section, we discuss the two types of transmission, which we have extensively considered in this work. The first is the quantum point contact (QPC)-type tunneling \cite{PhysRevB.41.7906} and the other one is the boxcar-type \cite{PhysRevLett.112.130601, PhysRevB.91.115425}.
 
\begin{figure} 
\centering
\includegraphics[width=0.60\linewidth]{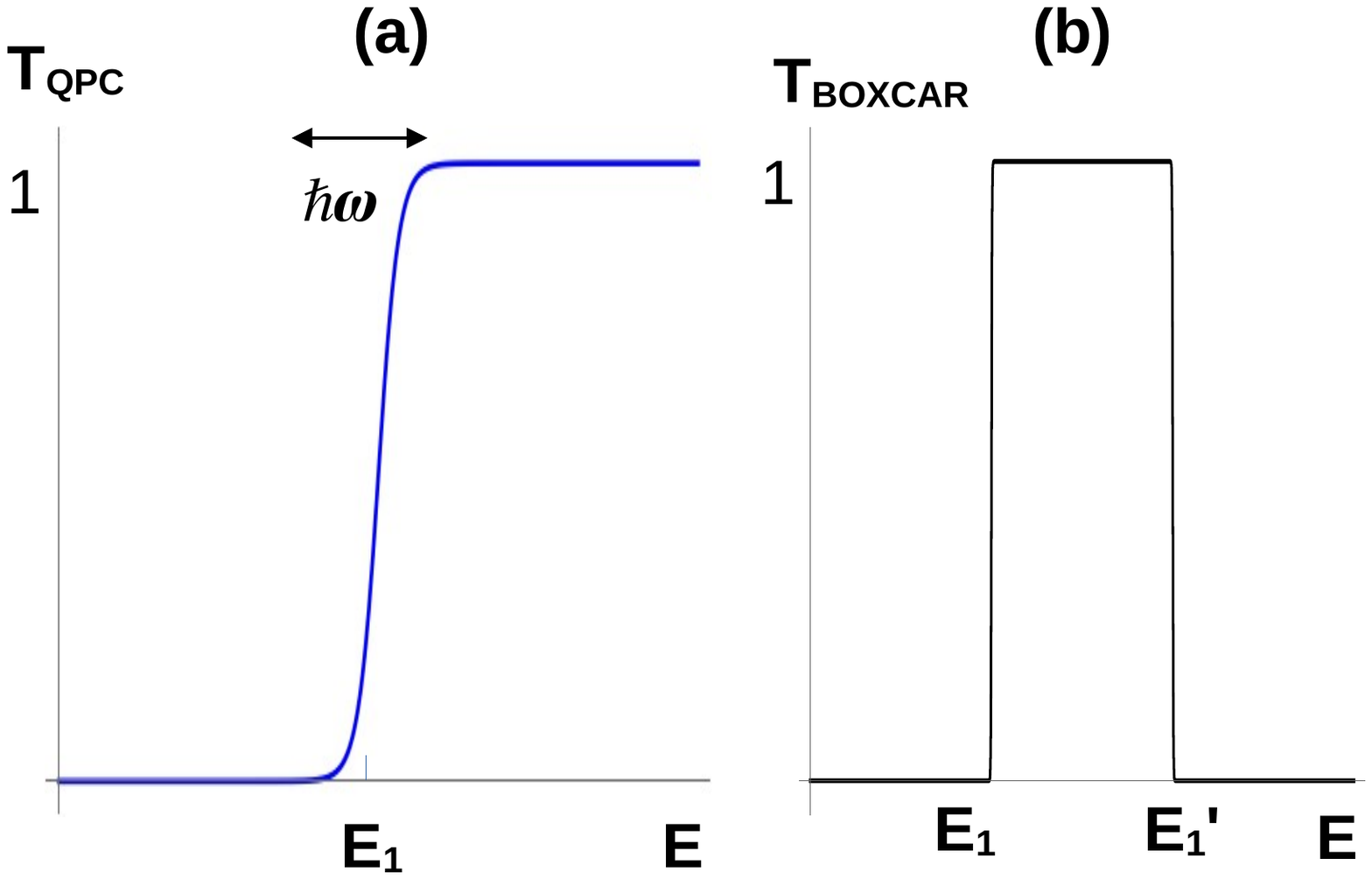}
\caption{(a) QPC type, (b) Boxcar-type transmission.}
\label{fig:1}
\end{figure}

 A quantum point contact can be modeled by a potential
 \begin{equation} \label{eq7}
     V(x,y) = V_0 - \frac{1}{2}m \omega_x^2 x^2 + \frac{1}{2}m \omega_y^2 y^2,
 \end{equation}
 where $m$ is the effective mass of the electron and the electron feels the potential $V(x, y)$ when the QPC is present in a setup. As derived in Ref. \cite{PhysRevB.41.7906}, the transmission probability ($T_{QPC}$) for this constriction is given as
 \begin{equation}\label{eq8}
     T_{QPC} = \sum_{n}\frac{1}{1 + \text{exp}\left[-2\pi(E - E_n)/\hbar \omega_x\right]},
 \end{equation}
where, $E_n = V_0 + \hbar \omega_y \left(n + 1/2\right)$, with $n= 0,1,2,...$. We consider $E_n$ in the 1D limit, where only a few subbands are below the Fermi energy $E_F$. We assume the length along $x$-direction ($l_x$) to be much larger than that of $y$-direction ($l_y$), therefore only $E_n$ is quantized along $y$-direction. one can consider $n = 1$ and $\omega_x, \omega_y = \omega$ and we have the transmission probability ($T_{QPC}$) as,
\begin{equation}\label{eq9}
    T_{QPC} = \frac{1}{1+e^{-2 \pi (E - E_1)/\hbar \omega}},
\end{equation}
 which has been shown in Fig. \ref{fig:1}(a), where
  below a certain energy value $E_1$, it vanishes, and above this, it is always one. Here, $\hbar \omega$ is the width of the step. In our work, we have capacitively connected a gate voltage $V_g$ to the QPC-type constriction, giving $V(x, y) \implies V(x, y) + eV_g$, which implies $E_1 \rightarrow E_1 +  eV_g$. We denote $E_1 + eV_g$ as $E_1''$.

  Similarly, the boxcar-type of transmission is 1 in the energy range $ E_1 < E < E_1'$, otherwise zero \cite{PhysRevLett.112.130601, PhysRevB.91.115425}, See Fig. \ref{fig:1}(b). This transmission function will be used while we discuss Whitney's approach to studying the thermoelectrics of a two-terminal QH and QSH setup. As discussed by Whitney, one can generate a boxcar-type of transmission by considering many quantum dots in a region between two terminals \cite{PhysRevLett.112.130601, PhysRevB.91.115425}. When there is only a single quantum dot with a single energy level, the transmission probability via it will be the resonant-tunneling type \cite{PhysRevLett.68.3468}. With one quantum dot, the electron will hop into it from one terminal and later hop out into the other terminal. But when one increases the number of quantum dots, each having a single energy level, then the electrons hop repeatedly between the quantum dots while going from one terminal to the other, which gives a boxcar-type transmission, which is $1$ in between two arbitrary energy levels $E_1$ and $E_1'$ as shown in Fig. \ref{fig:1}(b). When $E_1' \rightarrow \infty$, the boxcar type transmission becomes QPC-like, always 1 above $E_1$ and zero otherwise.

 However, as discussed in the introduction, implementing the boxcar-type transmission is more difficult than implementing the QPC-type transmission. In fact, QPC was the first type of tunneling used to study the quantum transport in 2DEGs \cite{PhysRevLett.60.848, wharam1988one, van1992thermo, PhysRevLett.68.3765} such as GaAs-AlGaAs heterostructure. Most of the ballistic or edge mode transport can be done much more conveniently with the help of QPC in mesoscopic samples. For the implementation of QPC, one needs to control electrostatically the local electrostatic potential, which is felt by the electron with the help of metallic split gates \cite{PhysRevB.34.5635, PhysRevLett.56.1198, beenakker1991quantum}. Here, in this work, with the help of only the QPC-type of tunneling, one can study the nonlinear transport, which can help reach the Whitney bounds.

\textit{\underline{Linear transport in QH and QSH setups: }}
In the linear response regime for the QH case, i.e., when $\mu_{\alpha} (= eV_{\alpha}) \ll k_B T$ and ($T_{\alpha} - T) \ll T$, one can see that the current (both charge and heat) varies linearly with the voltage and temperature biases applied across the sample  \cite{BENENTI20171}. 
  Since, terminal 3 is a VT probe i.e., $I_3 = J_3 = 0$, the charge and heat currents in terminal 1 can be written as, 
\begin{equation} \label{eq3}
    \begin{pmatrix}
        I_1 \\
        J_1
    \end{pmatrix} = \begin{pmatrix}
        L_{eV} & L_{e T}\\
        L_{hV} & L_{h T}
    \end{pmatrix} \begin{pmatrix}
        -V \\
        \tau
    \end{pmatrix}
\end{equation}
where the Onsager matrix elements are $L_{eV}, L_{e T}, L_{hV}, L_{h T}$, see Eq. (\ref{eq:25}) in SM. These Onsager matrix elements are used to find the transport coefficients such as conductance ($G$), Seebeck coefficient ($S$), Peltier coefficient ($\Pi$), and the thermal conductance ($K$) and are given as,
\begin{align} \label{eq4}
    G &= L_{eV},\quad S = \frac{L_{eT}}{L_{eV}},\quad \Pi = \frac{L_{hV}}{L_{eV}},\quad K = L_{hT} - \frac{L_{hV}L_{eT}}{L_{eV}}.
\end{align}

Using a similar procedure, one can also derive the Onsager matrix elements in the QSH case for charge current $I_{\alpha}^\sigma$ and heat current $J_{\alpha}^\sigma$ by imposing the VT probe condition ($I_3 = \sum_{\sigma \in \{\uparrow,\downarrow \}} I_3^{\sigma}= 0$ and $J_3 = \sum_{\sigma \in \{\uparrow,\downarrow \}} J_3^{\sigma} = 0$) and they can be written in the same form as Eq. (\ref{eq3}) (see, Sec. \ref{Sec:III(A1)} in SM as
\begin{equation} \label{eq5}
    \begin{pmatrix}
        I_1^{\sigma} \\
        J_1^{\sigma}
    \end{pmatrix} = \begin{pmatrix}
        L_{eV}^{\sigma} & L_{e T}^{\sigma} \\
        L_{hV}^{\sigma} & L_{h T}^{\sigma}
    \end{pmatrix} \begin{pmatrix}
        -V \\
        \tau
    \end{pmatrix}
\end{equation}
where the spin polarized Onsager matrix elements $L_{eV}^{\sigma}$, $L_{eT}^{\sigma}$, $L_{hV}^{\sigma}$ and $L_{hT}^{\sigma}$ are derived in  Eq. (\ref{eq:75}) of SM.
Using Eq. (\ref{eq5}), the spin polarized transport coefficients such as Conductance ($G^{\sigma}$), Seebeck Coefficient ($S^{\sigma}$), Peltier Coefficient ($\Pi^{\sigma}$), and thermal conductance ($K^{\sigma}$) are defined as,
\begin{align} \label{eq6}
    G^{\sigma} &= L_{eV}^{\sigma},\quad S^{\sigma} = \frac{L_{eT}^{\sigma}}{L_{eV}^{\sigma}},\quad \Pi^{\sigma} = \frac{L_{hV}^{\sigma}}{L_{eV}^{\sigma}},\quad K^{\sigma} = L_{hT}^{\sigma} - \frac{L_{hV}^{\sigma}L_{eT}^{\sigma}}{L_{eV}^{\sigma}}.
\end{align}

 The output power in terminal 1, $P = I_1 V$ is maximum at a voltage bias $V = L_{eT}\tau/4L_{eV}$ and is given as $P_{max} = L_{eT}^2\tau^2/4L_{eV}$, see Sec. \ref{Sec:II(A2)} for QH and \ref{Sec:III(A2)} for QSH in SM. The efficiency at maximum power, i.e., $\eta|_{P_{max}} = \frac{P_{max}}{J} = \frac{\eta_c}{2}\frac{ZT}{Z T + 2}$ (see, Eq. (\ref{eq:28}) for QH and Eq. (\ref{eq:79}) for QSH in SM, $J = J_1$ is the heat current from the hotter terminal 1. $Z T = \frac{G S^2 T}{K}$ is the figure of merit. Similarly, maximum efficiency, i.e.,  $\eta_{max} = \eta_c \frac{\sqrt{ZT + 1} - 1}{\sqrt{ZT + 1} + 1}$ can be derived (see Eq. (\ref{eq:34}) for QH and Eq. (\ref{eq:83}) for QSH in SM). 

\begin{figure}
     \centering
     \begin{subfigure}[b]{0.235\textwidth}
         \centering
         \includegraphics[width=\textwidth]{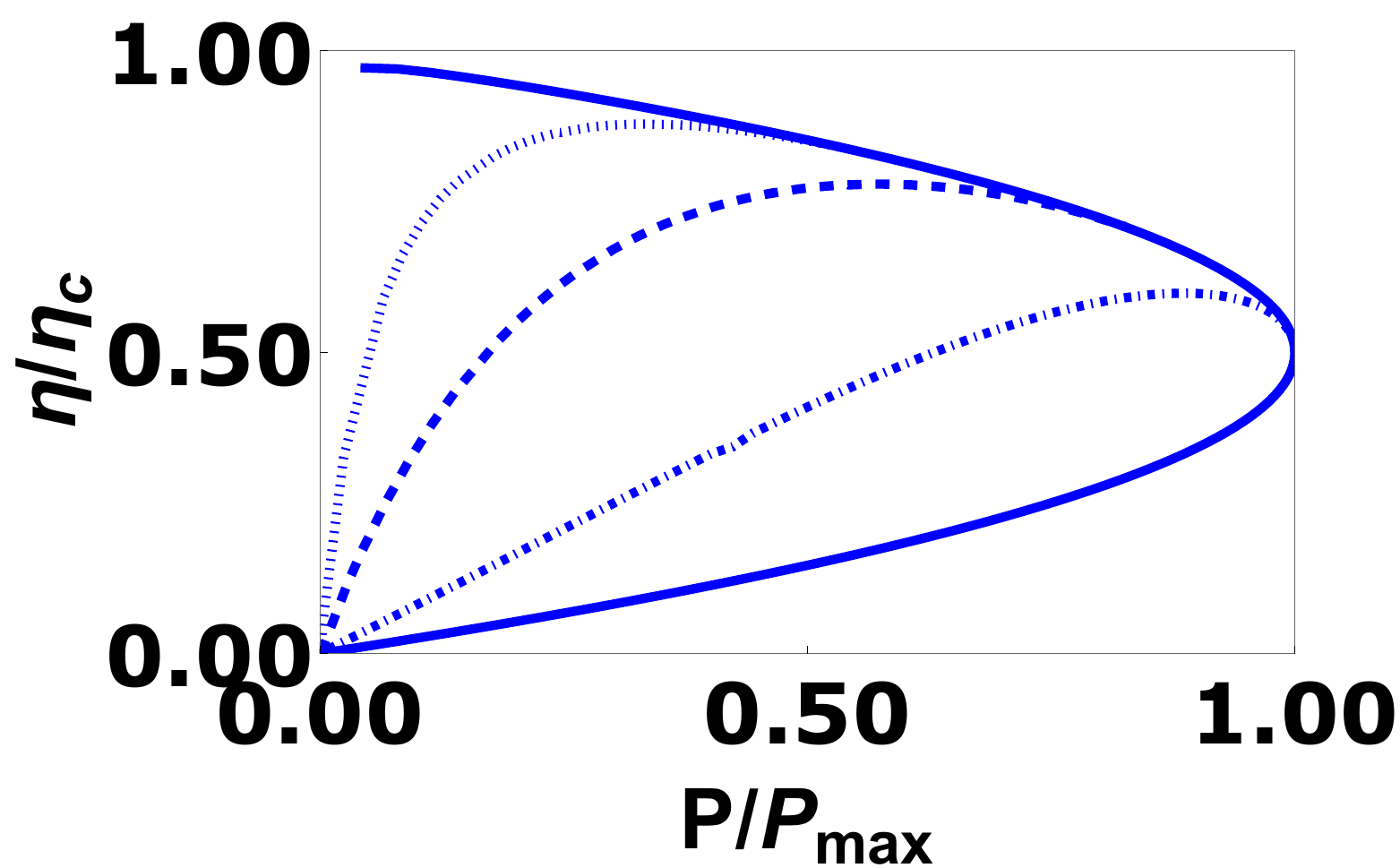}
         \caption{QH heat engine (linear)}
     \end{subfigure}
     \hspace{0.05cm}
     \begin{subfigure}[b]{0.235\textwidth}
         \centering
         \includegraphics[width=\textwidth]{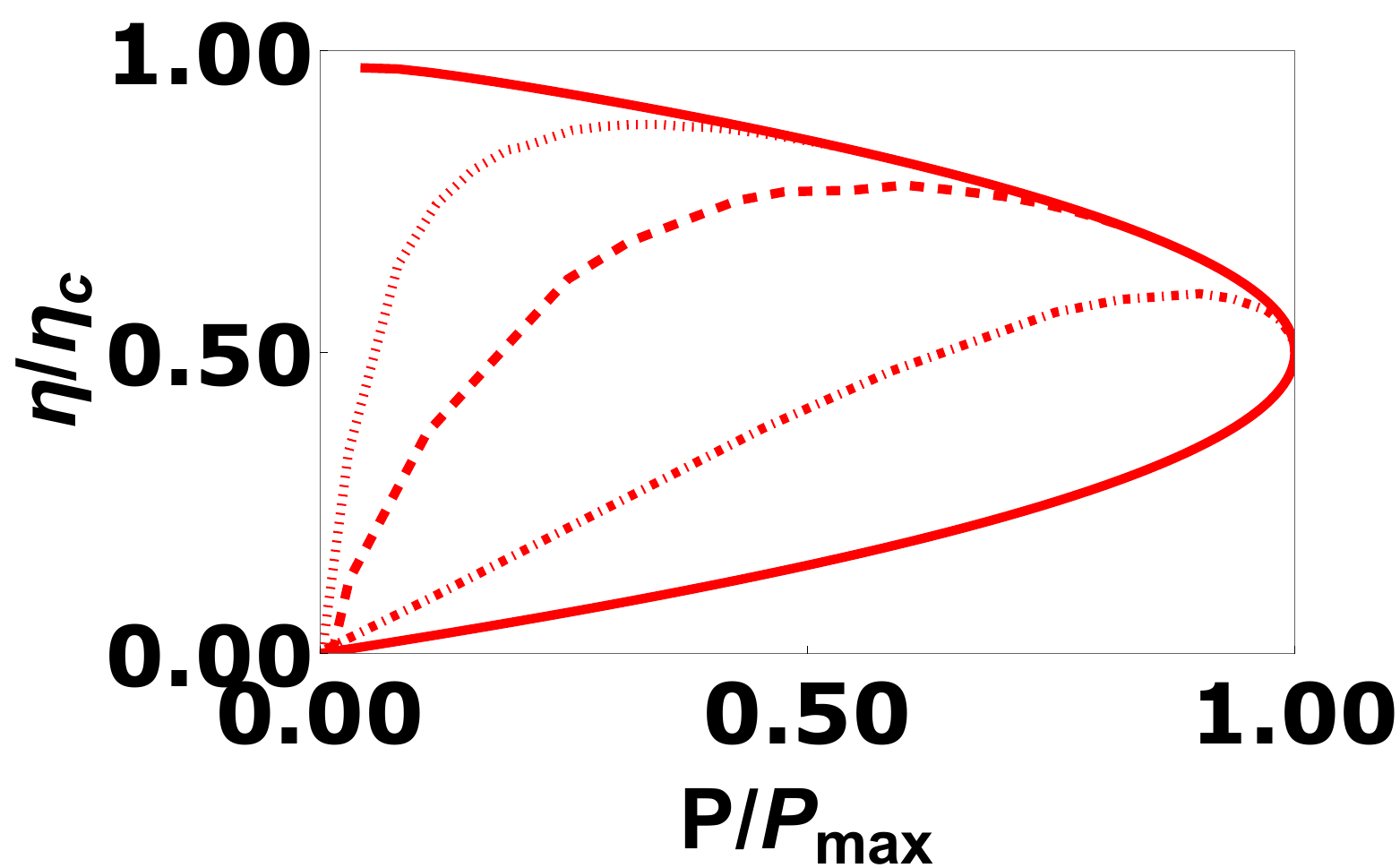}
         \caption{QSH heat engine (linear)}
     \end{subfigure}
     \\
     \begin{subfigure}[b]{0.235\textwidth}
         \centering
         \includegraphics[width=\textwidth]{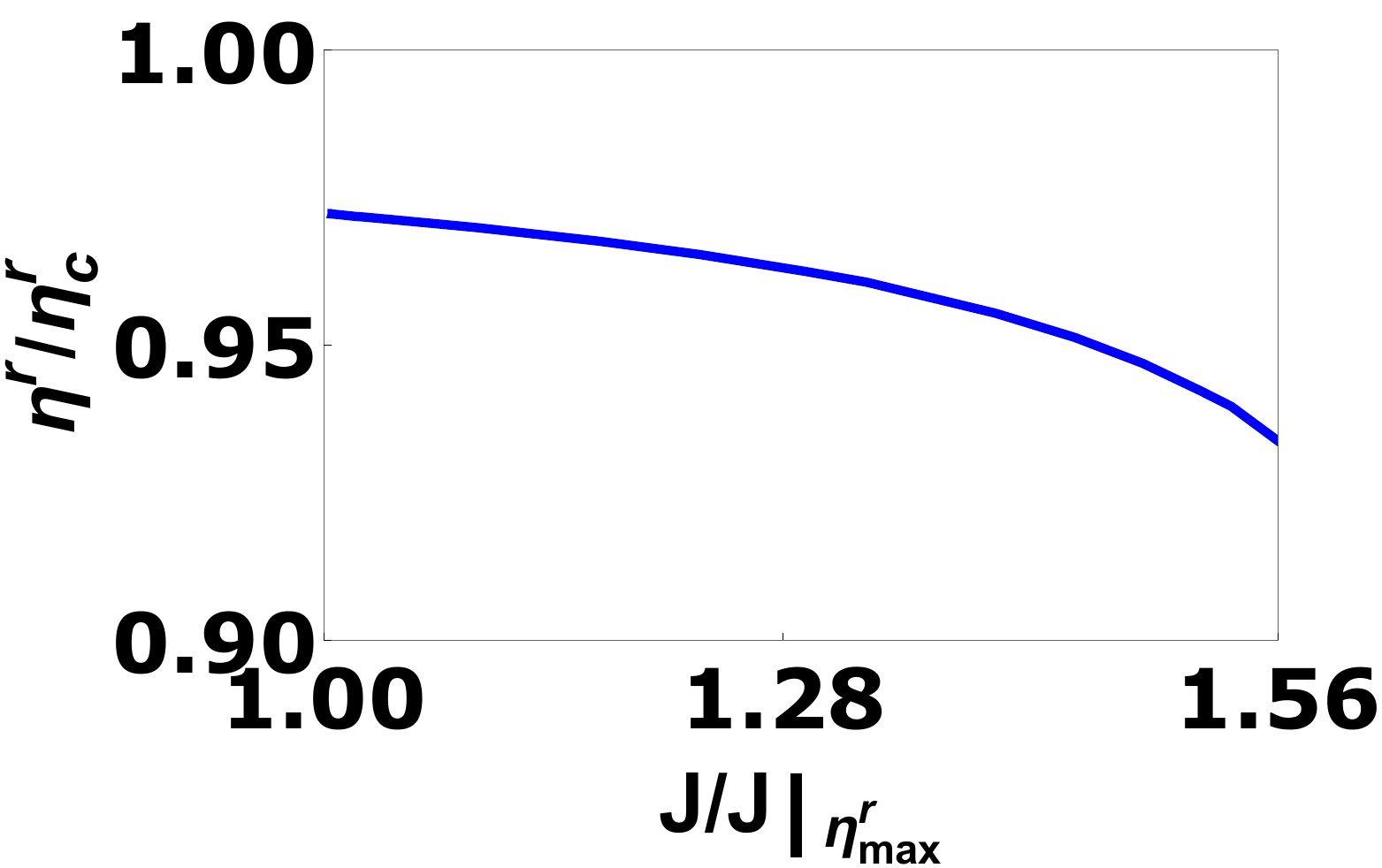}
         \caption{QH refrigerator (linear)}
     \end{subfigure}
     \hspace{0.05cm}
     \begin{subfigure}[b]{0.235\textwidth}
         \centering
         \includegraphics[width=\textwidth]{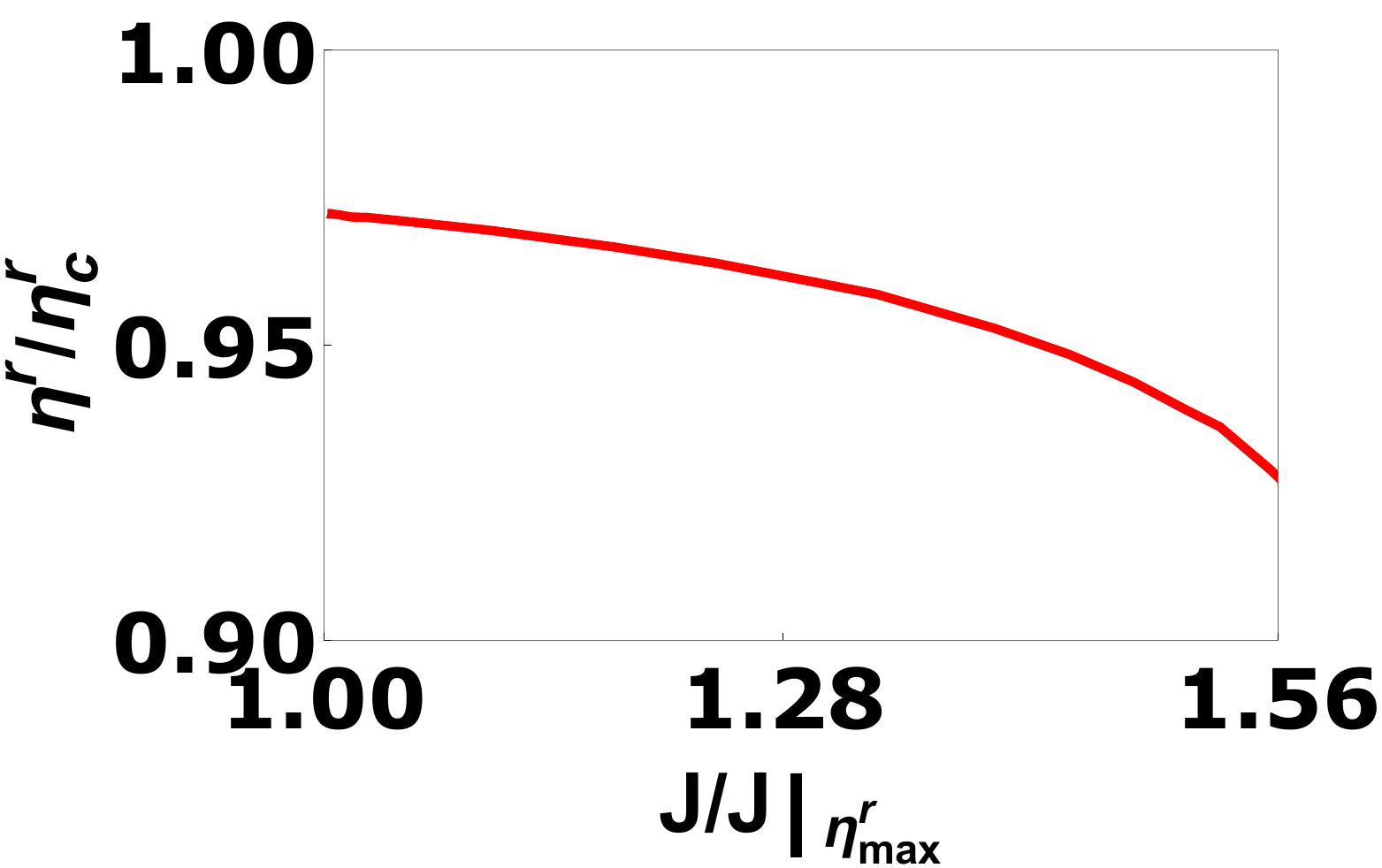}
         \caption{QSH refrigerator (linear)}
     \end{subfigure}
        \caption{Parametric plot of $\frac{\eta}{\eta_c}$ vs $\frac{P}{P_{max}}$ (a) QH, (b) QSH QHE. the blue (red) dot-dashed curve is for $V = 0.057 k_B T/e$, the blue (red) dashed curve for $V = 0.081 k_B T/e$ and the blue (red) dotted curve for $V = 0.096 k_B T/e$. The blue (red) line shows the bound to the power and efficiency. Parametric plot of $\frac{\eta^r}{\eta^r_c}$ vs $\frac{\textbf{J}}{\textbf{J}|_{\eta^r_{max}}}$ for (c) QH and (d) QSH QR, for QH (QSH) QR, blue (red) line is for $V = 0.115 k_B T/e$. Parameters taken are $\omega = 0.1 k_B T / \hbar$, $T_1 = 1.1K, T_2 = 1.0K$, $\mu = 0$, $E_1 = E_2 = k_B T$. }
         \label{Fig3}
       \end{figure}

In Fig. \ref{Fig3}(a, b), we parametrically plot the efficiency ($\eta/\eta_c$) and the power ($P/P_{max}$) in linear response regime for both QH and QSH QHE, calculated using Eqs. (\ref{eq3}-\ref{eq6}) {and observe that it achieves the Van den Broeck limit at $V = 0.057 k_B T/e$ and further increasing $V$, power decreases and efficiency increases (see blue (red) dashed curve for $V = 0.081 k_B T/e$ and blue (red) dotted curve for $V = 0.096 k_B T/e$) }. We considered threshold energies of QPC 1 and QPC 2 to be $E_1'' = E_1 + e V_g$ and $E_2'' = E_2 + e V_g$ respectively, with $E_1 = E_2$. If $E_1 \ne E_2$; the conclusions don't change, see, Secs. \ref{Sec:II(A4)} (\ref{Sec:II(B4)}) for QH and \ref{Sec:III(A3)} (\ref{Sec:II(B4)}) for QSH for the linear (non-linear) transport in SM.

{In Ref. \cite{PhysRevLett.114.146801}, it has been reported that a QH setup cannot work as a QR due to time-reversal symmetry breaking (TRS). Conversely, the same setup with QSH edge modes can act as both QHE and QR because TRS is preserved \cite{PhysRevE.97.022114}. In the QH setup as in \cite{PhysRevLett.114.146801}, the asymmetric parameter (AP), i.e., the Seebeck to Peltier coefficient ratio, is zero or infinity, affecting its refrigeration capability \cite{PhysRevLett.110.070603, PhysRevE.97.022114}. In the QSH setup, AP is always finite since TRS is preserved \cite{PhysRevE.97.022114}. However, this is not true for all QH setups. The QH setup in Fig. \ref{Fig3}(a) and the QSH setups in Fig. \ref{Fig3}(b) can act as both QHE and QR because their AP is finite (see Sec. \ref{Sec:II(A4)} of SM).}

For QR, we can calculate the cooling power at the maximum coefficient of performance ($\textbf{J}|_{\eta^r_{max}}$) and the maximum coefficient of performance ($\eta_{max}^r$) for both QH and QSH setups. The coefficient of performance (COP = $\textbf{J}/P$) is generally defined as the ratio of heat extracted from the cooler terminal ($\textbf{J} = -J_2$) to the power absorbed ($P$) by the setup. The maximum COP ($\eta^r_{max}$) is achieved using the condition $\frac{d(COP)}{dV} = 0$ and is given as $\eta^r_{max} = \eta^r_c \frac{\sqrt{Z T + 1} - 1}{\sqrt{Z T + 1} + 1}$, where $\eta^r_c$ is the Carnot coefficient of performance. As shown in Fig. \ref{Fig3}(c), $\eta^r_{max}$ reaches $0.97 \eta_c^r$ at $\textbf{J}|_{\eta^r_{max}}$ {at $V = 0.115 k_B T/e$}. Similar to the {QH setup}, we parametrically plot the cooling power $\textbf{J}$ vs coefficient of performance $\eta^r$ in Fig. \ref{Fig3}(d) for QSH setup.

 As shown in Sec. \ref{Sec:II(A2)}, for QH and Sec. \ref{Sec:III(A2)} for QSH setups in the SM both $\eta_{max}$ and $\eta|_{P_{max}}$ rely on $Z T$ for a QHE. Similarly, for QR, $\eta^r_{max}$ depend on $Z T$. To achieve Van den Broeck efficiency ($\frac{\eta_c}{2}$), $Z T$ must significantly exceed 1. The key lies in having a large Seebeck coefficient ($S$) and a small thermal conductance ($K$), both occurring in the same configuration. In both of our QH and QSH setups, we find that the Seebeck coefficient is notably large, while the thermal conductivity is remarkably small.

\begin{figure}
     \centering
     \begin{subfigure}[b]{0.235\textwidth}
         \centering
         \includegraphics[width=\textwidth]{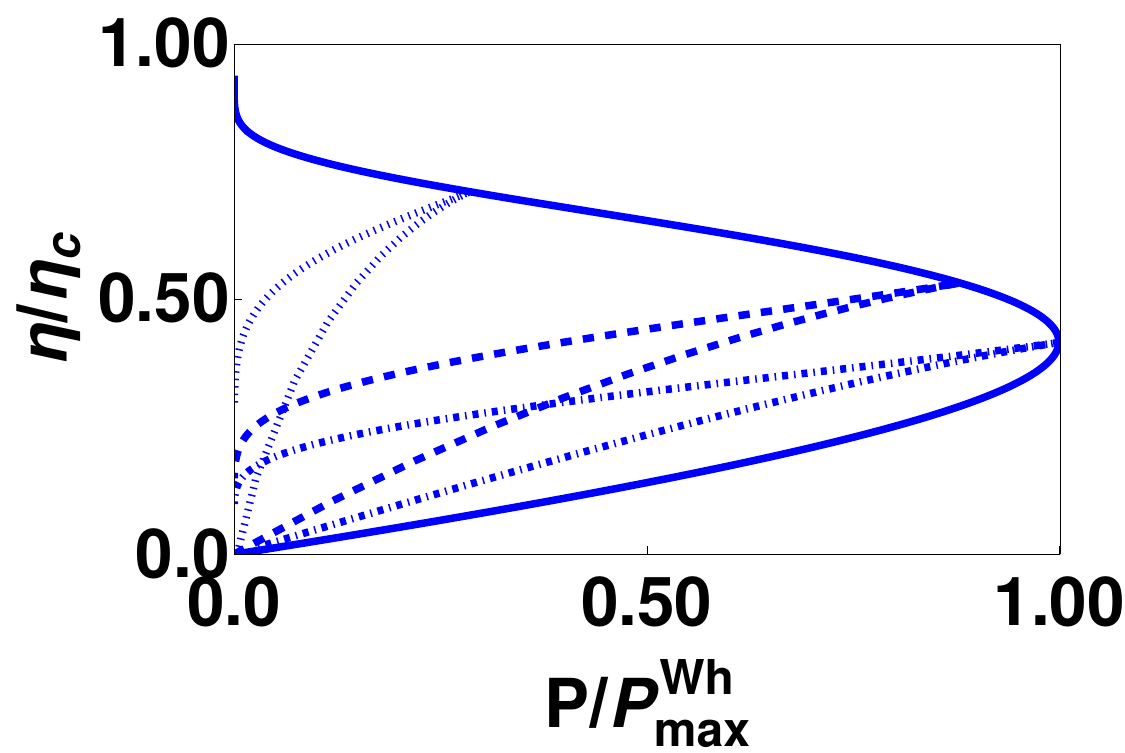}
         \caption{QH heat engine (nonlinear)}
     \end{subfigure}
     \hspace{0.05cm}
     \begin{subfigure}[b]{0.235\textwidth}
         \centering
         \includegraphics[width=\textwidth]{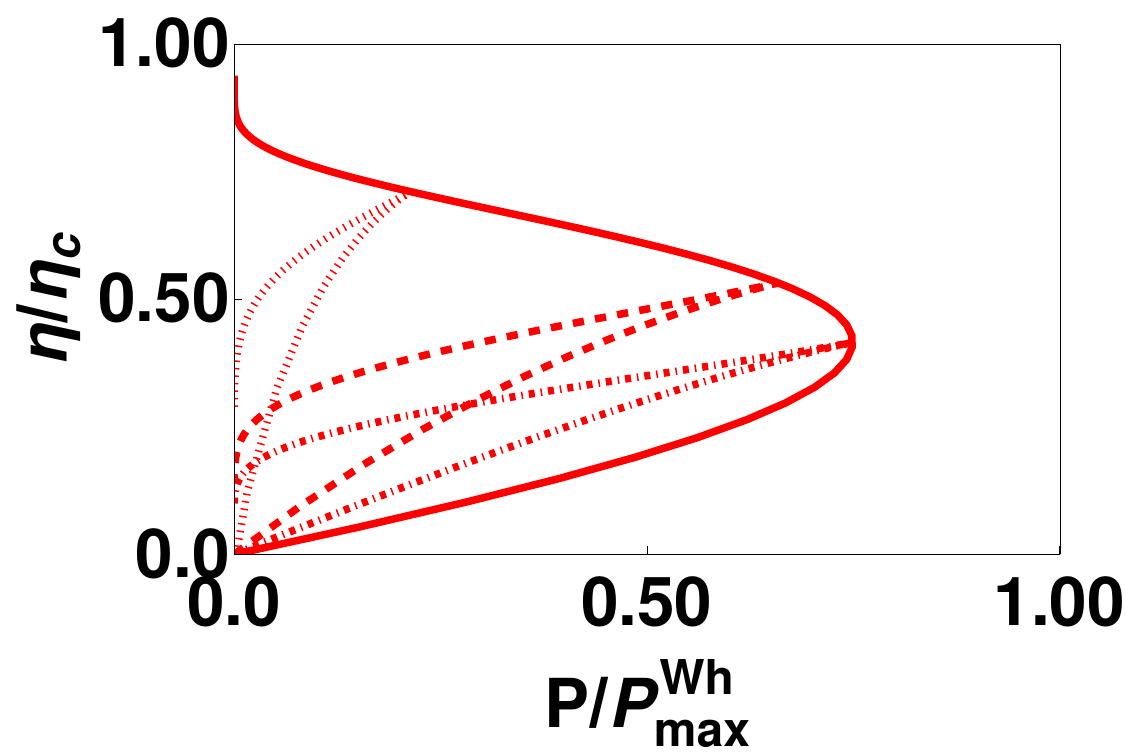}
         \caption{QSH heat engine (nonlinear)}
     \end{subfigure}
     \\
     \begin{subfigure}[b]{0.235\textwidth}
         \centering
         \includegraphics[width=\textwidth]{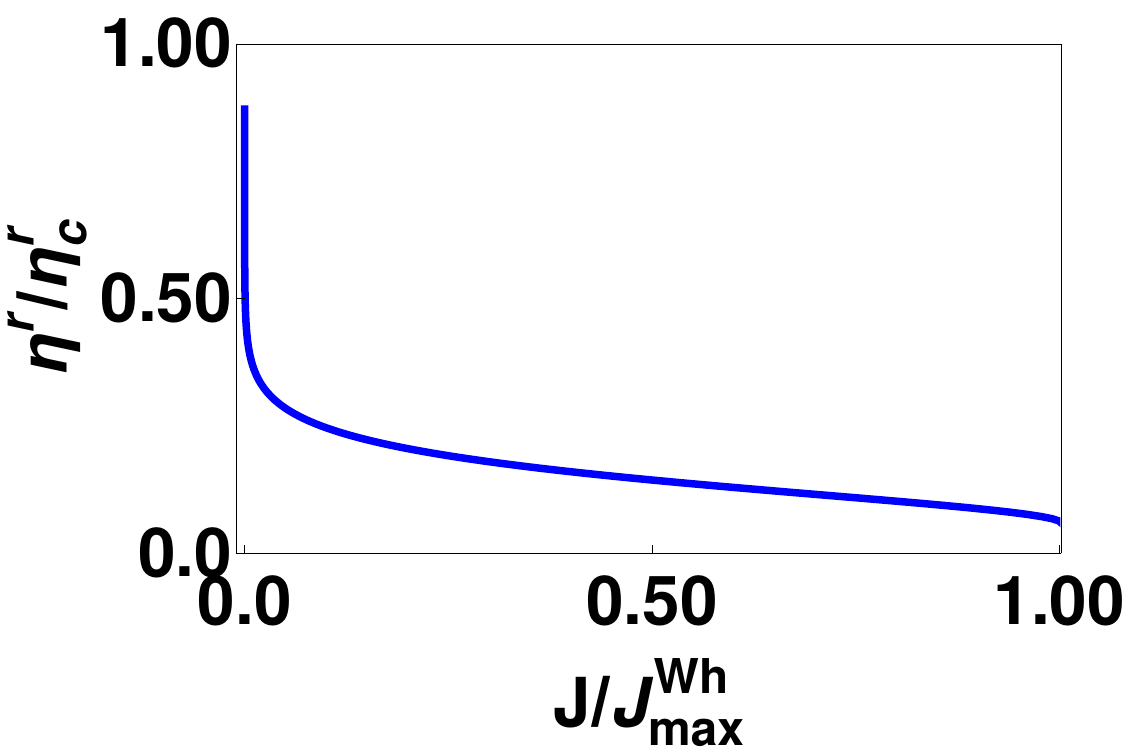}
         \caption{QSH refrigerator (nonlinear)}
     \end{subfigure}
     \hspace{0.05cm}
     \begin{subfigure}[b]{0.235\textwidth}
         \centering
         \includegraphics[width=\textwidth]{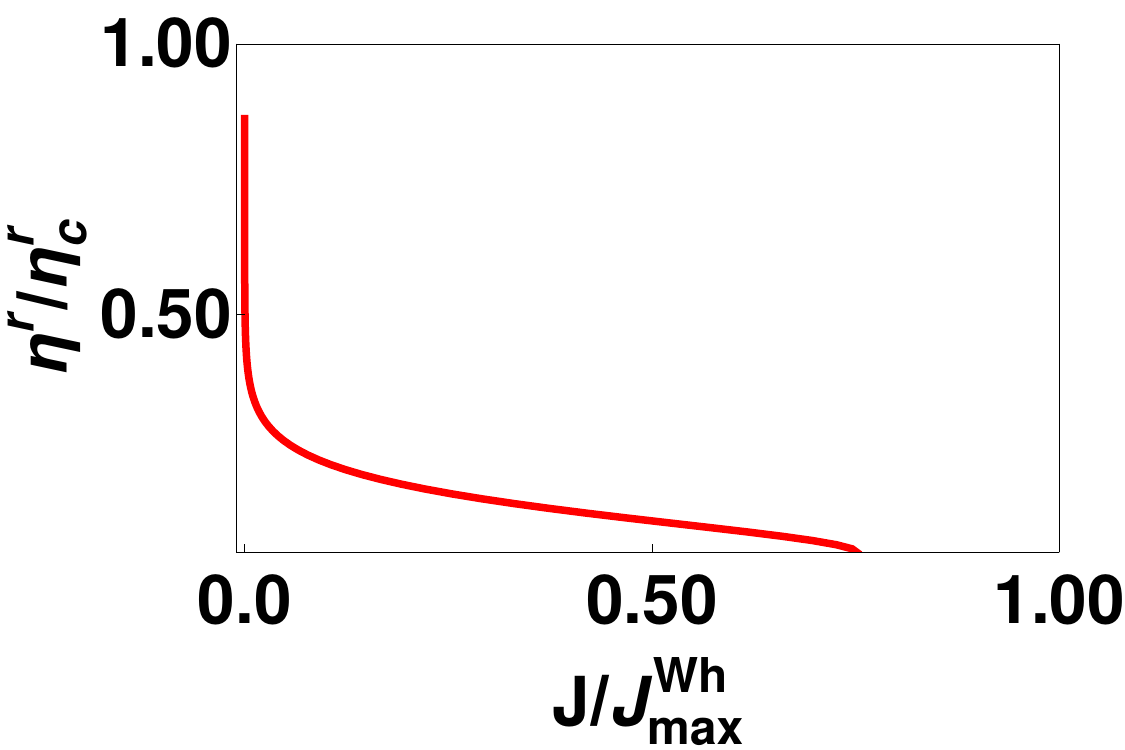}
         \caption{QSH refrigerator (nonlinear)}
     \end{subfigure}
        \caption{
Parametric plot of $\frac{\eta}{\eta_c}$ vs $\frac{P}{P_{max}^{Wh}}$ for (a) QH and (b) QSH, QHE. For QH (QSH) QHE, the blue (red) dot-dashed curve is for $V = 1.14 k_B T/e$, the blue (red) dashed curve for $V = 1.77 k_B T/e$ and the blue (red) dotted curve for $V = 3.7 k_B T/e$. The blue (red) line shows the bound to the power and efficiency. Parametric plot for $\frac{\eta^r}{\eta^r_c}$ vs $\frac{\textbf{J}}{\textbf{J}^{Wh}_{max}}$ for (c) QH and (d) QSH, QR.
         For QH (QSH) QR, blue (red) line is for $V = 20k_B T/e$. Parameters taken are $\omega = \frac{0.1 k_B T}{h}$, $T_1 = 2K, T_2 = 1K$, $\mu = 0$, $E_1 = E_2 = k_B T$.}
        \label{Fig4}
\end{figure}

  \textit{\underline{Non-linear Transport in QH and QSH Setups:}}
There has been ongoing work in this direction on nonlinear thermoelectric transport \cite{PhysRevLett.112.130601, PhysRevB.91.115425, PhysRevLett.93.106802, PhysRevB.90.115301, PhysRevB.99.245304, buttiker1993capacitance, christen1996gauge, PhysRevLett.110.026804, PhysRevB.88.045129, Meair_2013} in both QH and QSH setups. In a two-terminal QH setup, significant deviations from the Onsager-Casimir relation have been observed \cite{PhysRevLett.93.106802} {with a similar investigation done in QSH setup \cite{PhysRevB.90.115301}}. In a three-terminal QH setup \cite{PhysRevB.99.245304} with a voltage probe, efficient operation as a QR in the weakly non-linear regime has been demonstrated. Additionally, a normal metal Aharonov-Bohm heat engine exhibits enhanced thermoelectric performance in the nonlinear regime \cite{Haack_2021}, surpassing its linear regime counterpart \cite{PhysRevB.100.235442}.

  In the linear response regime, electron transport depends solely on the electron's kinetic energy. However, in the nonlinear regime, it relies on both the kinetic energy and the interaction potential ($U'$) of the sample where the electrons flow \cite{buttiker1993capacitance, christen1996gauge, PhysRevLett.110.026804, PhysRevB.88.045129, Meair_2013, PhysRevLett.77.143}. It implies that whereas in the linear regime, the QPC energy level $E_l$ shifts via the gate voltage: $E_l \longrightarrow E_l + e V_g$, in the nonlinear regime, the interaction potential $U'$ shifts the QPC energy level $E_l$ as $E_l \longrightarrow E_l + e U$, with $U = U' + u_g V_g$, where $u_g$ is the characteristic potential determining the response of $U$ to the gate voltage $V_g$. For the form of $U$, see Eq. (\ref{eq:48}) of SM. Introducing a gate voltage helps nullify the interaction potential ($U'$) in the nonlinear regime, as explained in Sec. \ref{Sec:II(B2)} of SM and Ref. \cite{Haack_2021}. From discussion in Sec. \ref{Sec:II(B2)} of SM, the gate voltage must be of the form $V_g = (-U' + V_g')/u_g$, where the first term inside the parentheses nullifies the interaction potential $U'$, while the second term raises the energy level $E_l$: $E_l \longrightarrow E_l + e V_g'$ for the nonlinear response regime. {We have given an estimation for the interaction potentials in two-terminal QH and three-terminal QH setups with a voltage-temperature probe in SM (see, Tables \ref{Table1} and \ref{Table2} of SM for different applied biases. Similarly, we provide an estimate of the interaction potential for two-terminal QSH and three-terminal QSH setups with voltage-temperature probe in SM (Tables \ref{Table3} and \ref{Table4} of SM.}

For nonlinear transport, we start with Eq. (\ref{eq1}) for QH and (\ref{eq2}) for QSH, respectively. Unlike linear response, solving this analytically is challenging, but numerical solutions are possible. In {the Mathematica code} \cite{My_mathematica_notebook}, for QHE, we numerically solve for $V_3$ and $\tau_3$ at a given $V$ and $\tau$, satisfying the VT probe condition ($I_3 = J_3 = 0$), then use these values to determine the current $I_1$ and $J_1$ (via Eqs. (\ref{eq1}) and (\ref{eq2}) for QH and QSH respectively), and finally compute power $P = V I_1$ and efficiency $\eta = \frac{P}{J_1}$. Similarly, for QR, we determine the cooling power $\textbf{J} = -J_2$ and the coefficient of performance $\eta^r = \frac{\textbf{J}}{P}$, where $P$ is the power absorbed by the setup.

In QH setups, as shown in Refs. \cite{PhysRevLett.112.130601, PhysRevB.91.115425}, there exists a bound to the maximum power ($P_{max} = P_{max}^{Wh}$), given by $0.0642 \pi^2 N k_B^2 (T_1 - T_2)^2 / h$. Similarly, the bound to the efficiency at maximum power ($\eta|_{P_{max}}$) is found at $P_{max}^{Wh}$, yielding $\eta|{P_{max}^{Wh}} = \eta_c / (1 + 0.936 (1 + T_2/T_1))$, where $T_1$ and $T_2$ are the temperatures of the hot and cold reservoirs, and $N$ is the number of edge modes. In our case, with $T_1 = 2K$, $T_1 = 1K$, and $N = 1$, we have $P_{max}^{Wh} = 0.632 k_B^2/h$ and $\eta|_{P_{max}^{Wh}} = 0.41 \eta_c$. {For power output $P \ll P_{max}^{Wh}$}, the bound to $\eta_{max}$ is given by $\eta|_{max}^{Wh} = \eta_c \left(1 - 0.478 \sqrt{T_2 P/T_1 P_{max}^{Wh}}\right)$. These quantities were derived for a two-terminal QH setup with a constriction and boxcar-type transmission (see, Sec. \ref{Sec:II(B1)} of SM and also Refs. \cite{PhysRevLett.112.130601, PhysRevB.91.115425}). We extend the same approach to a two-terminal QSH setup and prove that the bounds to $P_{max}$, $\eta|_{P_{max}}$, and $\eta_{max}$ are exactly similar to the QH case (see, Sec. \ref{Sec:II(B1)} of SM).

For a QR, as shown in Refs. \cite{PhysRevLett.112.130601, PhysRevB.91.115425}, the maximum cooling power ($J_{max}$) within the nonlinear regime is bounded by $J_{max}^{Wh}$, given by $\pi^2 N k_B^2 T_1^2/6h$, valid for the QH setup. The coefficient of performance ($\eta^r$) vanishes at {$J_{max}^{Wh}$}. Similarly, for cooling power $\textbf{J} \ll$ $J_{max}^{Wh}$, $\eta^r_{max}$ approaches $\left(\frac{T_1}{T_2} - 1\right)^{-1} \left(1 - 1.09 \sqrt{T_1 \textbf{J}/(T_1 - T_2)J_{max}^{Wh}}\right)$. Extending this approach to the QSH, we find that the bounds to $J_{max}$ and $\eta^r$ are the same as in the QH case (see, Sec. \ref{Sec:III(B1)} of SM).

\begin{table}[h!]
\centering
\caption{Thermoelectric performance of the QH and QSH setup in the linear and nonlinear response regime. $P$ is in units of $P_{max}$ for linear, and $P_{max}^{Wh}$ for nonlinear, $\textbf{J}$ is in units of $\textbf{J}|_{\eta^r_{max}}$ for linear and $J_{max}^{Wh}$ for nonlinear. }
\label{Table5}
\scalebox{0.90}{\begin{tabular}{ll|lll|ll|}
\cline{3-7}
                                                 &       & \multicolumn{3}{l|}{QHE}                                       & \multicolumn{2}{l|}{QR}          \\ \hline
\multicolumn{1}{|l|}{Regime}                     & Setup & \multicolumn{1}{l|}{$P$}    & \multicolumn{1}{l|}{$\frac{\eta|_{P_{max}}}{\eta_c}$} & $\frac{\eta_{max}}{\eta_c}$  & \multicolumn{1}{l|}{$\textbf{J}$}    & $\frac{\eta^r_{max}}{\eta^r_c}$ \\ \hline
\multicolumn{1}{|l|}{\multirow{2}{*}{Linear}}    & QH    & \multicolumn{1}{l|}{\begin{tabular}[c]{@{}l@{}}{1}\\ {[Fig. \ref{Fig3}(a)]}\end{tabular}}   & \multicolumn{1}{l|}{\begin{tabular}[c]{@{}l@{}}{0.50}\\ {[Fig. \ref{Fig3}(a)]}\end{tabular}}   & {\begin{tabular}[c]{@{}l@{}}{0.97}\\ {[Fig. \ref{Fig3}(a)]}\end{tabular}} & \multicolumn{1}{l|}{\begin{tabular}[c]{@{}l@{}}{1}\\ {[Fig. \ref{Fig3}(c)]}\end{tabular}}   & {\begin{tabular}[c]{@{}l@{}}{0.97}\\ {[Fig. \ref{Fig3}(c)]}\end{tabular}} \\ \cline{2-7} 
\multicolumn{1}{|l|}{}                           & QSH   & \multicolumn{1}{l|}{\begin{tabular}[c]{@{}l@{}}{1}\\ {[Fig. \ref{Fig3}(b)]}\end{tabular}}    & \multicolumn{1}{l|}{\begin{tabular}[c]{@{}l@{}}{0.50}\\ {[Fig. \ref{Fig3}(b)]}\end{tabular}}   & {\begin{tabular}[c]{@{}l@{}}{0.97}\\ {[Fig. \ref{Fig3}(b)]}\end{tabular}} & \multicolumn{1}{l|}{\begin{tabular}[c]{@{}l@{}}{1}\\ {[Fig. \ref{Fig3}(d)]}\end{tabular}}    & {\begin{tabular}[c]{@{}l@{}}{0.97}\\ {[Fig. \ref{Fig3}(d)]}\end{tabular}} \\ \hline
\multicolumn{1}{|l|}{\multirow{2}{*}{Nonlinear}} & QH    & \multicolumn{1}{l|}{\begin{tabular}[c]{@{}l@{}}{1}\\ {[Fig. \ref{Fig4}(a)]}\end{tabular}}    & \multicolumn{1}{l|}{\begin{tabular}[c]{@{}l@{}}{0.41}\\ {[Fig. \ref{Fig4}(a)]}\end{tabular}}   & {\begin{tabular}[c]{@{}l@{}}{0.93}\\ {[Fig. \ref{Fig4}(a)]}\end{tabular}} & \multicolumn{1}{l|}{\begin{tabular}[c]{@{}l@{}}{1}\\ {[Fig. \ref{Fig4}(c)]}\end{tabular}}    & {\begin{tabular}[c]{@{}l@{}}{0.86}\\ {[Fig. \ref{Fig4}(c)]}\end{tabular}} \\ \cline{2-7} 
\multicolumn{1}{|l|}{}                           & QSH   & \multicolumn{1}{l|}{\begin{tabular}[c]{@{}l@{}}{0.76}\\ {[Fig. \ref{Fig4}(b)]}\end{tabular}} & \multicolumn{1}{l|}{\begin{tabular}[c]{@{}l@{}}{0.41}\\ {[Fig. \ref{Fig4}(b)]}\end{tabular}}   & {\begin{tabular}[c]{@{}l@{}}{0.93}\\ {[Fig. \ref{Fig4}(b)]}\end{tabular}} & \multicolumn{1}{l|}{\begin{tabular}[c]{@{}l@{}}{0.75}\\ {[Fig. \ref{Fig4}(d)]}\end{tabular}} & {\begin{tabular}[c]{@{}l@{}}{0.86}\\ {[Fig. \ref{Fig4}(d)]}\end{tabular}} \\ \hline
\end{tabular}}
\end{table}

Adjusting the gate voltage to nullify the interaction potential shifts the energy level of the QPC-like constriction. This is explained in Sec. \ref{Sec:II(B2)} of SM, initially demonstrated in a two-terminal nonlinear QH setup and extended to a three-terminal configuration (Sec. \ref{Sec:II(B4)} of SM. A similar approach is applied to a two-terminal QSH setup in Sec. \ref{Sec:III(B2)} of SM{, later extended to a three-terminal configuration (Sec. \ref{Sec:III(B4)} of SM.} We examine {$P_{max}$, $\eta|_{P_{max}}$ and $\eta_{max}$} numerically. In Fig. \ref{Fig4}(a), for the QH setup, maximum power approaches the Whitney limit at a specific voltage bias $V = 1.14 k_B T/e$ (blue dot-dashed curve). With increasing voltage bias $V$, power decreases and efficiency increases {(see blue dashed line for $V = 1.77 k_B T/e$ and blue dotted for $V = 3.7 k_B T/e$ in Fig. \ref{Fig4}(a)).} The QHE is more efficient as $\eta|_{P_{max}}$ and $\eta_{max}$ reach $0.41 \eta_c$ and $0.93 \eta_c$ respectively. Similarly, in Fig. \ref{Fig4}(c), {$J_{max}$} reaches $J_{max}^{Wh}$, while the {$\eta_{max}^r$} is approximately 0.86 $\eta_c^r$. Details for generating the plots in Fig. \ref{Fig4}(a) and (c) are provided in Sec. \ref{Sec:II(B4)} of S/M.
  
  In the case of QSH {at $V = 1.14 k_B T/e$}, the maximum possible power achieved is less than the Whitney limit, but the efficiency at this power is $0.41 \eta_c$ (see, Fig. \ref{Fig4}(b)) {and the power decreases and efficiency increases as one increases the voltage bias further, see red dashed curve in Fig. \ref{Fig4}(b) for $V = 1.77k_B T/e$ and red dotted curve in Fig. \ref{Fig4}(b) for $V = 3.7k_B T/e$.} {$\eta_{max}$} approaches $0.93\eta_c$ similar to the chiral case (see, Fig. \ref{Fig4}(b)). In Fig. \ref{Fig4}(d), we observe that the maximum cooling power is less than the Whitney limit, whereas {$\eta^r$} is around 86$\%$ of $\eta_c^r$. The method to obtain the plots in Fig. \ref{Fig4}(b), (d) are discussed in Sec. \ref{Sec:III(B4)} of SM.

\textit{\underline{Experimental Realization and Conclusion}}: In three-terminal QH and QSH setups with QPC-like constrictions, we achieve optimal thermoelectric performance approaching efficiency benchmarks for QHE and coefficient of performance for QR, both in linear (Fig. \ref{Fig3}) and nonlinear (Fig. \ref{Fig4}) response regimes. For nonlinear response, maximum power generated approaches Whitney bound $P_{max}^{Wh}$ (see Fig. \ref{Fig4}(a), Table \ref{Table5}) and maximum cooling power approaches $J_{max}^{Wh}$ (see Fig. \ref{Fig4}(c), Table \ref{Table5}) for QH setup. In QSH, maximum power as QHE decreases from $P_{max}^{Wh}$, yet efficiency at maximum power remains $0.41 \eta_c$ (Van den Broeck efficiency equivalent) (see Fig. \ref{Fig4}(b), Table \ref{Table5}). Similarly, as QR, maximum cooling power is below $J_{max}^{Wh}$ (Fig. \ref{Fig4}(d), Table \ref{Table5}). This power suppression (both $P_{max}^{Wh}$ and $J_{max}^{Wh}$) arises as a result of helical edge mode transport with a VT probe as there is a finite spin current in three-terminal QSH setup, which is absent in chiral case. Designing a VT probe experimentally is slightly challenging due to potential heat flow between reservoir and surroundings, altering its temperature \cite{BENENTI20171}. 

In the linear response regime, an optimal quantum thermoelectric material is characterized by a figure of merit ($ZT$) significantly greater than 1. Theoretical proposals report $ZT$ values of 4 in ferromagnet-superconductor junctions (Ref. \cite{PhysRevLett.112.057001}) and up to 40-50 in other studies (Refs. \cite{PhysRevLett.114.067001}, \cite{mani2019designing}). Materials with $ZT \gg 1$ are highly desirable for optimal thermoelectric performance as both quantum heat engines and refrigerators. {To summarize, the main take home messages of this letter are: 1. Use of QPC instead of boxcar type transmission \cite{beenakker1991quantum, van1992thermo, PhysRevLett.60.848, PhysRevLett.68.3765}, which is much easier to implement, 2. Investigating thermoelectric transport in the non-linear response regime for QSH setups, which has never been attempted, 3. Investigation of QH and QSH setups as QHE and QR with a voltage-temperature probe in the non-linear response regime, which has again never been explored before and finally 4. An estimate of interaction potential and the applied gate voltage to nullify it in 2T QSH setup (see, Sec. \ref{Sec:III(B2)} of SMand 3T QH (see, Sec. \ref{Sec:II(B4)} of SM and QSH setups (see, Sec. \ref{Sec:III(B4)} of SM with VTP for the first time.}

Quantum heat engines and quantum refrigerators in our work can significantly impact energy conversion and cryogenic cooling. Cooling cryogenic devices to millikelvin temperatures is vital for studying quantum coherence and phase transitions in nano-sized materials (Refs. \cite{BENENTI20171, RevModPhys.78.217, Muhonen_2012}). Traditionally, cryogenics focused on cooling phonons in solids, but quantum refrigeration now allows more effective electron cooling (Ref. \cite{BENENTI20171}). Our work achieves maximum cooling power, reaching the Whitney bound, implying cooling to the lowest possible temperatures. 

In conclusion, this letter proposes achieving the Van den Broeck (VDB) limit in linear and the corresponding Whitney bound in the nonlinear response regimes through a setup applicable to three-terminal QH and QSH systems, extendable to multiterminal setups with multiple VT probes. Furthermore, this approach can be applied to various edge mode systems, including trivial helical \cite{Nichele_2016, mani2017probing, PhysRevB.108.115301} and anomalous modes \cite{chang2015high, PhysRevLett.113.137201, Checkelsky_2014}.

\bibliography{apssamp}

\providecommand{\noopsort}[1]{}\providecommand{\singleletter}[1]{#1}%
\begin{thebibliography}{67}%
\makeatletter
\providecommand \@ifxundefined [1]{%
 \@ifx{#1\undefined}
}%
\providecommand \@ifnum [1]{%
 \ifnum #1\expandafter \@firstoftwo
 \else \expandafter \@secondoftwo
 \fi
}%
\providecommand \@ifx [1]{%
 \ifx #1\expandafter \@firstoftwo
 \else \expandafter \@secondoftwo
 \fi
}%
\providecommand \natexlab [1]{#1}%
\providecommand \enquote  [1]{``#1''}%
\providecommand \bibnamefont  [1]{#1}%
\providecommand \bibfnamefont [1]{#1}%
\providecommand \citenamefont [1]{#1}%
\providecommand \href@noop [0]{\@secondoftwo}%
\providecommand \href [0]{\begingroup \@sanitize@url \@href}%
\providecommand \@href[1]{\@@startlink{#1}\@@href}%
\providecommand \@@href[1]{\endgroup#1\@@endlink}%
\providecommand \@sanitize@url [0]{\catcode `\\12\catcode `\$12\catcode
  `\&12\catcode `\#12\catcode `\^12\catcode `\_12\catcode `\%12\relax}%
\providecommand \@@startlink[1]{}%
\providecommand \@@endlink[0]{}%
\providecommand \url  [0]{\begingroup\@sanitize@url \@url }%
\providecommand \@url [1]{\endgroup\@href {#1}{\urlprefix }}%
\providecommand \urlprefix  [0]{URL }%
\providecommand \Eprint [0]{\href }%
\providecommand \doibase [0]{https://doi.org/}%
\providecommand \selectlanguage [0]{\@gobble}%
\providecommand \bibinfo  [0]{\@secondoftwo}%
\providecommand \bibfield  [0]{\@secondoftwo}%
\providecommand \translation [1]{[#1]}%
\providecommand \BibitemOpen [0]{}%
\providecommand \bibitemStop [0]{}%
\providecommand \bibitemNoStop [0]{.\EOS\space}%
\providecommand \EOS [0]{\spacefactor3000\relax}%
\providecommand \BibitemShut  [1]{\csname bibitem#1\endcsname}%
\let\auto@bib@innerbib\@empty
\bibitem [{\citenamefont {Klitzing}\ \emph {et~al.}(1980)\citenamefont
  {Klitzing}, \citenamefont {Dorda},\ and\ \citenamefont
  {Pepper}}]{PhysRevLett.45.494}%
  \BibitemOpen
  \bibfield  {author} {\bibinfo {author} {\bibfnamefont {K.~v.}\ \bibnamefont
  {Klitzing}}, \bibinfo {author} {\bibfnamefont {G.}~\bibnamefont {Dorda}},\
  and\ \bibinfo {author} {\bibfnamefont {M.}~\bibnamefont {Pepper}},\
  }\bibfield  {title} {\bibinfo {title} {New method for high-accuracy
  determination of the fine-structure constant based on quantized hall
  resistance},\ }\href {https://doi.org/10.1103/PhysRevLett.45.494} {\bibfield
  {journal} {\bibinfo  {journal} {Phys. Rev. Lett.}\ }\textbf {\bibinfo
  {volume} {45}},\ \bibinfo {pages} {494} (\bibinfo {year} {1980})}\BibitemShut
  {NoStop}%
\bibitem [{\citenamefont {Halperin}(1982)}]{PhysRevB.25.2185}%
  \BibitemOpen
  \bibfield  {author} {\bibinfo {author} {\bibfnamefont {B.~I.}\ \bibnamefont
  {Halperin}},\ }\bibfield  {title} {\bibinfo {title} {Quantized hall
  conductance, current-carrying edge states, and the existence of extended
  states in a two-dimensional disordered potential},\ }\href
  {https://doi.org/10.1103/PhysRevB.25.2185} {\bibfield  {journal} {\bibinfo
  {journal} {Phys. Rev. B}\ }\textbf {\bibinfo {volume} {25}},\ \bibinfo
  {pages} {2185} (\bibinfo {year} {1982})}\BibitemShut {NoStop}%
\bibitem [{\citenamefont {B\"uttiker}(1988)}]{PhysRevB.38.9375}%
  \BibitemOpen
  \bibfield  {author} {\bibinfo {author} {\bibfnamefont {M.}~\bibnamefont
  {B\"uttiker}},\ }\bibfield  {title} {\bibinfo {title} {Absence of
  backscattering in the quantum hall effect in multiprobe conductors},\ }\href
  {https://doi.org/10.1103/PhysRevB.38.9375} {\bibfield  {journal} {\bibinfo
  {journal} {Phys. Rev. B}\ }\textbf {\bibinfo {volume} {38}},\ \bibinfo
  {pages} {9375} (\bibinfo {year} {1988})}\BibitemShut {NoStop}%
\bibitem [{\citenamefont {Bernevig}\ \emph {et~al.}(2006)\citenamefont
  {Bernevig}, \citenamefont {Hughes},\ and\ \citenamefont {Zhang}}]{zhang}%
  \BibitemOpen
  \bibfield  {author} {\bibinfo {author} {\bibfnamefont {B.~A.}\ \bibnamefont
  {Bernevig}}, \bibinfo {author} {\bibfnamefont {T.~L.}\ \bibnamefont
  {Hughes}},\ and\ \bibinfo {author} {\bibfnamefont {S.-C.}\ \bibnamefont
  {Zhang}},\ }\bibfield  {title} {\bibinfo {title} {Quantum spin hall effect
  and topological phase transition in hgte quantum wells},\ }\href
  {https://doi.org/10.1126/science.1133734} {\bibfield  {journal} {\bibinfo
  {journal} {Science}\ }\textbf {\bibinfo {volume} {314}},\ \bibinfo {pages}
  {1757} (\bibinfo {year} {2006})}\BibitemShut {NoStop}%
\bibitem [{\citenamefont {K\"onig}\ \emph {et~al.}(2007)\citenamefont
  {K\"onig}, \citenamefont {Wiedmann}, \citenamefont {Br\"une}, \citenamefont
  {Roth}, \citenamefont {Buhmann}, \citenamefont {Molenkamp}, \citenamefont
  {Qi},\ and\ \citenamefont {Zhang}}]{Konig}%
  \BibitemOpen
  \bibfield  {author} {\bibinfo {author} {\bibfnamefont {M.}~\bibnamefont
  {K\"onig}}, \bibinfo {author} {\bibfnamefont {S.}~\bibnamefont {Wiedmann}},
  \bibinfo {author} {\bibfnamefont {C.}~\bibnamefont {Br\"une}}, \bibinfo
  {author} {\bibfnamefont {A.}~\bibnamefont {Roth}}, \bibinfo {author}
  {\bibfnamefont {H.}~\bibnamefont {Buhmann}}, \bibinfo {author} {\bibfnamefont
  {L.}~\bibnamefont {Molenkamp}}, \bibinfo {author} {\bibfnamefont
  {X.}~\bibnamefont {Qi}},\ and\ \bibinfo {author} {\bibfnamefont
  {S.}~\bibnamefont {Zhang}},\ }\bibfield  {title} {\bibinfo {title} {Quantum
  spin hall insulator state in hgte quantum wells},\ }\href
  {https://doi.org/10.1126/science.1148047} {\bibfield  {journal} {\bibinfo
  {journal} {Science}\ }\textbf {\bibinfo {volume} {318}},\ \bibinfo {pages}
  {776770} (\bibinfo {year} {2007})}\BibitemShut {NoStop}%
\bibitem [{\citenamefont {Roth}\ \emph {et~al.}(2009)\citenamefont {Roth},
  \citenamefont {Br\"une}, \citenamefont {Buhmann}, \citenamefont {Molenkamp},
  \citenamefont {Maciejko}, \citenamefont {Qi},\ and\ \citenamefont
  {Zhang}}]{Roth_2009}%
  \BibitemOpen
  \bibfield  {author} {\bibinfo {author} {\bibfnamefont {A.}~\bibnamefont
  {Roth}}, \bibinfo {author} {\bibfnamefont {C.}~\bibnamefont {Br\"une}},
  \bibinfo {author} {\bibfnamefont {H.}~\bibnamefont {Buhmann}}, \bibinfo
  {author} {\bibfnamefont {L.~W.}\ \bibnamefont {Molenkamp}}, \bibinfo {author}
  {\bibfnamefont {J.}~\bibnamefont {Maciejko}}, \bibinfo {author}
  {\bibfnamefont {X.-L.}\ \bibnamefont {Qi}},\ and\ \bibinfo {author}
  {\bibfnamefont {S.-C.}\ \bibnamefont {Zhang}},\ }\bibfield  {title} {\bibinfo
  {title} {Nonlocal transport in the quantum spin hall state},\ }\href
  {https://doi.org/10.1126/science.1174736} {\bibfield  {journal} {\bibinfo
  {journal} {Science}\ }\textbf {\bibinfo {volume} {325}},\ \bibinfo {pages}
  {294} (\bibinfo {year} {2009})}\BibitemShut {NoStop}%
\bibitem [{\citenamefont {Datta}(1995)}]{datta_1995}%
  \BibitemOpen
  \bibfield  {author} {\bibinfo {author} {\bibfnamefont {S.}~\bibnamefont
  {Datta}},\ }\href {https://doi.org/10.1017/CBO9780511805776} {\emph {\bibinfo
  {title} {Electronic Transport in Mesoscopic Systems}}},\ Cambridge Studies in
  Semiconductor Physics and Microelectronic Engineering\ (\bibinfo  {publisher}
  {Cambridge University Press},\ \bibinfo {year} {1995})\BibitemShut {NoStop}%
\bibitem [{\citenamefont {Shen}(2017)}]{Shen_2017}%
  \BibitemOpen
  \bibfield  {author} {\bibinfo {author} {\bibfnamefont {S.-Q.}\ \bibnamefont
  {Shen}},\ }\href {https://doi.org/10.1007/978-981-10-4606-3} {\emph {\bibinfo
  {title} {Topological Insulators}}}\ (\bibinfo  {publisher} {Springer
  Singapore},\ \bibinfo {year} {2017})\BibitemShut {NoStop}%
\bibitem [{\citenamefont {S\'anchez}\ \emph
  {et~al.}(2015{\natexlab{a}})\citenamefont {S\'anchez}, \citenamefont
  {Sothmann},\ and\ \citenamefont {Jordan}}]{PhysRevLett.114.146801}%
  \BibitemOpen
  \bibfield  {author} {\bibinfo {author} {\bibfnamefont {R.}~\bibnamefont
  {S\'anchez}}, \bibinfo {author} {\bibfnamefont {B.}~\bibnamefont
  {Sothmann}},\ and\ \bibinfo {author} {\bibfnamefont {A.~N.}\ \bibnamefont
  {Jordan}},\ }\bibfield  {title} {\bibinfo {title} {Chiral thermoelectrics
  with quantum hall edge states},\ }\href
  {https://doi.org/10.1103/PhysRevLett.114.146801} {\bibfield  {journal}
  {\bibinfo  {journal} {Phys. Rev. Lett.}\ }\textbf {\bibinfo {volume} {114}},\
  \bibinfo {pages} {146801} (\bibinfo {year} {2015}{\natexlab{a}})}\BibitemShut
  {NoStop}%
\bibitem [{\citenamefont {Brandner}\ and\ \citenamefont
  {Seifert}(2013)}]{Brandner_2013}%
  \BibitemOpen
  \bibfield  {author} {\bibinfo {author} {\bibfnamefont {K.}~\bibnamefont
  {Brandner}}\ and\ \bibinfo {author} {\bibfnamefont {U.}~\bibnamefont
  {Seifert}},\ }\bibfield  {title} {\bibinfo {title} {Multi-terminal
  thermoelectric transport in a magnetic field: bounds on onsager coefficients
  and efficiency},\ }\href {https://doi.org/10.1088/1367-2630/15/10/105003}
  {\bibfield  {journal} {\bibinfo  {journal} {New Journal of Physics}\ }\textbf
  {\bibinfo {volume} {15}},\ \bibinfo {pages} {105003} (\bibinfo {year}
  {2013})}\BibitemShut {NoStop}%
\bibitem [{\citenamefont {Haack}\ and\ \citenamefont
  {Giazotto}(2021)}]{Haack_2021}%
  \BibitemOpen
  \bibfield  {author} {\bibinfo {author} {\bibfnamefont {G.}~\bibnamefont
  {Haack}}\ and\ \bibinfo {author} {\bibfnamefont {F.}~\bibnamefont
  {Giazotto}},\ }\bibfield  {title} {\bibinfo {title} {Nonlinear regime for
  enhanced performance of an aharonov--bohm heat engine},\ }\href@noop {}
  {\bibfield  {journal} {\bibinfo  {journal} {AVS Quantum Science}\ }\textbf
  {\bibinfo {volume} {3}},\ \bibinfo {pages} {046801} (\bibinfo {year}
  {2021})}\BibitemShut {NoStop}%
\bibitem [{\citenamefont {Haack}\ and\ \citenamefont
  {Giazotto}(2019)}]{PhysRevB.100.235442}%
  \BibitemOpen
  \bibfield  {author} {\bibinfo {author} {\bibfnamefont {G.}~\bibnamefont
  {Haack}}\ and\ \bibinfo {author} {\bibfnamefont {F.}~\bibnamefont
  {Giazotto}},\ }\bibfield  {title} {\bibinfo {title} {Efficient and tunable
  aharonov-bohm quantum heat engine},\ }\href
  {https://doi.org/10.1103/PhysRevB.100.235442} {\bibfield  {journal} {\bibinfo
   {journal} {Phys. Rev. B}\ }\textbf {\bibinfo {volume} {100}},\ \bibinfo
  {pages} {235442} (\bibinfo {year} {2019})}\BibitemShut {NoStop}%
\bibitem [{\citenamefont {Whitney}(2014)}]{PhysRevLett.112.130601}%
  \BibitemOpen
  \bibfield  {author} {\bibinfo {author} {\bibfnamefont {R.~S.}\ \bibnamefont
  {Whitney}},\ }\bibfield  {title} {\bibinfo {title} {Most efficient quantum
  thermoelectric at finite power output},\ }\href
  {https://doi.org/10.1103/PhysRevLett.112.130601} {\bibfield  {journal}
  {\bibinfo  {journal} {Phys. Rev. Lett.}\ }\textbf {\bibinfo {volume} {112}},\
  \bibinfo {pages} {130601} (\bibinfo {year} {2014})}\BibitemShut {NoStop}%
\bibitem [{\citenamefont {Whitney}(2015)}]{PhysRevB.91.115425}%
  \BibitemOpen
  \bibfield  {author} {\bibinfo {author} {\bibfnamefont {R.~S.}\ \bibnamefont
  {Whitney}},\ }\bibfield  {title} {\bibinfo {title} {Finding the quantum
  thermoelectric with maximal efficiency and minimal entropy production at
  given power output},\ }\href {https://doi.org/10.1103/PhysRevB.91.115425}
  {\bibfield  {journal} {\bibinfo  {journal} {Phys. Rev. B}\ }\textbf {\bibinfo
  {volume} {91}},\ \bibinfo {pages} {115425} (\bibinfo {year}
  {2015})}\BibitemShut {NoStop}%
\bibitem [{\citenamefont {Mani}\ and\ \citenamefont
  {Benjamin}(2018)}]{PhysRevE.97.022114}%
  \BibitemOpen
  \bibfield  {author} {\bibinfo {author} {\bibfnamefont {A.}~\bibnamefont
  {Mani}}\ and\ \bibinfo {author} {\bibfnamefont {C.}~\bibnamefont
  {Benjamin}},\ }\bibfield  {title} {\bibinfo {title} {Helical thermoelectrics
  and refrigeration},\ }\href {https://doi.org/10.1103/PhysRevE.97.022114}
  {\bibfield  {journal} {\bibinfo  {journal} {Phys. Rev. E}\ }\textbf {\bibinfo
  {volume} {97}},\ \bibinfo {pages} {022114} (\bibinfo {year}
  {2018})}\BibitemShut {NoStop}%
\bibitem [{\citenamefont {Sartipi}\ and\ \citenamefont
  {Vahedi}(2023)}]{PhysRevB.108.195435}%
  \BibitemOpen
  \bibfield  {author} {\bibinfo {author} {\bibfnamefont {Z.}~\bibnamefont
  {Sartipi}}\ and\ \bibinfo {author} {\bibfnamefont {J.}~\bibnamefont
  {Vahedi}},\ }\bibfield  {title} {\bibinfo {title} {Optimal performance of
  voltage-probe quantum heat engines},\ }\href
  {https://doi.org/10.1103/PhysRevB.108.195435} {\bibfield  {journal} {\bibinfo
   {journal} {Phys. Rev. B}\ }\textbf {\bibinfo {volume} {108}},\ \bibinfo
  {pages} {195435} (\bibinfo {year} {2023})}\BibitemShut {NoStop}%
\bibitem [{\citenamefont {S\'anchez}\ and\ \citenamefont
  {Serra}(2011)}]{PhysRevB.84.201307}%
  \BibitemOpen
  \bibfield  {author} {\bibinfo {author} {\bibfnamefont {D.}~\bibnamefont
  {S\'anchez}}\ and\ \bibinfo {author} {\bibfnamefont {L.~m.~c.}\ \bibnamefont
  {Serra}},\ }\bibfield  {title} {\bibinfo {title} {Thermoelectric transport of
  mesoscopic conductors coupled to voltage and thermal probes},\ }\href
  {https://doi.org/10.1103/PhysRevB.84.201307} {\bibfield  {journal} {\bibinfo
  {journal} {Phys. Rev. B}\ }\textbf {\bibinfo {volume} {84}},\ \bibinfo
  {pages} {201307} (\bibinfo {year} {2011})}\BibitemShut {NoStop}%
\bibitem [{\citenamefont {Gresta}\ \emph {et~al.}(2019)\citenamefont {Gresta},
  \citenamefont {Real},\ and\ \citenamefont
  {Arrachea}}]{PhysRevLett.123.186801}%
  \BibitemOpen
  \bibfield  {author} {\bibinfo {author} {\bibfnamefont {D.}~\bibnamefont
  {Gresta}}, \bibinfo {author} {\bibfnamefont {M.}~\bibnamefont {Real}},\ and\
  \bibinfo {author} {\bibfnamefont {L.}~\bibnamefont {Arrachea}},\ }\bibfield
  {title} {\bibinfo {title} {Optimal thermoelectricity with quantum spin hall
  edge states},\ }\href {https://doi.org/10.1103/PhysRevLett.123.186801}
  {\bibfield  {journal} {\bibinfo  {journal} {Phys. Rev. Lett.}\ }\textbf
  {\bibinfo {volume} {123}},\ \bibinfo {pages} {186801} (\bibinfo {year}
  {2019})}\BibitemShut {NoStop}%
\bibitem [{\citenamefont {Ozaeta}\ \emph {et~al.}(2014)\citenamefont {Ozaeta},
  \citenamefont {Virtanen}, \citenamefont {Bergeret},\ and\ \citenamefont
  {Heikkil\"a}}]{PhysRevLett.112.057001}%
  \BibitemOpen
  \bibfield  {author} {\bibinfo {author} {\bibfnamefont {A.}~\bibnamefont
  {Ozaeta}}, \bibinfo {author} {\bibfnamefont {P.}~\bibnamefont {Virtanen}},
  \bibinfo {author} {\bibfnamefont {F.~S.}\ \bibnamefont {Bergeret}},\ and\
  \bibinfo {author} {\bibfnamefont {T.~T.}\ \bibnamefont {Heikkil\"a}},\
  }\bibfield  {title} {\bibinfo {title} {Predicted very large thermoelectric
  effect in ferromagnet-superconductor junctions in the presence of a
  spin-splitting magnetic field},\ }\href
  {https://doi.org/10.1103/PhysRevLett.112.057001} {\bibfield  {journal}
  {\bibinfo  {journal} {Phys. Rev. Lett.}\ }\textbf {\bibinfo {volume} {112}},\
  \bibinfo {pages} {057001} (\bibinfo {year} {2014})}\BibitemShut {NoStop}%
\bibitem [{\citenamefont {Giazotto}\ \emph {et~al.}(2015)\citenamefont
  {Giazotto}, \citenamefont {Heikkil\"a},\ and\ \citenamefont
  {Bergeret}}]{PhysRevLett.114.067001}%
  \BibitemOpen
  \bibfield  {author} {\bibinfo {author} {\bibfnamefont {F.}~\bibnamefont
  {Giazotto}}, \bibinfo {author} {\bibfnamefont {T.~T.}\ \bibnamefont
  {Heikkil\"a}},\ and\ \bibinfo {author} {\bibfnamefont {F.~S.}\ \bibnamefont
  {Bergeret}},\ }\bibfield  {title} {\bibinfo {title} {Very large thermophase
  in ferromagnetic josephson junctions},\ }\href
  {https://doi.org/10.1103/PhysRevLett.114.067001} {\bibfield  {journal}
  {\bibinfo  {journal} {Phys. Rev. Lett.}\ }\textbf {\bibinfo {volume} {114}},\
  \bibinfo {pages} {067001} (\bibinfo {year} {2015})}\BibitemShut {NoStop}%
\bibitem [{\citenamefont {Mani}\ \emph {et~al.}(2019)\citenamefont {Mani},
  \citenamefont {Pal},\ and\ \citenamefont {Benjamin}}]{mani2019designing}%
  \BibitemOpen
  \bibfield  {author} {\bibinfo {author} {\bibfnamefont {A.}~\bibnamefont
  {Mani}}, \bibinfo {author} {\bibfnamefont {S.}~\bibnamefont {Pal}},\ and\
  \bibinfo {author} {\bibfnamefont {C.}~\bibnamefont {Benjamin}},\ }\bibfield
  {title} {\bibinfo {title} {Designing a highly efficient graphene quantum spin
  heat engine},\ }\href@noop {} {\bibfield  {journal} {\bibinfo  {journal}
  {Scientific reports}\ }\textbf {\bibinfo {volume} {9}},\ \bibinfo {pages}
  {6018} (\bibinfo {year} {2019})}\BibitemShut {NoStop}%
\bibitem [{\citenamefont {Mani}\ and\ \citenamefont
  {Benjamin}(2017{\natexlab{a}})}]{PhysRevE.96.032118}%
  \BibitemOpen
  \bibfield  {author} {\bibinfo {author} {\bibfnamefont {A.}~\bibnamefont
  {Mani}}\ and\ \bibinfo {author} {\bibfnamefont {C.}~\bibnamefont
  {Benjamin}},\ }\bibfield  {title} {\bibinfo {title} {Strained-graphene-based
  highly efficient quantum heat engine operating at maximum power},\ }\href
  {https://doi.org/10.1103/PhysRevE.96.032118} {\bibfield  {journal} {\bibinfo
  {journal} {Phys. Rev. E}\ }\textbf {\bibinfo {volume} {96}},\ \bibinfo
  {pages} {032118} (\bibinfo {year} {2017}{\natexlab{a}})}\BibitemShut
  {NoStop}%
\bibitem [{\citenamefont {Rameshti}\ and\ \citenamefont
  {Moghaddam}(2015)}]{PhysRevB.91.155407}%
  \BibitemOpen
  \bibfield  {author} {\bibinfo {author} {\bibfnamefont {B.~Z.}\ \bibnamefont
  {Rameshti}}\ and\ \bibinfo {author} {\bibfnamefont {A.~G.}\ \bibnamefont
  {Moghaddam}},\ }\bibfield  {title} {\bibinfo {title} {Spin-dependent seebeck
  effect and spin caloritronics in magnetic graphene},\ }\href
  {https://doi.org/10.1103/PhysRevB.91.155407} {\bibfield  {journal} {\bibinfo
  {journal} {Phys. Rev. B}\ }\textbf {\bibinfo {volume} {91}},\ \bibinfo
  {pages} {155407} (\bibinfo {year} {2015})}\BibitemShut {NoStop}%
\bibitem [{\citenamefont {S\'anchez}\ \emph
  {et~al.}(2015{\natexlab{b}})\citenamefont {S\'anchez}, \citenamefont
  {Sothmann},\ and\ \citenamefont {Jordan}}]{Sanchez_2015}%
  \BibitemOpen
  \bibfield  {author} {\bibinfo {author} {\bibfnamefont {R.}~\bibnamefont
  {S\'anchez}}, \bibinfo {author} {\bibfnamefont {B.}~\bibnamefont
  {Sothmann}},\ and\ \bibinfo {author} {\bibfnamefont {A.~N.}\ \bibnamefont
  {Jordan}},\ }\bibfield  {title} {\bibinfo {title} {Heat diode and engine
  based on quantum hall edge states},\ }\href
  {https://doi.org/10.1088/1367-2630/17/7/075006} {\bibfield  {journal}
  {\bibinfo  {journal} {New Journal of Physics}\ }\textbf {\bibinfo {volume}
  {17}},\ \bibinfo {pages} {075006} (\bibinfo {year}
  {2015}{\natexlab{b}})}\BibitemShut {NoStop}%
\bibitem [{\citenamefont {Chen}\ and\ \citenamefont
  {Yan}(1989)}]{chen1989effect}%
  \BibitemOpen
  \bibfield  {author} {\bibinfo {author} {\bibfnamefont {L.}~\bibnamefont
  {Chen}}\ and\ \bibinfo {author} {\bibfnamefont {Z.}~\bibnamefont {Yan}},\
  }\bibfield  {title} {\bibinfo {title} {The effect of heat-transfer law on
  performance of a two-heat-source endoreversible cycle},\ }\href@noop {}
  {\bibfield  {journal} {\bibinfo  {journal} {The Journal of Chemical Physics}\
  }\textbf {\bibinfo {volume} {90}},\ \bibinfo {pages} {3740} (\bibinfo {year}
  {1989})}\BibitemShut {NoStop}%
\bibitem [{\citenamefont {Shiraishi}\ \emph {et~al.}(2016)\citenamefont
  {Shiraishi}, \citenamefont {Saito},\ and\ \citenamefont
  {Tasaki}}]{PhysRevLett.117.190601}%
  \BibitemOpen
  \bibfield  {author} {\bibinfo {author} {\bibfnamefont {N.}~\bibnamefont
  {Shiraishi}}, \bibinfo {author} {\bibfnamefont {K.}~\bibnamefont {Saito}},\
  and\ \bibinfo {author} {\bibfnamefont {H.}~\bibnamefont {Tasaki}},\
  }\bibfield  {title} {\bibinfo {title} {Universal trade-off relation between
  power and efficiency for heat engines},\ }\href
  {https://doi.org/10.1103/PhysRevLett.117.190601} {\bibfield  {journal}
  {\bibinfo  {journal} {Phys. Rev. Lett.}\ }\textbf {\bibinfo {volume} {117}},\
  \bibinfo {pages} {190601} (\bibinfo {year} {2016})}\BibitemShut {NoStop}%
\bibitem [{\citenamefont {Holubec}\ and\ \citenamefont
  {Ryabov}(2016)}]{holubec2016maximum}%
  \BibitemOpen
  \bibfield  {author} {\bibinfo {author} {\bibfnamefont {V.}~\bibnamefont
  {Holubec}}\ and\ \bibinfo {author} {\bibfnamefont {A.}~\bibnamefont
  {Ryabov}},\ }\bibfield  {title} {\bibinfo {title} {Maximum efficiency of
  low-dissipation heat engines at arbitrary power},\ }\href@noop {} {\bibfield
  {journal} {\bibinfo  {journal} {Journal of Statistical Mechanics: Theory and
  Experiment}\ }\textbf {\bibinfo {volume} {2016}},\ \bibinfo {pages} {073204}
  (\bibinfo {year} {2016})}\BibitemShut {NoStop}%
\bibitem [{\citenamefont {Ma}\ \emph {et~al.}(2018)\citenamefont {Ma},
  \citenamefont {Xu}, \citenamefont {Dong},\ and\ \citenamefont
  {Sun}}]{PhysRevE.98.042112}%
  \BibitemOpen
  \bibfield  {author} {\bibinfo {author} {\bibfnamefont {Y.-H.}\ \bibnamefont
  {Ma}}, \bibinfo {author} {\bibfnamefont {D.}~\bibnamefont {Xu}}, \bibinfo
  {author} {\bibfnamefont {H.}~\bibnamefont {Dong}},\ and\ \bibinfo {author}
  {\bibfnamefont {C.-P.}\ \bibnamefont {Sun}},\ }\bibfield  {title} {\bibinfo
  {title} {Universal constraint for efficiency and power of a low-dissipation
  heat engine},\ }\href {https://doi.org/10.1103/PhysRevE.98.042112} {\bibfield
   {journal} {\bibinfo  {journal} {Phys. Rev. E}\ }\textbf {\bibinfo {volume}
  {98}},\ \bibinfo {pages} {042112} (\bibinfo {year} {2018})}\BibitemShut
  {NoStop}%
\bibitem [{\citenamefont {Kosloff}\ and\ \citenamefont
  {Levy}(2014)}]{kosloff2014quantum}%
  \BibitemOpen
  \bibfield  {author} {\bibinfo {author} {\bibfnamefont {R.}~\bibnamefont
  {Kosloff}}\ and\ \bibinfo {author} {\bibfnamefont {A.}~\bibnamefont {Levy}},\
  }\bibfield  {title} {\bibinfo {title} {Quantum heat engines and
  refrigerators: Continuous devices},\ }\href@noop {} {\bibfield  {journal}
  {\bibinfo  {journal} {Annual review of physical chemistry}\ }\textbf
  {\bibinfo {volume} {65}},\ \bibinfo {pages} {365} (\bibinfo {year}
  {2014})}\BibitemShut {NoStop}%
\bibitem [{\citenamefont {Wineland}\ and\ \citenamefont
  {Dehmelt}(1975)}]{wineland1975proposed}%
  \BibitemOpen
  \bibfield  {author} {\bibinfo {author} {\bibfnamefont {D.}~\bibnamefont
  {Wineland}}\ and\ \bibinfo {author} {\bibfnamefont {H.}~\bibnamefont
  {Dehmelt}},\ }\bibfield  {title} {\bibinfo {title} {Proposed
  1014$\delta$$\nu$/$\nu$ laser fluorescence spectroscopy on tl+ mono-ion
  oscillator},\ }\href@noop {} {\bibfield  {journal} {\bibinfo  {journal}
  {Bull. Am. Phys. Soc}\ }\textbf {\bibinfo {volume} {20}} (\bibinfo {year}
  {1975})}\BibitemShut {NoStop}%
\bibitem [{\citenamefont {H{\"a}nsch}\ and\ \citenamefont
  {Schawlow}(1975)}]{hansch1975cooling}%
  \BibitemOpen
  \bibfield  {author} {\bibinfo {author} {\bibfnamefont {T.~W.}\ \bibnamefont
  {H{\"a}nsch}}\ and\ \bibinfo {author} {\bibfnamefont {A.~L.}\ \bibnamefont
  {Schawlow}},\ }\bibfield  {title} {\bibinfo {title} {Cooling of gases by
  laser radiation},\ }\href@noop {} {\bibfield  {journal} {\bibinfo  {journal}
  {Optics Communications}\ }\textbf {\bibinfo {volume} {13}},\ \bibinfo {pages}
  {68} (\bibinfo {year} {1975})}\BibitemShut {NoStop}%
\bibitem [{\citenamefont {Benenti}\ \emph {et~al.}(2017)\citenamefont
  {Benenti}, \citenamefont {Casati}, \citenamefont {Saito},\ and\ \citenamefont
  {Whitney}}]{BENENTI20171}%
  \BibitemOpen
  \bibfield  {author} {\bibinfo {author} {\bibfnamefont {G.}~\bibnamefont
  {Benenti}}, \bibinfo {author} {\bibfnamefont {G.}~\bibnamefont {Casati}},
  \bibinfo {author} {\bibfnamefont {K.}~\bibnamefont {Saito}},\ and\ \bibinfo
  {author} {\bibfnamefont {R.}~\bibnamefont {Whitney}},\ }\bibfield  {title}
  {\bibinfo {title} {Fundamental aspects of steady-state conversion of heat to
  work at the nanoscale},\ }\href
  {https://doi.org/https://doi.org/10.1016/j.physrep.2017.05.008} {\bibfield
  {journal} {\bibinfo  {journal} {Physics Reports}\ }\textbf {\bibinfo {volume}
  {694}},\ \bibinfo {pages} {1} (\bibinfo {year} {2017})}\BibitemShut {NoStop}%
\bibitem [{\citenamefont {Curzon}\ and\ \citenamefont
  {Ahlborn}(1975)}]{curzon1975efficiency}%
  \BibitemOpen
  \bibfield  {author} {\bibinfo {author} {\bibfnamefont {F.~L.}\ \bibnamefont
  {Curzon}}\ and\ \bibinfo {author} {\bibfnamefont {B.}~\bibnamefont
  {Ahlborn}},\ }\bibfield  {title} {\bibinfo {title} {Efficiency of a carnot
  engine at maximum power output},\ }\href@noop {} {\bibfield  {journal}
  {\bibinfo  {journal} {American Journal of Physics}\ }\textbf {\bibinfo
  {volume} {43}},\ \bibinfo {pages} {22} (\bibinfo {year} {1975})}\BibitemShut
  {NoStop}%
\bibitem [{\citenamefont {Van~den Broeck}(2005)}]{PhysRevLett.95.190602}%
  \BibitemOpen
  \bibfield  {author} {\bibinfo {author} {\bibfnamefont {C.}~\bibnamefont
  {Van~den Broeck}},\ }\bibfield  {title} {\bibinfo {title} {Thermodynamic
  efficiency at maximum power},\ }\href
  {https://doi.org/10.1103/PhysRevLett.95.190602} {\bibfield  {journal}
  {\bibinfo  {journal} {Phys. Rev. Lett.}\ }\textbf {\bibinfo {volume} {95}},\
  \bibinfo {pages} {190602} (\bibinfo {year} {2005})}\BibitemShut {NoStop}%
\bibitem [{\citenamefont {Whitney}(2016)}]{Whitney_2016}%
  \BibitemOpen
  \bibfield  {author} {\bibinfo {author} {\bibfnamefont {R.}~\bibnamefont
  {Whitney}},\ }\bibfield  {title} {\bibinfo {title} {Quantum coherent
  three-terminal thermoelectrics: Maximum efficiency at given power output},\
  }\href {https://doi.org/10.3390/e18060208} {\bibfield  {journal} {\bibinfo
  {journal} {Entropy}\ }\textbf {\bibinfo {volume} {18}},\ \bibinfo {pages}
  {208} (\bibinfo {year} {2016})}\BibitemShut {NoStop}%
\bibitem [{\citenamefont {Yamamoto}\ and\ \citenamefont
  {Hatano}(2015)}]{PhysRevE.92.042165}%
  \BibitemOpen
  \bibfield  {author} {\bibinfo {author} {\bibfnamefont {K.}~\bibnamefont
  {Yamamoto}}\ and\ \bibinfo {author} {\bibfnamefont {N.}~\bibnamefont
  {Hatano}},\ }\bibfield  {title} {\bibinfo {title} {Thermodynamics of the
  mesoscopic thermoelectric heat engine beyond the linear-response regime},\
  }\href {https://doi.org/10.1103/PhysRevE.92.042165} {\bibfield  {journal}
  {\bibinfo  {journal} {Phys. Rev. E}\ }\textbf {\bibinfo {volume} {92}},\
  \bibinfo {pages} {042165} (\bibinfo {year} {2015})}\BibitemShut {NoStop}%
\bibitem [{\citenamefont {van Wees}\ \emph {et~al.}(1988)\citenamefont {van
  Wees}, \citenamefont {van Houten}, \citenamefont {Beenakker}, \citenamefont
  {Williamson}, \citenamefont {Kouwenhoven}, \citenamefont {van~der Marel},\
  and\ \citenamefont {Foxon}}]{PhysRevLett.60.848}%
  \BibitemOpen
  \bibfield  {author} {\bibinfo {author} {\bibfnamefont {B.~J.}\ \bibnamefont
  {van Wees}}, \bibinfo {author} {\bibfnamefont {H.}~\bibnamefont {van
  Houten}}, \bibinfo {author} {\bibfnamefont {C.~W.~J.}\ \bibnamefont
  {Beenakker}}, \bibinfo {author} {\bibfnamefont {J.~G.}\ \bibnamefont
  {Williamson}}, \bibinfo {author} {\bibfnamefont {L.~P.}\ \bibnamefont
  {Kouwenhoven}}, \bibinfo {author} {\bibfnamefont {D.}~\bibnamefont {van~der
  Marel}},\ and\ \bibinfo {author} {\bibfnamefont {C.~T.}\ \bibnamefont
  {Foxon}},\ }\bibfield  {title} {\bibinfo {title} {Quantized conductance of
  point contacts in a two-dimensional electron gas},\ }\href
  {https://doi.org/10.1103/PhysRevLett.60.848} {\bibfield  {journal} {\bibinfo
  {journal} {Phys. Rev. Lett.}\ }\textbf {\bibinfo {volume} {60}},\ \bibinfo
  {pages} {848} (\bibinfo {year} {1988})}\BibitemShut {NoStop}%
\bibitem [{\citenamefont {Wharam}\ \emph {et~al.}(1988)\citenamefont {Wharam},
  \citenamefont {Thornton}, \citenamefont {Newbury}, \citenamefont {Pepper},
  \citenamefont {Ahmed}, \citenamefont {Frost}, \citenamefont {Hasko},
  \citenamefont {Peacock}, \citenamefont {Ritchie},\ and\ \citenamefont
  {Jones}}]{wharam1988one}%
  \BibitemOpen
  \bibfield  {author} {\bibinfo {author} {\bibfnamefont {D.}~\bibnamefont
  {Wharam}}, \bibinfo {author} {\bibfnamefont {T.~J.}\ \bibnamefont
  {Thornton}}, \bibinfo {author} {\bibfnamefont {R.}~\bibnamefont {Newbury}},
  \bibinfo {author} {\bibfnamefont {M.}~\bibnamefont {Pepper}}, \bibinfo
  {author} {\bibfnamefont {H.}~\bibnamefont {Ahmed}}, \bibinfo {author}
  {\bibfnamefont {J.}~\bibnamefont {Frost}}, \bibinfo {author} {\bibfnamefont
  {D.}~\bibnamefont {Hasko}}, \bibinfo {author} {\bibfnamefont
  {D.}~\bibnamefont {Peacock}}, \bibinfo {author} {\bibfnamefont
  {D.}~\bibnamefont {Ritchie}},\ and\ \bibinfo {author} {\bibfnamefont
  {G.}~\bibnamefont {Jones}},\ }\bibfield  {title} {\bibinfo {title}
  {One-dimensional transport and the quantisation of the ballistic
  resistance},\ }\href@noop {} {\bibfield  {journal} {\bibinfo  {journal}
  {Journal of Physics C: solid state physics}\ }\textbf {\bibinfo {volume}
  {21}},\ \bibinfo {pages} {L209} (\bibinfo {year} {1988})}\BibitemShut
  {NoStop}%
\bibitem [{\citenamefont {Van~Houten}\ \emph {et~al.}(1992)\citenamefont
  {Van~Houten}, \citenamefont {Molenkamp}, \citenamefont {Beenakker},\ and\
  \citenamefont {Foxon}}]{van1992thermo}%
  \BibitemOpen
  \bibfield  {author} {\bibinfo {author} {\bibfnamefont {H.}~\bibnamefont
  {Van~Houten}}, \bibinfo {author} {\bibfnamefont {L.}~\bibnamefont
  {Molenkamp}}, \bibinfo {author} {\bibfnamefont {C.}~\bibnamefont
  {Beenakker}},\ and\ \bibinfo {author} {\bibfnamefont {C.}~\bibnamefont
  {Foxon}},\ }\bibfield  {title} {\bibinfo {title} {Thermo-electric properties
  of quantum point contacts},\ }\href@noop {} {\bibfield  {journal} {\bibinfo
  {journal} {Semiconductor Science and Technology}\ }\textbf {\bibinfo {volume}
  {7}},\ \bibinfo {pages} {B215} (\bibinfo {year} {1992})}\BibitemShut
  {NoStop}%
\bibitem [{\citenamefont {Molenkamp}\ \emph {et~al.}(1992)\citenamefont
  {Molenkamp}, \citenamefont {Gravier}, \citenamefont {van Houten},
  \citenamefont {Buijk}, \citenamefont {Mabesoone},\ and\ \citenamefont
  {Foxon}}]{PhysRevLett.68.3765}%
  \BibitemOpen
  \bibfield  {author} {\bibinfo {author} {\bibfnamefont {L.~W.}\ \bibnamefont
  {Molenkamp}}, \bibinfo {author} {\bibfnamefont {T.}~\bibnamefont {Gravier}},
  \bibinfo {author} {\bibfnamefont {H.}~\bibnamefont {van Houten}}, \bibinfo
  {author} {\bibfnamefont {O.~J.~A.}\ \bibnamefont {Buijk}}, \bibinfo {author}
  {\bibfnamefont {M.~A.~A.}\ \bibnamefont {Mabesoone}},\ and\ \bibinfo {author}
  {\bibfnamefont {C.~T.}\ \bibnamefont {Foxon}},\ }\bibfield  {title} {\bibinfo
  {title} {Peltier coefficient and thermal conductance of a quantum point
  contact},\ }\href {https://doi.org/10.1103/PhysRevLett.68.3765} {\bibfield
  {journal} {\bibinfo  {journal} {Phys. Rev. Lett.}\ }\textbf {\bibinfo
  {volume} {68}},\ \bibinfo {pages} {3765} (\bibinfo {year}
  {1992})}\BibitemShut {NoStop}%
\bibitem [{\citenamefont {Pedersen}\ \emph {et~al.}(1998)\citenamefont
  {Pedersen}, \citenamefont {van Langen},\ and\ \citenamefont
  {B\"uttiker}}]{PhysRevB.57.1838}%
  \BibitemOpen
  \bibfield  {author} {\bibinfo {author} {\bibfnamefont {M.~H.}\ \bibnamefont
  {Pedersen}}, \bibinfo {author} {\bibfnamefont {S.~A.}\ \bibnamefont {van
  Langen}},\ and\ \bibinfo {author} {\bibfnamefont {M.}~\bibnamefont
  {B\"uttiker}},\ }\bibfield  {title} {\bibinfo {title} {Charge fluctuations in
  quantum point contacts and chaotic cavities in the presence of transport},\
  }\href {https://doi.org/10.1103/PhysRevB.57.1838} {\bibfield  {journal}
  {\bibinfo  {journal} {Phys. Rev. B}\ }\textbf {\bibinfo {volume} {57}},\
  \bibinfo {pages} {1838} (\bibinfo {year} {1998})}\BibitemShut {NoStop}%
\bibitem [{\citenamefont {Mishra}\ \emph {et~al.}(2023)\citenamefont {Mishra},
  \citenamefont {Das},\ and\ \citenamefont {Benjamin}}]{das2023majorana}%
  \BibitemOpen
  \bibfield  {author} {\bibinfo {author} {\bibfnamefont {S.}~\bibnamefont
  {Mishra}}, \bibinfo {author} {\bibfnamefont {R.}~\bibnamefont {Das}},\ and\
  \bibinfo {author} {\bibfnamefont {C.}~\bibnamefont {Benjamin}},\ }\href@noop
  {} {\bibinfo {title} {Majorana fermion induced power-law scaling in the
  violation of wiedemann-franz law}} (\bibinfo {year} {2023}),\ \Eprint
  {https://arxiv.org/abs/2309.05492} {arXiv:2309.05492 [cond-mat.mes-hall]}
  \BibitemShut {NoStop}%
\bibitem [{\citenamefont {Buccheri}\ \emph {et~al.}(2022)\citenamefont
  {Buccheri}, \citenamefont {Nava}, \citenamefont {Egger}, \citenamefont
  {Sodano},\ and\ \citenamefont {Giuliano}}]{PhysRevB.105.L081403}%
  \BibitemOpen
  \bibfield  {author} {\bibinfo {author} {\bibfnamefont {F.}~\bibnamefont
  {Buccheri}}, \bibinfo {author} {\bibfnamefont {A.}~\bibnamefont {Nava}},
  \bibinfo {author} {\bibfnamefont {R.}~\bibnamefont {Egger}}, \bibinfo
  {author} {\bibfnamefont {P.}~\bibnamefont {Sodano}},\ and\ \bibinfo {author}
  {\bibfnamefont {D.}~\bibnamefont {Giuliano}},\ }\bibfield  {title} {\bibinfo
  {title} {Violation of the wiedemann-franz law in the topological kondo
  model},\ }\href {https://doi.org/10.1103/PhysRevB.105.L081403} {\bibfield
  {journal} {\bibinfo  {journal} {Phys. Rev. B}\ }\textbf {\bibinfo {volume}
  {105}},\ \bibinfo {pages} {L081403} (\bibinfo {year} {2022})}\BibitemShut
  {NoStop}%
\bibitem [{\citenamefont {B\"uttiker}(1990)}]{PhysRevB.41.7906}%
  \BibitemOpen
  \bibfield  {author} {\bibinfo {author} {\bibfnamefont {M.}~\bibnamefont
  {B\"uttiker}},\ }\bibfield  {title} {\bibinfo {title} {Quantized transmission
  of a saddle-point constriction},\ }\href
  {https://doi.org/10.1103/PhysRevB.41.7906} {\bibfield  {journal} {\bibinfo
  {journal} {Phys. Rev. B}\ }\textbf {\bibinfo {volume} {41}},\ \bibinfo
  {pages} {7906} (\bibinfo {year} {1990})}\BibitemShut {NoStop}%
\bibitem [{\citenamefont {Jalabert}\ \emph {et~al.}(1992)\citenamefont
  {Jalabert}, \citenamefont {Stone},\ and\ \citenamefont
  {Alhassid}}]{PhysRevLett.68.3468}%
  \BibitemOpen
  \bibfield  {author} {\bibinfo {author} {\bibfnamefont {R.~A.}\ \bibnamefont
  {Jalabert}}, \bibinfo {author} {\bibfnamefont {A.~D.}\ \bibnamefont
  {Stone}},\ and\ \bibinfo {author} {\bibfnamefont {Y.}~\bibnamefont
  {Alhassid}},\ }\bibfield  {title} {\bibinfo {title} {Statistical theory of
  coulomb blockade oscillations: Quantum chaos in quantum dots},\ }\href
  {https://doi.org/10.1103/PhysRevLett.68.3468} {\bibfield  {journal} {\bibinfo
   {journal} {Phys. Rev. Lett.}\ }\textbf {\bibinfo {volume} {68}},\ \bibinfo
  {pages} {3468} (\bibinfo {year} {1992})}\BibitemShut {NoStop}%
\bibitem [{\citenamefont {Zheng}\ \emph {et~al.}(1986)\citenamefont {Zheng},
  \citenamefont {Wei}, \citenamefont {Tsui},\ and\ \citenamefont
  {Weimann}}]{PhysRevB.34.5635}%
  \BibitemOpen
  \bibfield  {author} {\bibinfo {author} {\bibfnamefont {H.~Z.}\ \bibnamefont
  {Zheng}}, \bibinfo {author} {\bibfnamefont {H.~P.}\ \bibnamefont {Wei}},
  \bibinfo {author} {\bibfnamefont {D.~C.}\ \bibnamefont {Tsui}},\ and\
  \bibinfo {author} {\bibfnamefont {G.}~\bibnamefont {Weimann}},\ }\bibfield
  {title} {\bibinfo {title} {Gate-controlled transport in narrow
  gaas/${\mathrm{al}}_{\mathrm{x}}$${\mathrm{ga}}_{1\mathrm{\ensuremath{-}}\mathrm{x}}$as
  heterostructures},\ }\href {https://doi.org/10.1103/PhysRevB.34.5635}
  {\bibfield  {journal} {\bibinfo  {journal} {Phys. Rev. B}\ }\textbf {\bibinfo
  {volume} {34}},\ \bibinfo {pages} {5635} (\bibinfo {year}
  {1986})}\BibitemShut {NoStop}%
\bibitem [{\citenamefont {Thornton}\ \emph {et~al.}(1986)\citenamefont
  {Thornton}, \citenamefont {Pepper}, \citenamefont {Ahmed}, \citenamefont
  {Andrews},\ and\ \citenamefont {Davies}}]{PhysRevLett.56.1198}%
  \BibitemOpen
  \bibfield  {author} {\bibinfo {author} {\bibfnamefont {T.~J.}\ \bibnamefont
  {Thornton}}, \bibinfo {author} {\bibfnamefont {M.}~\bibnamefont {Pepper}},
  \bibinfo {author} {\bibfnamefont {H.}~\bibnamefont {Ahmed}}, \bibinfo
  {author} {\bibfnamefont {D.}~\bibnamefont {Andrews}},\ and\ \bibinfo {author}
  {\bibfnamefont {G.~J.}\ \bibnamefont {Davies}},\ }\bibfield  {title}
  {\bibinfo {title} {One-dimensional conduction in the 2d electron gas of a
  gaas-algaas heterojunction},\ }\href
  {https://doi.org/10.1103/PhysRevLett.56.1198} {\bibfield  {journal} {\bibinfo
   {journal} {Phys. Rev. Lett.}\ }\textbf {\bibinfo {volume} {56}},\ \bibinfo
  {pages} {1198} (\bibinfo {year} {1986})}\BibitemShut {NoStop}%
\bibitem [{\citenamefont {Beenakker}\ and\ \citenamefont {van
  Houten}(1991)}]{beenakker1991quantum}%
  \BibitemOpen
  \bibfield  {author} {\bibinfo {author} {\bibfnamefont {C.}~\bibnamefont
  {Beenakker}}\ and\ \bibinfo {author} {\bibfnamefont {H.}~\bibnamefont {van
  Houten}},\ }\bibfield  {title} {\bibinfo {title} {Quantum transport in
  semiconductor nanostructures},\ }in\ \href@noop {} {\emph {\bibinfo
  {booktitle} {Solid state physics}}},\ Vol.~\bibinfo {volume} {44}\ (\bibinfo
  {publisher} {Elsevier},\ \bibinfo {year} {1991})\ pp.\ \bibinfo {pages}
  {1--228}\BibitemShut {NoStop}%
\bibitem [{\citenamefont {Brandner}\ \emph {et~al.}(2013)\citenamefont
  {Brandner}, \citenamefont {Saito},\ and\ \citenamefont
  {Seifert}}]{PhysRevLett.110.070603}%
  \BibitemOpen
  \bibfield  {author} {\bibinfo {author} {\bibfnamefont {K.}~\bibnamefont
  {Brandner}}, \bibinfo {author} {\bibfnamefont {K.}~\bibnamefont {Saito}},\
  and\ \bibinfo {author} {\bibfnamefont {U.}~\bibnamefont {Seifert}},\
  }\bibfield  {title} {\bibinfo {title} {Strong bounds on onsager coefficients
  and efficiency for three-terminal thermoelectric transport in a magnetic
  field},\ }\href {https://doi.org/10.1103/PhysRevLett.110.070603} {\bibfield
  {journal} {\bibinfo  {journal} {Phys. Rev. Lett.}\ }\textbf {\bibinfo
  {volume} {110}},\ \bibinfo {pages} {070603} (\bibinfo {year}
  {2013})}\BibitemShut {NoStop}%
\bibitem [{\citenamefont {S\'anchez}\ and\ \citenamefont
  {B\"uttiker}(2004)}]{PhysRevLett.93.106802}%
  \BibitemOpen
  \bibfield  {author} {\bibinfo {author} {\bibfnamefont {D.}~\bibnamefont
  {S\'anchez}}\ and\ \bibinfo {author} {\bibfnamefont {M.}~\bibnamefont
  {B\"uttiker}},\ }\bibfield  {title} {\bibinfo {title} {Magnetic-field
  asymmetry of nonlinear mesoscopic transport},\ }\href
  {https://doi.org/10.1103/PhysRevLett.93.106802} {\bibfield  {journal}
  {\bibinfo  {journal} {Phys. Rev. Lett.}\ }\textbf {\bibinfo {volume} {93}},\
  \bibinfo {pages} {106802} (\bibinfo {year} {2004})}\BibitemShut {NoStop}%
\bibitem [{\citenamefont {Hwang}\ \emph {et~al.}(2014)\citenamefont {Hwang},
  \citenamefont {L\'opez}, \citenamefont {Lee},\ and\ \citenamefont
  {S\'anchez}}]{PhysRevB.90.115301}%
  \BibitemOpen
  \bibfield  {author} {\bibinfo {author} {\bibfnamefont {S.-Y.}\ \bibnamefont
  {Hwang}}, \bibinfo {author} {\bibfnamefont {R.}~\bibnamefont {L\'opez}},
  \bibinfo {author} {\bibfnamefont {M.}~\bibnamefont {Lee}},\ and\ \bibinfo
  {author} {\bibfnamefont {D.}~\bibnamefont {S\'anchez}},\ }\bibfield  {title}
  {\bibinfo {title} {Nonlinear spin-thermoelectric transport in two-dimensional
  topological insulators},\ }\href {https://doi.org/10.1103/PhysRevB.90.115301}
  {\bibfield  {journal} {\bibinfo  {journal} {Phys. Rev. B}\ }\textbf {\bibinfo
  {volume} {90}},\ \bibinfo {pages} {115301} (\bibinfo {year}
  {2014})}\BibitemShut {NoStop}%
\bibitem [{\citenamefont {S\'anchez}\ \emph {et~al.}(2019)\citenamefont
  {S\'anchez}, \citenamefont {S\'anchez}, \citenamefont {L\'opez},\ and\
  \citenamefont {Sothmann}}]{PhysRevB.99.245304}%
  \BibitemOpen
  \bibfield  {author} {\bibinfo {author} {\bibfnamefont {D.}~\bibnamefont
  {S\'anchez}}, \bibinfo {author} {\bibfnamefont {R.}~\bibnamefont
  {S\'anchez}}, \bibinfo {author} {\bibfnamefont {R.}~\bibnamefont {L\'opez}},\
  and\ \bibinfo {author} {\bibfnamefont {B.}~\bibnamefont {Sothmann}},\
  }\bibfield  {title} {\bibinfo {title} {Nonlinear chiral refrigerators},\
  }\href {https://doi.org/10.1103/PhysRevB.99.245304} {\bibfield  {journal}
  {\bibinfo  {journal} {Phys. Rev. B}\ }\textbf {\bibinfo {volume} {99}},\
  \bibinfo {pages} {245304} (\bibinfo {year} {2019})}\BibitemShut {NoStop}%
\bibitem [{\citenamefont {Buttiker}(1993)}]{buttiker1993capacitance}%
  \BibitemOpen
  \bibfield  {author} {\bibinfo {author} {\bibfnamefont {M.}~\bibnamefont
  {Buttiker}},\ }\bibfield  {title} {\bibinfo {title} {Capacitance, admittance,
  and rectification properties of small conductors},\ }\href@noop {} {\bibfield
   {journal} {\bibinfo  {journal} {Journal of Physics: Condensed Matter}\
  }\textbf {\bibinfo {volume} {5}},\ \bibinfo {pages} {9361} (\bibinfo {year}
  {1993})}\BibitemShut {NoStop}%
\bibitem [{\citenamefont {Christen}\ and\ \citenamefont
  {B{\"u}ttiker}(1996)}]{christen1996gauge}%
  \BibitemOpen
  \bibfield  {author} {\bibinfo {author} {\bibfnamefont {T.}~\bibnamefont
  {Christen}}\ and\ \bibinfo {author} {\bibfnamefont {M.}~\bibnamefont
  {B{\"u}ttiker}},\ }\bibfield  {title} {\bibinfo {title} {Gauge-invariant
  nonlinear electric transport in mesoscopic conductors},\ }\href@noop {}
  {\bibfield  {journal} {\bibinfo  {journal} {Europhysics letters}\ }\textbf
  {\bibinfo {volume} {35}},\ \bibinfo {pages} {523} (\bibinfo {year}
  {1996})}\BibitemShut {NoStop}%
\bibitem [{\citenamefont {S\'anchez}\ and\ \citenamefont
  {L\'opez}(2013)}]{PhysRevLett.110.026804}%
  \BibitemOpen
  \bibfield  {author} {\bibinfo {author} {\bibfnamefont {D.}~\bibnamefont
  {S\'anchez}}\ and\ \bibinfo {author} {\bibfnamefont {R.}~\bibnamefont
  {L\'opez}},\ }\bibfield  {title} {\bibinfo {title} {Scattering theory of
  nonlinear thermoelectric transport},\ }\href
  {https://doi.org/10.1103/PhysRevLett.110.026804} {\bibfield  {journal}
  {\bibinfo  {journal} {Phys. Rev. Lett.}\ }\textbf {\bibinfo {volume} {110}},\
  \bibinfo {pages} {026804} (\bibinfo {year} {2013})}\BibitemShut {NoStop}%
\bibitem [{\citenamefont {L\'opez}\ and\ \citenamefont
  {S\'anchez}(2013)}]{PhysRevB.88.045129}%
  \BibitemOpen
  \bibfield  {author} {\bibinfo {author} {\bibfnamefont {R.}~\bibnamefont
  {L\'opez}}\ and\ \bibinfo {author} {\bibfnamefont {D.}~\bibnamefont
  {S\'anchez}},\ }\bibfield  {title} {\bibinfo {title} {Nonlinear heat
  transport in mesoscopic conductors: Rectification, peltier effect, and
  wiedemann-franz law},\ }\href {https://doi.org/10.1103/PhysRevB.88.045129}
  {\bibfield  {journal} {\bibinfo  {journal} {Phys. Rev. B}\ }\textbf {\bibinfo
  {volume} {88}},\ \bibinfo {pages} {045129} (\bibinfo {year}
  {2013})}\BibitemShut {NoStop}%
\bibitem [{\citenamefont {Meair}\ and\ \citenamefont
  {Jacquod}(2013)}]{Meair_2013}%
  \BibitemOpen
  \bibfield  {author} {\bibinfo {author} {\bibfnamefont {J.}~\bibnamefont
  {Meair}}\ and\ \bibinfo {author} {\bibfnamefont {P.}~\bibnamefont
  {Jacquod}},\ }\bibfield  {title} {\bibinfo {title} {Scattering theory of
  nonlinear thermoelectricity in quantum coherent conductors},\ }\href
  {https://doi.org/10.1088/0953-8984/25/8/082201} {\bibfield  {journal}
  {\bibinfo  {journal} {Journal of Physics: Condensed Matter}\ }\textbf
  {\bibinfo {volume} {25}},\ \bibinfo {pages} {082201} (\bibinfo {year}
  {2013})}\BibitemShut {NoStop}%
\bibitem [{\citenamefont {Christen}\ and\ \citenamefont
  {B\"uttiker}(1996)}]{PhysRevLett.77.143}%
  \BibitemOpen
  \bibfield  {author} {\bibinfo {author} {\bibfnamefont {T.}~\bibnamefont
  {Christen}}\ and\ \bibinfo {author} {\bibfnamefont {M.}~\bibnamefont
  {B\"uttiker}},\ }\bibfield  {title} {\bibinfo {title} {Low frequency
  admittance of a quantum point contact},\ }\href
  {https://doi.org/10.1103/PhysRevLett.77.143} {\bibfield  {journal} {\bibinfo
  {journal} {Phys. Rev. Lett.}\ }\textbf {\bibinfo {volume} {77}},\ \bibinfo
  {pages} {143} (\bibinfo {year} {1996})}\BibitemShut {NoStop}%
\bibitem [{My_(2024)}]{My_mathematica_notebook}%
  \BibitemOpen
  \href {https://github.com/Sachiraj/thermoelectricity.git} {\bibinfo {title}
  {{Mathematica notebook for the numerical calculation can be found in
  https://github.com/Sachiraj/thermoelectricity.git}}} (\bibinfo {year}
  {2024})\BibitemShut {NoStop}%
\bibitem [{\citenamefont {Giazotto}\ \emph {et~al.}(2006)\citenamefont
  {Giazotto}, \citenamefont {Heikkil\"a}, \citenamefont {Luukanen},
  \citenamefont {Savin},\ and\ \citenamefont {Pekola}}]{RevModPhys.78.217}%
  \BibitemOpen
  \bibfield  {author} {\bibinfo {author} {\bibfnamefont {F.}~\bibnamefont
  {Giazotto}}, \bibinfo {author} {\bibfnamefont {T.~T.}\ \bibnamefont
  {Heikkil\"a}}, \bibinfo {author} {\bibfnamefont {A.}~\bibnamefont
  {Luukanen}}, \bibinfo {author} {\bibfnamefont {A.~M.}\ \bibnamefont
  {Savin}},\ and\ \bibinfo {author} {\bibfnamefont {J.~P.}\ \bibnamefont
  {Pekola}},\ }\bibfield  {title} {\bibinfo {title} {Opportunities for
  mesoscopics in thermometry and refrigeration: Physics and applications},\
  }\href {https://doi.org/10.1103/RevModPhys.78.217} {\bibfield  {journal}
  {\bibinfo  {journal} {Rev. Mod. Phys.}\ }\textbf {\bibinfo {volume} {78}},\
  \bibinfo {pages} {217} (\bibinfo {year} {2006})}\BibitemShut {NoStop}%
\bibitem [{\citenamefont {Muhonen}\ \emph {et~al.}(2012)\citenamefont
  {Muhonen}, \citenamefont {Meschke},\ and\ \citenamefont
  {Pekola}}]{Muhonen_2012}%
  \BibitemOpen
  \bibfield  {author} {\bibinfo {author} {\bibfnamefont {J.~T.}\ \bibnamefont
  {Muhonen}}, \bibinfo {author} {\bibfnamefont {M.}~\bibnamefont {Meschke}},\
  and\ \bibinfo {author} {\bibfnamefont {J.~P.}\ \bibnamefont {Pekola}},\
  }\bibfield  {title} {\bibinfo {title} {Micrometre-scale refrigerators},\
  }\href {https://doi.org/10.1088/0034-4885/75/4/046501} {\bibfield  {journal}
  {\bibinfo  {journal} {Reports on Progress in Physics}\ }\textbf {\bibinfo
  {volume} {75}},\ \bibinfo {pages} {046501} (\bibinfo {year}
  {2012})}\BibitemShut {NoStop}%
\bibitem [{\citenamefont {Nichele}\ \emph {et~al.}(2016)\citenamefont
  {Nichele}, \citenamefont {Suominen}, \citenamefont {Kjaergaard},
  \citenamefont {Marcus}, \citenamefont {Sajadi}, \citenamefont {Folk},
  \citenamefont {Qu}, \citenamefont {Beukman}, \citenamefont {de~Vries},
  \citenamefont {van Veen}, \citenamefont {Nadj-Perge}, \citenamefont
  {Kouwenhoven}, \citenamefont {Nguyen}, \citenamefont {Kiselev}, \citenamefont
  {Yi}, \citenamefont {Sokolich}, \citenamefont {Manfra}, \citenamefont
  {Spanton},\ and\ \citenamefont {Moler}}]{Nichele_2016}%
  \BibitemOpen
  \bibfield  {author} {\bibinfo {author} {\bibfnamefont {F.}~\bibnamefont
  {Nichele}}, \bibinfo {author} {\bibfnamefont {H.~J.}\ \bibnamefont
  {Suominen}}, \bibinfo {author} {\bibfnamefont {M.}~\bibnamefont
  {Kjaergaard}}, \bibinfo {author} {\bibfnamefont {C.~M.}\ \bibnamefont
  {Marcus}}, \bibinfo {author} {\bibfnamefont {E.}~\bibnamefont {Sajadi}},
  \bibinfo {author} {\bibfnamefont {J.~A.}\ \bibnamefont {Folk}}, \bibinfo
  {author} {\bibfnamefont {F.}~\bibnamefont {Qu}}, \bibinfo {author}
  {\bibfnamefont {A.~J.~A.}\ \bibnamefont {Beukman}}, \bibinfo {author}
  {\bibfnamefont {F.~K.}\ \bibnamefont {de~Vries}}, \bibinfo {author}
  {\bibfnamefont {J.}~\bibnamefont {van Veen}}, \bibinfo {author}
  {\bibfnamefont {S.}~\bibnamefont {Nadj-Perge}}, \bibinfo {author}
  {\bibfnamefont {L.~P.}\ \bibnamefont {Kouwenhoven}}, \bibinfo {author}
  {\bibfnamefont {B.-M.}\ \bibnamefont {Nguyen}}, \bibinfo {author}
  {\bibfnamefont {A.~A.}\ \bibnamefont {Kiselev}}, \bibinfo {author}
  {\bibfnamefont {W.}~\bibnamefont {Yi}}, \bibinfo {author} {\bibfnamefont
  {M.}~\bibnamefont {Sokolich}}, \bibinfo {author} {\bibfnamefont {M.~J.}\
  \bibnamefont {Manfra}}, \bibinfo {author} {\bibfnamefont {E.~M.}\
  \bibnamefont {Spanton}},\ and\ \bibinfo {author} {\bibfnamefont {K.~A.}\
  \bibnamefont {Moler}},\ }\bibfield  {title} {\bibinfo {title} {Edge transport
  in the trivial phase of inas/gasb},\ }\href
  {https://doi.org/10.1088/1367-2630/18/8/083005} {\bibfield  {journal}
  {\bibinfo  {journal} {New Journal of Physics}\ }\textbf {\bibinfo {volume}
  {18}},\ \bibinfo {pages} {083005} (\bibinfo {year} {2016})}\BibitemShut
  {NoStop}%
\bibitem [{\citenamefont {Mani}\ and\ \citenamefont
  {Benjamin}(2017{\natexlab{b}})}]{mani2017probing}%
  \BibitemOpen
  \bibfield  {author} {\bibinfo {author} {\bibfnamefont {A.}~\bibnamefont
  {Mani}}\ and\ \bibinfo {author} {\bibfnamefont {C.}~\bibnamefont
  {Benjamin}},\ }\bibfield  {title} {\bibinfo {title} {Probing helicity and the
  topological origins of helicity via non-local hanbury-brown and twiss
  correlations},\ }\href@noop {} {\bibfield  {journal} {\bibinfo  {journal}
  {Scientific Reports}\ }\textbf {\bibinfo {volume} {7}},\ \bibinfo {pages}
  {6954} (\bibinfo {year} {2017}{\natexlab{b}})}\BibitemShut {NoStop}%
\bibitem [{\citenamefont {Mishra}\ and\ \citenamefont
  {Benjamin}(2023)}]{PhysRevB.108.115301}%
  \BibitemOpen
  \bibfield  {author} {\bibinfo {author} {\bibfnamefont {S.}~\bibnamefont
  {Mishra}}\ and\ \bibinfo {author} {\bibfnamefont {C.}~\bibnamefont
  {Benjamin}},\ }\bibfield  {title} {\bibinfo {title} {Finite-temperature
  quantum noise correlations as a probe for topological helical edge modes},\
  }\href {https://doi.org/10.1103/PhysRevB.108.115301} {\bibfield  {journal}
  {\bibinfo  {journal} {Phys. Rev. B}\ }\textbf {\bibinfo {volume} {108}},\
  \bibinfo {pages} {115301} (\bibinfo {year} {2023})}\BibitemShut {NoStop}%
\bibitem [{\citenamefont {Chang}\ \emph {et~al.}(2015)\citenamefont {Chang},
  \citenamefont {Zhao}, \citenamefont {Kim}, \citenamefont {Zhang},
  \citenamefont {Assaf}, \citenamefont {Heiman}, \citenamefont {Zhang},
  \citenamefont {Liu}, \citenamefont {Chan},\ and\ \citenamefont
  {Moodera}}]{chang2015high}%
  \BibitemOpen
  \bibfield  {author} {\bibinfo {author} {\bibfnamefont {C.-Z.}\ \bibnamefont
  {Chang}}, \bibinfo {author} {\bibfnamefont {W.}~\bibnamefont {Zhao}},
  \bibinfo {author} {\bibfnamefont {D.~Y.}\ \bibnamefont {Kim}}, \bibinfo
  {author} {\bibfnamefont {H.}~\bibnamefont {Zhang}}, \bibinfo {author}
  {\bibfnamefont {B.~A.}\ \bibnamefont {Assaf}}, \bibinfo {author}
  {\bibfnamefont {D.}~\bibnamefont {Heiman}}, \bibinfo {author} {\bibfnamefont
  {S.-C.}\ \bibnamefont {Zhang}}, \bibinfo {author} {\bibfnamefont
  {C.}~\bibnamefont {Liu}}, \bibinfo {author} {\bibfnamefont {M.~H.}\
  \bibnamefont {Chan}},\ and\ \bibinfo {author} {\bibfnamefont {J.~S.}\
  \bibnamefont {Moodera}},\ }\bibfield  {title} {\bibinfo {title}
  {High-precision realization of robust quantum anomalous hall state in a hard
  ferromagnetic topological insulator},\ }\href@noop {} {\bibfield  {journal}
  {\bibinfo  {journal} {Nature materials}\ }\textbf {\bibinfo {volume} {14}},\
  \bibinfo {pages} {473} (\bibinfo {year} {2015})}\BibitemShut {NoStop}%
\bibitem [{\citenamefont {Kou}\ \emph {et~al.}(2014)\citenamefont {Kou},
  \citenamefont {Guo}, \citenamefont {Fan}, \citenamefont {Pan}, \citenamefont
  {Lang}, \citenamefont {Jiang}, \citenamefont {Shao}, \citenamefont {Nie},
  \citenamefont {Murata}, \citenamefont {Tang}, \citenamefont {Wang},
  \citenamefont {He}, \citenamefont {Lee}, \citenamefont {Lee},\ and\
  \citenamefont {Wang}}]{PhysRevLett.113.137201}%
  \BibitemOpen
  \bibfield  {author} {\bibinfo {author} {\bibfnamefont {X.}~\bibnamefont
  {Kou}}, \bibinfo {author} {\bibfnamefont {S.-T.}\ \bibnamefont {Guo}},
  \bibinfo {author} {\bibfnamefont {Y.}~\bibnamefont {Fan}}, \bibinfo {author}
  {\bibfnamefont {L.}~\bibnamefont {Pan}}, \bibinfo {author} {\bibfnamefont
  {M.}~\bibnamefont {Lang}}, \bibinfo {author} {\bibfnamefont {Y.}~\bibnamefont
  {Jiang}}, \bibinfo {author} {\bibfnamefont {Q.}~\bibnamefont {Shao}},
  \bibinfo {author} {\bibfnamefont {T.}~\bibnamefont {Nie}}, \bibinfo {author}
  {\bibfnamefont {K.}~\bibnamefont {Murata}}, \bibinfo {author} {\bibfnamefont
  {J.}~\bibnamefont {Tang}}, \bibinfo {author} {\bibfnamefont {Y.}~\bibnamefont
  {Wang}}, \bibinfo {author} {\bibfnamefont {L.}~\bibnamefont {He}}, \bibinfo
  {author} {\bibfnamefont {T.-K.}\ \bibnamefont {Lee}}, \bibinfo {author}
  {\bibfnamefont {W.-L.}\ \bibnamefont {Lee}},\ and\ \bibinfo {author}
  {\bibfnamefont {K.~L.}\ \bibnamefont {Wang}},\ }\bibfield  {title} {\bibinfo
  {title} {Scale-invariant quantum anomalous hall effect in magnetic
  topological insulators beyond the two-dimensional limit},\ }\href
  {https://doi.org/10.1103/PhysRevLett.113.137201} {\bibfield  {journal}
  {\bibinfo  {journal} {Phys. Rev. Lett.}\ }\textbf {\bibinfo {volume} {113}},\
  \bibinfo {pages} {137201} (\bibinfo {year} {2014})}\BibitemShut {NoStop}%
\bibitem [{\citenamefont {Checkelsky}\ \emph {et~al.}(2014)\citenamefont
  {Checkelsky}, \citenamefont {Yoshimi}, \citenamefont {Tsukazaki},
  \citenamefont {Takahashi}, \citenamefont {Kozuka}, \citenamefont {Falson},
  \citenamefont {Kawasaki},\ and\ \citenamefont {Tokura}}]{Checkelsky_2014}%
  \BibitemOpen
  \bibfield  {author} {\bibinfo {author} {\bibfnamefont {J.~G.}\ \bibnamefont
  {Checkelsky}}, \bibinfo {author} {\bibfnamefont {R.}~\bibnamefont {Yoshimi}},
  \bibinfo {author} {\bibfnamefont {A.}~\bibnamefont {Tsukazaki}}, \bibinfo
  {author} {\bibfnamefont {K.~S.}\ \bibnamefont {Takahashi}}, \bibinfo {author}
  {\bibfnamefont {Y.}~\bibnamefont {Kozuka}}, \bibinfo {author} {\bibfnamefont
  {J.}~\bibnamefont {Falson}}, \bibinfo {author} {\bibfnamefont
  {M.}~\bibnamefont {Kawasaki}},\ and\ \bibinfo {author} {\bibfnamefont
  {Y.}~\bibnamefont {Tokura}},\ }\bibfield  {title} {\bibinfo {title}
  {Trajectory of the anomalous hall effect towards the quantized state in a
  ferromagnetic topological insulator},\ }\href
  {https://doi.org/10.1038/nphys3053} {\bibfield  {journal} {\bibinfo
  {journal} {Nature Physics}\ }\textbf {\bibinfo {volume} {10}},\ \bibinfo
  {pages} {731} (\bibinfo {year} {2014})}\BibitemShut {NoStop}%
\end{thebibliography}%

\clearpage
\onecolumngrid

\makeatletter
\renewcommand\paragraph{\@startsection{paragraph}{4}{\z@}%
            {-2.5ex\@plus -1ex \@minus -.25ex}%
            {1.25ex \@plus .25ex}%
            {\normalfont\normalsize\bfseries}}
\makeatother
\setcounter{secnumdepth}{4}
\setcounter{tocdepth}{4} 

\newcommand{\mathsym}[1]{{}}
\newcommand{\unicode}[1]{{}}

\newcounter{mathematicapage}
\definecolor{orange}{RGB}{255,127,0}
\definecolor{blue2}{RGB}{33,114,173}

\section*{Supplementary Material for ``Reaching Van Den Broeck limit in linear response and Whitney limit in nonlinear response in edge mode
quantum thermoelectrics and refrigeration"}
\begin{center}
    Sachiraj Mishra$^{1,2}$, Colin Benjamin$^{1,2}$\\
    $^1$\textit{School of Physical Sciences, National Institute of Science Education and Research, HBNI, Jatni-752050, India}\\
$^2$\textit{Homi Bhabha National Institute, Training School Complex, AnushaktiNagar, Mumbai, 400094, India}
\end{center}

\maketitle
In this supplementary material, we discuss various aspects of linear and nonlinear thermoelectricity for both quantum Hall (QH) and quantum spin Hall (QSH) systems with voltage-temperature probe. We also discuss the technical details, and methods, and summarize the calculations. In Sec. \ref{Sec:II}, we derive the formulae for charge and heat current for chiral edge mode transport in a generic multiterminal QH setup and discuss the linear transport regime (See Sec. \ref{Sec:II(A)}), and derive charge and heat current in terms of Onsager coefficients. In Sec. \ref{Sec:II(A1)}, we discuss our setup and the transmission probabilities and also derive the Onsager coefficients with a voltage-temperature probe. In Sec. \ref{Sec:II(A2)}, we derive the formula for maximum power ($P_{max}$), efficiency at maximum power ($\eta|_{P_{max}}$) and maximum efficiency ($\eta_{max}$) for QH setup as a quantum heat engine. Further in Sec. \ref{Sec:II(A3)}, we derive the formula for the maximum coefficient of performance ($\eta^r_{max}$) and maximum cooling power $J|_{\eta^r_{max}}$ and in Sec. \ref{Sec:II(A4)}, we discuss why our QH setup can work as a quantum refrigerator even if time-reversal symmetry is broken, whereas a QH setup discussed in \cite{PhysRevLett.114.146801} cannot work as a refrigerator. In Sec. \ref{Sec:II(B)}, we discuss the transport in the nonlinear transport regime for a QH setup, whereas in Sec. \ref{Sec:II(B1)}, we discuss the QH setup as a nonlinear quantum heat engine and quantum refrigerator using {Boxcar-type constriction}. Then in Sec. \ref{Sec:II(B2)}, we discuss the 2T QH setup both as a quantum heat engine and refrigerator using {QPC-type constriction and also estimate the interaction potential and the applied gate voltage to nullify it}, where we extend the same study to a 3T QH setup in Sec. \ref{Sec:II(B3)}. Finally, in Sec. \ref{Sec:II(B4)}, we study the nonlinear thermoelectricity both as a quantum heat engine and refrigerator for the 3T QH setup with a voltage-temperature probe, {which also includes the estimation of interaction potential of two QPC-type constrictions and their respective applied gate voltages}. In Sec. \ref{Sec:III}, we discuss the charge and heat currents for Helical edge mode transport in a generic multiterminal QSH setup in the linear response regime, and Sec. \ref{Sec:III(A)} discusses the transport in the linear response regime in a generic multiterminal QSH setup. Next, we discuss our QSH setup and its transmission probabilities and in Sec. \ref{Sec:III(A1)} the Onsager coefficients are derived with a voltage-temperature probe. In Sec. \ref{Sec:III(A2)}, we derive $P_{max}$, $\eta|_{P_{max}}$ and $\eta_{max}$ for the QSH setup as a quantum heat engine. In Sec. \ref{Sec:III(A3)}, we further discuss the cooling power and coefficient of performance in our 3T QSH setup as a quantum refrigerator. In Sec. \ref{Sec:III(B)}, we discuss the nonlinear transport regime in QSH setup, whereas in Sec. \ref{Sec:III(B1)}, we discuss the QSH setup as a nonlinear quantum heat engine and quantum refrigerator using {Boxcar-type constriction}. Then in Sec. \ref{Sec:III(B2)}, we discuss the 2T QSH setup both as a quantum heat engine and refrigerator using {QPC-type constriction}, {which includes a discussion on the estimation of interaction potential and its nullification by the applied gate voltage}, and we also extend the same study to a 3T QSH setup in Sec. \ref{Sec:III(B3)}. Finally, in Sec. \ref{Sec:III(B4)}, we study the nonlinear thermoelectricity both as a quantum heat engine and refrigerator for the 3T QSH setup with a voltage-temperature probe, {which also delves into the interaction potential of two QPC-type constrictions and the applied gate voltage to nullify it}.

\section{Charge and heat currents in chiral QH setup} \label{Sec:II}
For a multiterminal QH setup, the charge and heat current in terminal $\alpha$ can be derived using Landauer-Buttiker formalism. First, we focus on the charge current in terminal $\alpha$, which has a Fermi-Dirac distribution $f_{\alpha}(E) = \left(1+e^{\frac{E - \mu_{\alpha}}{k_B T_{\alpha}}}\right)^{-1}$, where $\mu_{\alpha}$ and $T_{\alpha}$ are the equilibrium chemical potential and temperature in terminal $\alpha$ respectively. The current outgoing from terminal $\alpha$ is given as,
\begin{equation} \label{eq:4}
   I^{\text{out}}_{\alpha} = \frac{e}{L}\sum_k v f_{\alpha}(E),\quad \text{where the summation is over all `$k$' states of electrons.}
\end{equation}

Electrons can enter terminal $\alpha$ from any other terminal $\beta$ including itself. The fraction of particles incident from terminal $\beta$ in edge channel $m$ that scatter into terminal $\alpha$ in edge channel $n$ is $\sum_{\beta, m, n}|s_{\alpha n, \beta m}|^2 f_{\beta}$, where $s_{\alpha n, \beta m}$ is the amplitude for an electron to scatter from terminal $\beta$ in the edge channel $m$ into terminal $\alpha$ in the edge channel $n$. Thus, the incoming current in terminal $\alpha$ is given as,
\begin{equation} \label{eq:5}
    I_{\alpha}^{\text{in}} = \frac{e}{L}\sum_k v \sum_{\beta, m, n} |s_{\alpha n, \beta m}|^2 f_{\beta}(E).
\end{equation}

Thus, the total current in terminal $\alpha$ is,
\begin{align} \label{eq:6}
\begin{split}
    I_{\alpha} = I_{\alpha}^{\text{out}} - I_{\alpha}^{\text{in}} &= \frac{e}{L}\sum_k v \left(f_{\alpha}(E)-\sum_{\beta, m, n} |s_{\alpha n, \beta m}|^2\right)f_{\beta}(E) = \frac{e}{L}\sum_k v \sum_{\beta, m, n} \left(\delta_{\alpha \beta} \delta_{mn} - |s_{\alpha n, \beta m}|^2\right)f_{\beta}(E).
    \end{split}
\end{align}
Converting the summation into integral, i.e., $\sum_{k} =  2 \times \frac{L}{2\pi} \int dk$, Eq. (\ref{eq:6}) becomes,
\begin{equation} \label{eq:7}
    I_{\alpha} = 2e\int_{-\infty}^{\infty} \frac{dk}{2\pi}v(k) \sum_{\beta, m, n} \left(\delta_{\alpha \beta} \delta_{mn} - |s_{\alpha n, \beta m}|^2\right)f_{\beta}(E).
\end{equation}
where the extra factor of $2$ has been taken because of spin degeneracy.
Inserting $v(k) = \frac{1}{\hbar}\frac{dE}{ dk}$ in Eq. (\ref{eq:7}), we get,
\begin{align} \label{eq:8}
\begin{split}
    I_{\alpha} &= \frac{2e}{h}\int_{-\infty}^{\infty}dE \sum_{\beta}f_{\beta}(E) [N_{\alpha}\delta_{\alpha \beta} - Tr(s_{\alpha \beta}^{\dagger} s_{\alpha \beta})],
    \end{split}
\end{align}

where $N_{\alpha}$ is the number of edge modes in terminal $\alpha$ and $s_{\alpha \beta}^{\dagger}s_{\alpha \beta} = |s_{\alpha \beta}|^2 = \mathcal{T}_{\alpha \beta}$ is the transmission probability for an electron to transmit from terminal $\beta$ to terminal $\alpha$.
From probability conservation, i.e., $\sum_{\beta}\mathcal{T}_{\alpha \beta} = N_{\alpha}$, Eq. (\ref{eq:8}) can also be written as, 
\begin{equation}\label{eq:9}
    I_{\alpha} = \frac{2e}{h}\int_{-\infty}^{\infty}dE \sum_{\beta}\mathcal{T}_{\alpha \beta}(f_{\alpha} - f_{\beta}).
\end{equation}
This is the general formula for electric charge current irrespective of whether transport is linear or nonlinear.
Now, the heat current can be found using the first law of thermodynamics. According to this law, the heat in terminal $\alpha$ with chemical potential $\mu_{\alpha}$, wherein volume is kept constant, is given as,
\begin{equation}\label{eq:10}
    dQ_{\alpha} = dE_{\alpha} - \mu_{\alpha} dN_{\alpha}.
\end{equation}
where, $dE_{\alpha}$ is the change in the internal energy and $dN_{\alpha}$ is the change in particle number in terminal $\alpha$.
Taking the time derivative of the above equation, we get $J_{\alpha} = J_{\alpha E} - \mu_{\alpha} J_{\alpha N}$, where $J_{\alpha}$ is the heat current, $J_{ \alpha E}$ is the energy current and $J_{\alpha N}$ is the particle current in terminal $\alpha$, given as
\begin{equation}\label{eq:11}
    J_{\alpha E} = \frac{2}{h}\int_{-\infty}^{\infty}dE E \sum_{\beta}f_{\beta}(E)[N_{\alpha} \delta_{\alpha \beta} - Tr(s_{\alpha \beta}^{\dagger}s_{\alpha \beta})], \quad \text{and}\quad  J_{\alpha N} = \frac{2}{h}\int_{-\infty}^{\infty}dE \sum_{\beta}f_{\beta}(E)[N_{\alpha} \delta_{\alpha \beta} - Tr(s_{\alpha \beta}^{\dagger}s_{\alpha \beta})].
\end{equation}

The heat current is thus \cite{BENENTI20171},
\begin{equation} \label{eq:12}
    J_{\alpha} = \frac{2}{h}\int_{-\infty}^{\infty}dE (E - \mu_{\alpha}) \sum_{\beta}f_{\beta}(E) [N_{\alpha}\delta_{\alpha \beta} - Tr(s_{\alpha \beta}^{\dagger} s_{\alpha \beta})]= \frac{2}{h}\int_{-\infty}^{\infty}dE (E - \mu_{\alpha}) \sum_{\beta}\mathcal{T}_{\alpha \beta}(f_{\alpha} - f_{\beta}).
\end{equation}

This is the general formula for heat current irrespective of whether transport is linear or nonlinear.

\subsection{Transport in the linear response regime in the QH setup}\label{Sec:II(A)}
We can write charge and heat current as in Eqs. (\ref{eq:9}) and (\ref{eq:12}) in the linear response regime. We denote the chemical potential and temperature of reservoir $i$ as $\mu_i$ and $T_i$ respectively, we is given as,
\begin{align} \label{eq:13}
\begin{split}
    \mu_{i} = \mu + e V_{i},\quad
    T_{i} = T + \tau_{i}, \quad i \in \text{all terminals},
\end{split}
\end{align}
with Fermi-Dirac distribution $f_i = \left(1 + e^{\frac{E - \mu_i}{k_B T_i}}\right)^{-1}$. $\mu$ and $T$ are the equilibrium chemical potential and temperature respectively for terminal $i$ with Fermi-Dirac distribution $f = \left(1 + e^{\frac{E - \mu}{k_B T}}\right)^{-1}$ and $V_i$ and $\tau_i$ are the voltage bias and temperature bias applied in terminals $i$. Now, doing a Taylor series expansion of 
$f_{i}$ up to the first order, we get,
\begin{equation} \label{eq:14}
    f_{i}= f + \left. \frac{\partial f_i}{\partial \mu_{i}}\right\vert_{\mu_i = \mu, T_i = T}e V_{i} + \left. \frac{\partial f_i}{\partial T_{i}}\right\vert_{\mu_i = \mu, T_i = T} \tau_{i}.
\end{equation}
Using the property of $f_i$, 
\begin{align} \label{eq:15}
\begin{split}
   \left. \frac{\partial f_i}{\partial \mu_{i}}\right\vert_{\mu_i = \mu, T_i = T} &= \frac{-\partial f}{\partial E} \quad \text{and},
   \left. \frac{\partial f_i}{\partial T_{i}}\right\vert_{\mu_i = \mu, T_i = T} = -\left(\frac{E - \mu}{k_B T}\right)\frac{\partial f}{\partial E} \implies  f_{i}= f - \frac{\partial f}{\partial E}\left(e V_{i} + \frac{E - \mu}{k_BT} \tau_i\right).
   \end{split}
\end{align}

We can use Eq. (\ref{eq:15}) in Eqs. (\ref{eq:9}) and (\ref{eq:12}) to get the Onsager coefficients. The charge current as in Eq. (\ref{eq:9}) becomes,
\begin{equation} \label{eq:16}
\begin{split}
    I_{\alpha} = \frac{2e}{h}\int_{-\infty}^{\infty} dE \sum_{\beta} \left(f - \frac{\partial f}{\partial E}\left(e V_{\beta} + \frac{E - \mu}{T} \tau_{\beta}\right)\right)
    [N_{\alpha} \delta_{\alpha \beta} - \mathcal{T}_{\alpha \beta}],
    \end{split}
\end{equation}
 and, the heat current as in Eq. (\ref{eq:12}) becomes,
 \begin{equation} \label{eq:17}
     \begin{split}
         J_{\alpha} = \frac{2}{h} \int_{\infty}^{\infty} dE (E - \mu) \sum_{\beta}\left(f - \frac{\partial f}{\partial E}\left(e V_{\beta} + \frac{E - \mu}{T} \tau_{\beta}\right)\right) [N_{\alpha} \delta_{\alpha \beta} - \mathcal{T}_{\alpha \beta}],
     \end{split}
 \end{equation}
 where going from Eq. (\ref{eq:12}) to Eq. (\ref{eq:17}), we have taken $\mu_{\alpha} \simeq \mu$ and $\mathcal{T}_{\alpha \beta} = Tr(s_{\alpha \beta}^{\dagger}s_{\alpha \beta})$. Now in Eqs. (\ref{eq:16}) and (\ref{eq:17}), the equilibrium distribution $f$ makes zero contribution, since without any bias, there will be zero charge current and heat current, which is why Eqs. (\ref{eq:16}) and (\ref{eq:17}) reduce to,
 \begin{align} \label{eq:18}
 \begin{split}
     I_{\alpha} = \sum_{\beta} (G_{\alpha \beta}V_{\beta} + L_{\alpha \beta} \tau_{\beta}),\quad \text{and} \quad
     J_{\alpha} = \sum_{\beta} (\Pi_{\alpha \beta}V_{\beta} + K_{\alpha \beta} \tau_{\beta}).
     \end{split}
 \end{align}
 
 \begin{align} \label{eq:19}
     \begin{split}
 \text{where,}\quad        G_{\alpha \beta} &= \frac{2e^2}{h} \int_{-\infty}^{\infty}dE \left(- \frac{\partial f}{\partial E}\right)[N_{\alpha} \delta_{\alpha \beta} - \mathcal{T}_{\alpha \beta}],\quad 
         L_{\alpha \beta} = \frac{2e}{h T} \int_{-\infty}^{\infty} dE (E - \mu) \left(-\frac{\partial f}{\partial E}\right) [N_{\alpha}\delta_{\alpha \beta} - \mathcal{T}_{\alpha \beta}],\\
         \Pi_{\alpha \beta} &= \frac{2e}{h} \int_{-\infty}^{\infty} dE (E - \mu) \left(-\frac{\partial f}{\partial E}\right) [N_{\alpha}\delta_{\alpha \beta} - \mathcal{T}_{\alpha \beta}],\quad
         K_{\alpha \beta} = \frac{2}{h T} \int_{-\infty}^{\infty}dE (E - \mu)^2 \left(-\frac{\partial f}{\partial E}\right)[N_{\alpha}\delta_{\alpha \beta} - \mathcal{T}_{\alpha \beta}].
     \end{split}
 \end{align}\\
 The quantities as written in Eq. (\ref{eq:19}) are Onsager coefficients in linear response regime. 

In the subsequent subsections, we derive Onsager coefficients for our QH setup (See, Sec. \ref{Sec:II(A1)}) with a three-terminal voltage-temperature probe. In Sec. \ref{Sec:II(A2)}, we derive the general formulae for power and efficiency as a quantum heat engine. Further in Sec. \ref{Sec:II(A3)}, we discuss the cooling power and coefficient of performance as a quantum refrigerator and in Sec. \ref{Sec:II(A4)}, we discuss why our QH setup can work as a quantum refrigerator even if time-reversal symmetry is preserved although, the QH setup used in \cite{PhysRevLett.114.146801} cannot be used as a quantum refrigerator.

\begin{figure} 
\centering
\includegraphics[width=0.60\linewidth]{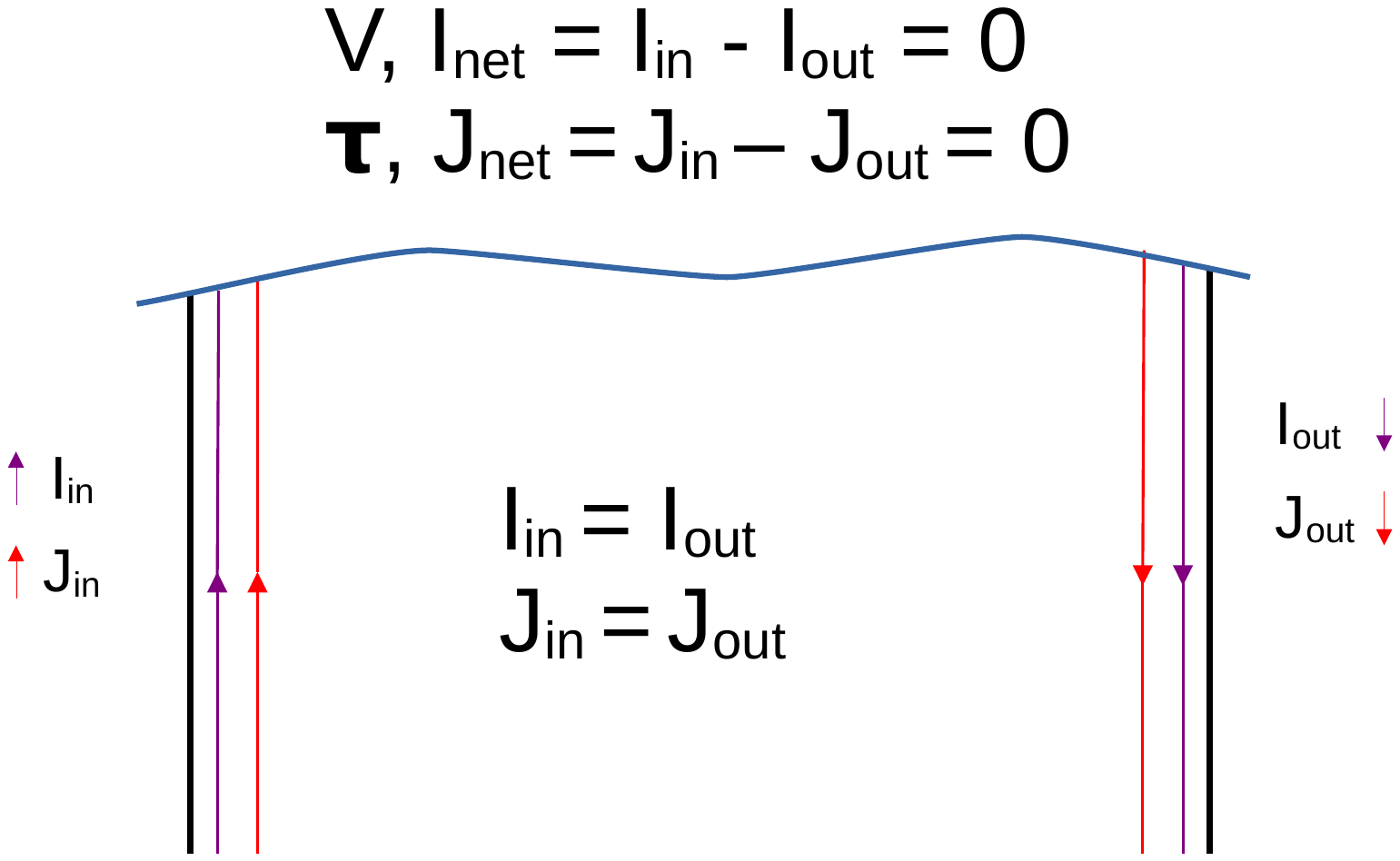}
\caption{A voltage-temperature probe. $V$ and $\tau'$ are the applied voltage bias and temperature bias at the reservoir (shown by wavy blue curve), which achieves $I_{net} = 0$ and $J_{net} = 0$ simultaneously. The purple arrows indicate the direction of the charge currents and the red arrows indicate the direction of the heat currents.}
\label{fig:2}
\end{figure}

\subsubsection{Derivation of Onsager coefficients in a three-terminal QH setup in the linear response regime}\label{Sec:II(A1)}
\begin{figure} 
\centering
\includegraphics[width=0.60\linewidth]{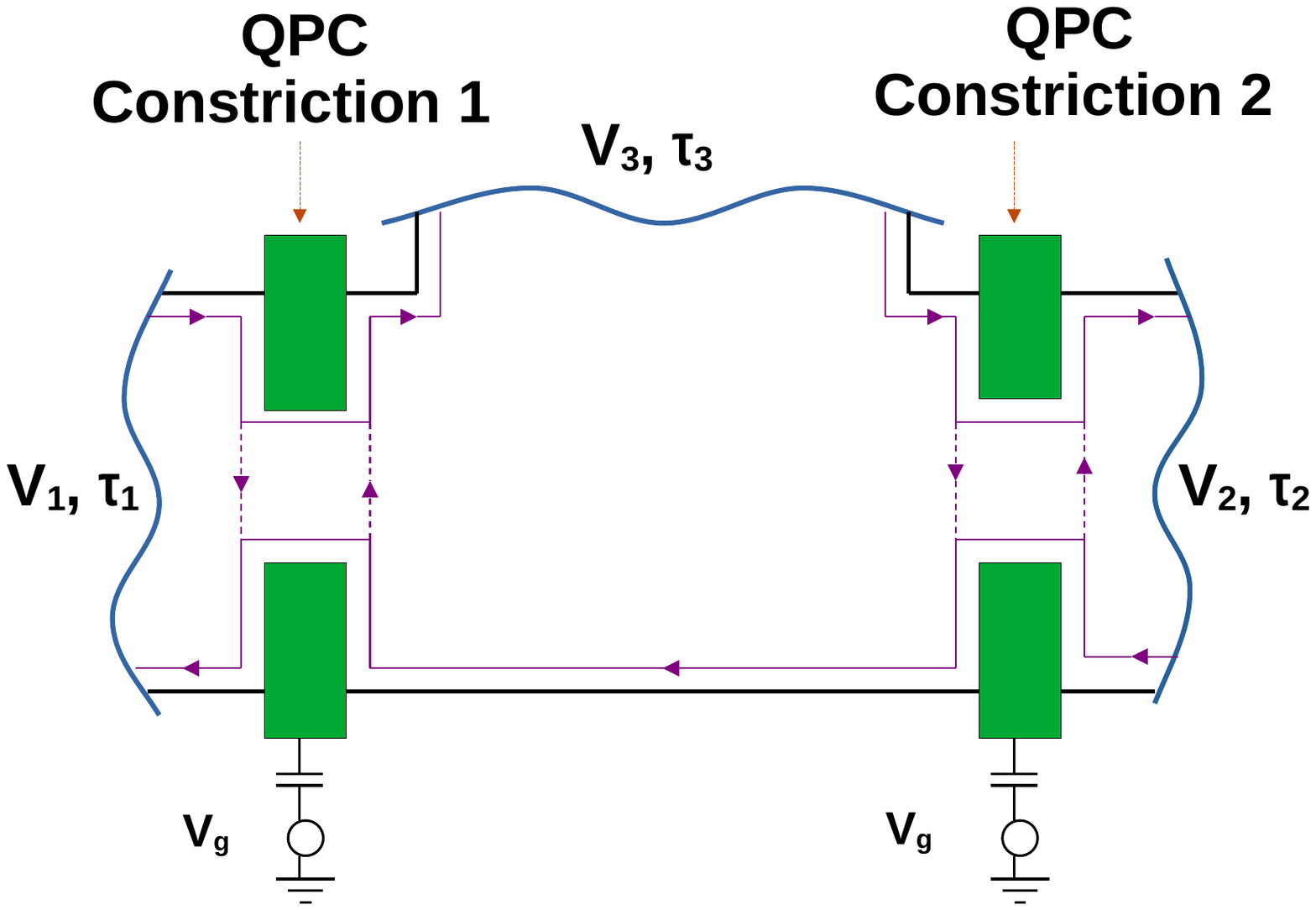}
\caption{Three terminal QH sample with chiral edge modes and two QPC-type constrictions. The edge mode of an electron, which can scatter from constriction, is shown by the purple solid (dashed) line. Terminal 3 is a voltage-temperature probe.}
\label{fig:3}
\end{figure}

  For this three-terminal QH setup (see, Fig. \ref{fig:3}), the transmission probabilities $\mathcal{T}_{\alpha \beta}$ can also be found. We discuss one example i.e., $\mathcal{T}_{12}$, which is the transmission probability for an electron to scatter from terminal 2 to terminal 1. For this process, the electron injects from terminal 2 with probability 1 and transmits via the constriction 2 and 1 with probabilities $\mathcal{T}_{2}$ and $\mathcal{T}_{1}$ respectively. Thus, the total probability for the electron to reach terminal 1 from terminal 2 is $\mathcal{T}_{1} \mathcal{T}_{2}$. Similarly, one can determine other transmission probabilities and they are given as, 

 \begin{align}\label{eq:20}
\begin{split}
\mathcal{T}_{11} &= 1-\mathcal{T}_1,\quad \mathcal{T}_{12} = \mathcal{T}_1 \mathcal{T}_2,\quad
    \mathcal{T}_{13} = \mathcal{T}_1(1-\mathcal{T}_2), \quad \mathcal{T}_{21} = 0,\quad
    \mathcal{T}_{22} = (1-\mathcal{T}_2),\quad
    \mathcal{T}_{23} = \mathcal{T}_2, \quad \mathcal{T}_{31} = \mathcal{T}_1,\\
    \mathcal{T}_{32} &= (1-\mathcal{T}_1)\mathcal{T}_2,\quad
    \mathcal{T}_{33} = (1-\mathcal{T}_1)(1-\mathcal{T}_2).
    \end{split}
\end{align}

For our QH setup, one can impose the voltage-temperature probe condition in terminal 3 in our setup as shown in Fig. \ref{fig:2}. In a generic multiterminal QH setup, the voltage-temperature probe is that terminal, where the net charge current ($I_{net} = I_{out} - I_{in}$) and net heat current ($J_{net} = J_{out} - J_{in}$) are zero simultaneously as shown in Fig. \ref{fig:3}, which can be achieved at an arbitrary voltage bias $V$ and temperature bias $\tau$.

Here, we assume $\tau_1 = \tau$ and $\tau_2 = 0$. For the voltage-temperature probe in our setup (see Fig. \ref{fig:3}), we solve for both $V_3$ and $\tau_3$.
According to Landauer-Buttiker formalism as shown in Eq. (\ref{eq:18}), the charge and heat currents in terminal 3 are given as,

\begin{align} \label{eq:21}
\begin{split}
    I_3 &= \sum_{\beta} G_{3 \beta}V_{\beta} + \sum_{\beta} L_{3 \beta} \tau_{\beta},\quad
    J_3 = \sum_{\beta} \Pi_{3 \beta} V_{\beta} + \sum_{\beta}K_{3 \beta} \tau_{\beta}.
    \end{split}
\end{align}
Putting $I_3 = J_3 = 0$, we get the solution for $V_3$ and $\tau_3$. They are given as,
\begin{align} \label{eq:22}
\begin{split}
    V_3 &= \frac{(-K_{33} L_{31} + L_{33}K_{31}) \tau - (-K_{33}G_{31}+L_{33}\Pi_{31})V}{X},\quad
    \tau_3 = \frac{(-G_{33} K_{31} + \Pi_{33}L_{31}) \tau - (-\Pi_{33}G_{31}+G_{33}\Pi_{31})V}{X}.
    \end{split}
\end{align}
 where, $X = G_{33}K_{33}-L_{33} \Pi_{33}$.
Now, the current conservation in this setup implies $I_1^e = - I_2^e$. The charge and heat currents in terminal 1 are given as,
\begin{align} \label{eq:23}
\begin{split}
    I_1 &= \sum_{\beta} G_{1 \beta}V_{\beta} + \sum_{\beta} L_{1 \beta} \tau_{\beta},\quad
    J_1 = \sum_{\beta} \Pi_{1 \beta} V_{\beta} + \sum_{\beta}K_{1 \beta} \tau_{\beta}.
    \end{split}
\end{align}
Now, using $V_3$ and $\tau_3$ from Eq. (\ref{eq:22}), Eq. (\ref{eq:23}) can be rewritten as,
\begin{equation} \label{eq:24}
    \begin{pmatrix}
        I_1 \\
        J_1
    \end{pmatrix} = \begin{pmatrix}
        L_{eV} && L_{e T}\\
        L_{hV} && L_{h T}
    \end{pmatrix} \begin{pmatrix}
        -V\\
        \tau
    \end{pmatrix}
\end{equation}
where,

    \begin{align} \label{eq:25}
        \begin{split}
            L_{eV} &= G_{11} + \frac{G_{13}(L_{33} \Pi_{31} - G_{31}K_{33})}{X} - \frac{L_{13}(G_{33}\Pi_{31}-G_{31}\Pi_{33})}{X},\\
            L_{e T} &= L_{11} + \frac{G_{13}(L_{33} K_{31} - L_{31}K_{33})}{X} - \frac{L_{13}(G_{33}K_{31}-L_{31}\Pi_{33})}{X},\\
            L_{hV} &= \Pi_{11} + \frac{\Pi_{13}(L_{33} \Pi_{31} - G_{31}K_{33})}{X} - \frac{K_{13}(G_{33}\Pi_{31}-G_{31}\Pi_{33})}{X},\\
              L_{h T} &= K_{11} + \frac{\Pi_{13}(L_{33} K_{31} - L_{31}K_{33})}{X} - \frac{K_{13}(G_{33}K_{31}-L_{31}\Pi_{33})}{X}.
             \end{split}
    \end{align}

Here, Eq. (\ref{eq:25}) is the expression for Onsager coefficients. These coefficients determine the transport parameters such as charge conductance $G$, Seebeck coefficient $S$, Peltier coefficient $\Pi$, and thermal conductance $K$ and they are given as
\begin{align} \label{eq:26}
    G &= L_{eV},\quad S = \frac{L_{eT}}{L_{eV}},\quad \Pi = \frac{L_{hV}}{L_{eV}},\quad K = L_{hT} - \frac{L_{hV}L_{eT}}{L_{eV}}.
\end{align}

\subsubsection{Power and efficiency in 3T QH setup as a quantum heat engine} \label{Sec:II(A2)}
Now, the charge power generated in terminal 1 is given as \cite{BENENTI20171},
\begin{align} \label{eq:27}
\begin{split}
    P &= VI_1  
     = V(-L_{eV} V + L_{e T} \tau).
    \end{split}
\end{align}
The maximum charge power can be obtained by finding $V$, for which $\frac{\partial P}{\partial V} = 0$, i.e., $V = \frac{ L_{e T} \tau}{2 L_{eV}}$. Thus maximum power is $P_{max} = \frac{L_{e T}^2 \tau^2}{4 L_{eV}}.$

Similarly, the efficiency at maximum power ($\eta|_{P_{max}}$) is,
\begin{equation} \label{eq:28}
    \eta|_{P_{max}} = \frac{P_{max}}{J_1} = \frac{ T \eta_c}{2}\frac{L_{e T}^2}{2 L_{h T} L_{eV} - L_{hV} L_{e T}} = \frac{\eta_c}{2}\frac{ZT}{ZT + 2}.
\end{equation}
 Here, $Z T$ is the figure of merit, which is defined as
\begin{equation}\label{eq:29}
    Z T = \frac{L_{hV}L_{eT}}{L_{eV}L_{hT}-L_{eT}L_{hV}} = \frac{G S^2 T}{K}.
\end{equation}
For $Z T \rightarrow \infty$, $\eta|_{P_{max}} = \frac{\eta_c}{2}$. The maximum value of efficiency at maximum power ($\eta|_{P_{max}}$) one can attain is half of Carnot efficiency, which is defined as Curzon-Ahlborn efficiency valid for quantum heat engine in the steady state regime. In thermodynamics \cite{curzon1975efficiency}, Curzon and Ahlborn proved that the highest value of efficiency at maximum power is $\eta_{CA}$ = $1- \sqrt{T_{2}/T_{1}}$. This can be rewritten as $\eta_{CA} = 1 - \sqrt{ 1 - \eta_c}$, where $\eta_c = 1 - T_{2}/T_{1}$ is the Carnot efficiency. $T_1 = T + \tau$ and $T_2 = T$ are the temperatures of the hotter and cooler reservoirs. Now the binomial expansion of $\sqrt{1-\eta_c}$ is,

\begin{equation}\label{eq:30}
    \sqrt{1-\eta_c} = 1 - \frac{\eta_c}{2} -\frac{\eta_c^2}{8}-...
\end{equation}

Inserting the Binomial expression of $\sqrt{1 - \eta_c}$ from Eq. (\ref{eq:30}) in $\eta_{CA}$, we get,

\begin{equation}\label{eq:31}
    \eta_{VDB} = 1 - \left(1 - \frac{\eta_c}{2} - \frac{\eta_c^2}{8}-...\right) = \frac{\eta_c}{2} + \frac{\eta_c^2}{8} + ...
\end{equation}

Now, the first order term in the binomial expansion of $\eta_{VDB}$ in Eq. (\ref{eq:31}) is exactly the upper bound on the efficiency at maximum power of steady state heat engine in the linear response regime, which is called Van den Broeck limit. This can be achieved when the figure of merit ($Z T$) is very large.

Similarly, one can find maximum efficiency. The general expression for efficiency is given as,
\begin{equation} \label{eq:32}
    \eta = \frac{V I_1}{J} = \frac{V(-L_{eV}V + L_{e T}\tau)}{-L_{hV} V + L_{h T} \tau}.
\end{equation}
The voltage bias $V_{max}$ required to achieve maximum efficiency is for $\frac{\partial \eta}{\partial V} = 0$,
\begin{equation} \label{eq:33}
    V_{max} = \frac{L_{h T}}{L_{h V}}\left(1 - \sqrt{\frac{L_{eV}L_{h T} - L_{eT} L_{h V}}{L_{eV} L_{h T}}}\right)\tau
\end{equation}
which can be derived from the condition $\frac{d \eta}{d V} = 0$ and one can verify that $\frac{d^2 \eta}{dV^2}<0$. Thus, the maximum efficiency is given as \cite{BENENTI20171}
\begin{equation} \label{eq:34}
    \eta_{max} = \eta_c x \frac{\sqrt{Z T + 1} - 1}{\sqrt{Z T + 1} + 1}.
\end{equation}
where, $x = \frac{L_{eT}}{L_{hV}} = \frac{T S}{\Pi}$ is the asymmetric parameter (AP) of the setup. When $x = 1$ and at the limit $Z T \rightarrow \infty$, the maximum efficiency approaches Carnot efficiency, which is desirable from any thermoelectric material.

\subsubsection{Cooling power and coefficient of performance in 3T QH setup as a quantum refrigerator} \label{Sec:II(A3)}

One can similarly derive the maximum coefficient of performance ($\eta^r_{max}$) and cooling power at maximum coefficient of performance ($J|_{\eta^r_{max}}$) for a QH setup to act as a quantum refrigerator since the asymmetric parameter, i.e., the ratio of Seebeck to the Peltier coefficient is 1, which plays a major role in the setup's performance as a heat engine. Here, the heat will be absorbed from the cooler terminals (terminals 2 and 3 with voltage probe condition, and only terminal 2 with voltage-temperature probe) and dumped into the hotter terminal (terminal 1 in either voltage probe or voltage-temperature probe). For a quantum refrigerator, the coefficient of performance ($\eta^r$) is defined as the ratio of heat taken from the cooler terminal ($J^Q$) and the power absorbed ($P$) by the setup, i.e.,   $\eta^r = \frac{J^Q}{P}$. Here, $J^Q = -(J_2 + J_3)$ for voltage probe and -$J_2$ for voltage-temperature probe and can be derived from Eq. (\ref{eq:18}). Similarly, $P = I_1 V$ can also be derived from Eq. (\ref{eq:18}). Now, $\eta^r_{max}$ can be found using the condition $\frac{d \eta^r}{dV}$ = 0, which gives
\begin{equation}\label{eq:35}
    V|_{\eta^r_{max}} = \frac{L_{hT}}{L_{hV}}\left(1 + \sqrt{\frac{L_{eV} L_{h T} - L_{hV}L_{e T}}{L_{eV}L_{h T}}}\right)
\end{equation}

where, $V|_{\eta^r_{max}}$ is the voltage required to achieve $\eta^r_{max}$. From here onwards one can derive $\eta^r_{max}$ by using Eq. (\ref{eq:18}) in $\eta^r = \frac{J^Q}{P}$, which yields
\begin{equation}\label{eq:36}
    \eta_{max}^r = \frac{\eta_c^r}{x} \frac{\sqrt{Z T + 1} - 1}{\sqrt{Z T + 1} + 1} 
\end{equation}
where $\eta_c^r$ = Carnot COP = $\frac{T}{\Delta T} = \eta_c^{-1}$ and $x$ is the asymmetric parameter (AP). when $x = 1$ and $Z T \rightarrow \infty$, $\eta_{max}^r$ approaches $\eta_c^r$. Similarly, the cooling power at maximum coefficient of performance ($J|_{\eta^r_{max}}$) is given as,
\begin{equation}\label{eq:37}
    J|_{\eta^r_{max}} = L_{hV} V_{max} + L_{hT} \tau = L_{hT} \left(\sqrt{\frac{L_{eV} L_{h T} - L_{hV}L_{e T}}{L_{eV}L_{h T}}}\right)\tau.
\end{equation}

\begin{figure} 
\centering
\includegraphics[width=0.60\linewidth]{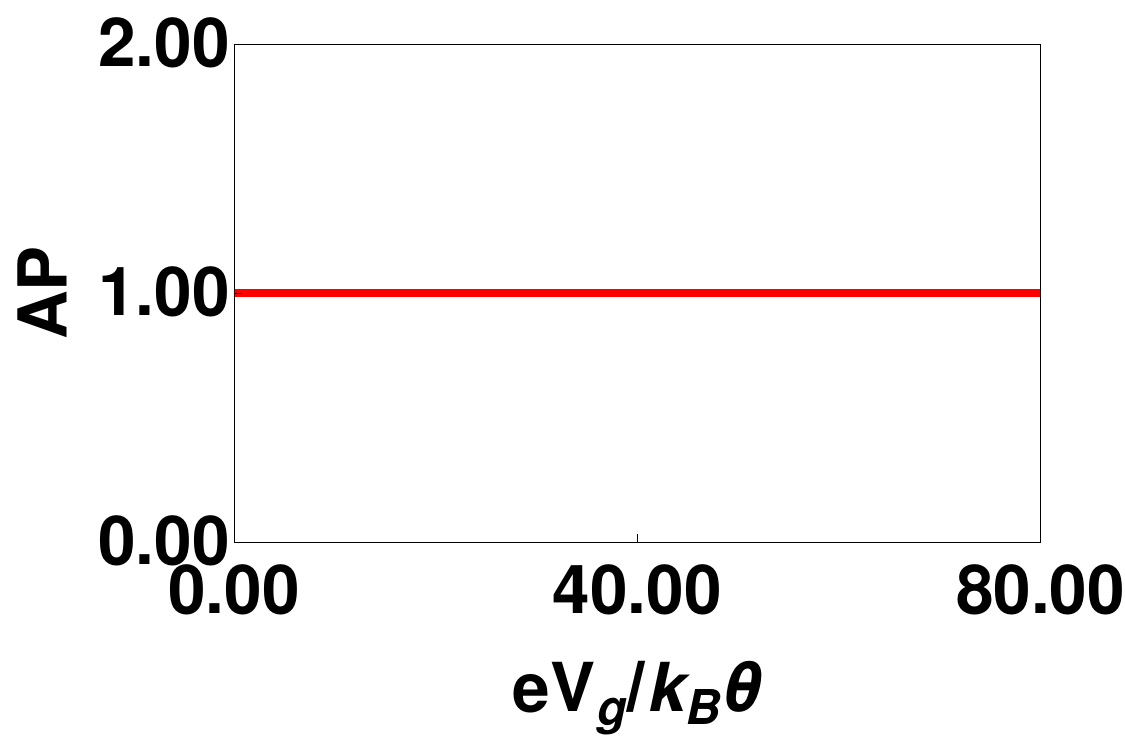}
\caption{Asymmetric parameter as a function of gate voltage $\frac{eV_g}{k_B T}$ of our setup. Parameters taken are $\omega = 0.1 k_B T / \hbar$, $T = 1.0K$, $\mu = 0$, $E_1 = k_B T$.}
\label{fig:4}
\end{figure}

\subsubsection{Why the QH setup in \cite{PhysRevLett.114.146801} can not work as a quantum refrigerator, whereas our setup does?}\label{Sec:II(A4)}
In this section, we discuss why our QH setup can work as a quantum refrigerator, whereas the QH setup, which is considered in \cite{PhysRevLett.114.146801} cannot work. The ability of a QH setup to work as a quantum refrigerator is closely related to its AP, which is evident from Eq. (\ref{eq:36}). As shown in \cite{PhysRevLett.110.070603}, when the AP deviates from one, then the coefficient of performance is reduced and can go to zero as AP is further increased, although it can still work as a quantum heat engine. There is a possibility that the coefficient of performance can achieve the Carnot limit $\eta_c^r$ at AP = 1 only \cite{PhysRevLett.110.070603}. Now, in the setup considered in \cite{PhysRevLett.114.146801}, the AP is either zero or infinity, resulting from time-reversal symmetry breaking. In this case, according to Ref. \cite{PhysRevLett.110.070603}, the coefficient of performance should be very small, which means the setup's ability to work as a quantum refrigerator is reduced. But in our setup, the AP is always one, shown in Fig. \ref{fig:4}, enabling us to use it as a quantum refrigerator. One can also see this analytically using Eq. (\ref{eq:25}). From the $L_{e T}$ expression in Eq. (\ref{eq:25}), one can see that
\begin{equation}\label{eq:38}
    T L_{e T} = \Pi_{11} + \frac{G_{13}(\Pi_{33} K_{31} - \Pi_{31}K_{33})}{X} - \frac{\Pi_{13}(G_{33}K_{31}-\Pi_{31}\Pi_{33})}{X}
\end{equation}
Comparing Eq. (\ref{eq:38}) with $L_{hV}$ of Eq. (\ref{eq:25}), one can see that both of them are almost same, which makes AP to be one regardless of any parameter.

\begin{figure}
     \centering
     \begin{subfigure}[b]{0.45\textwidth}
         \centering
         \includegraphics[width=\textwidth]{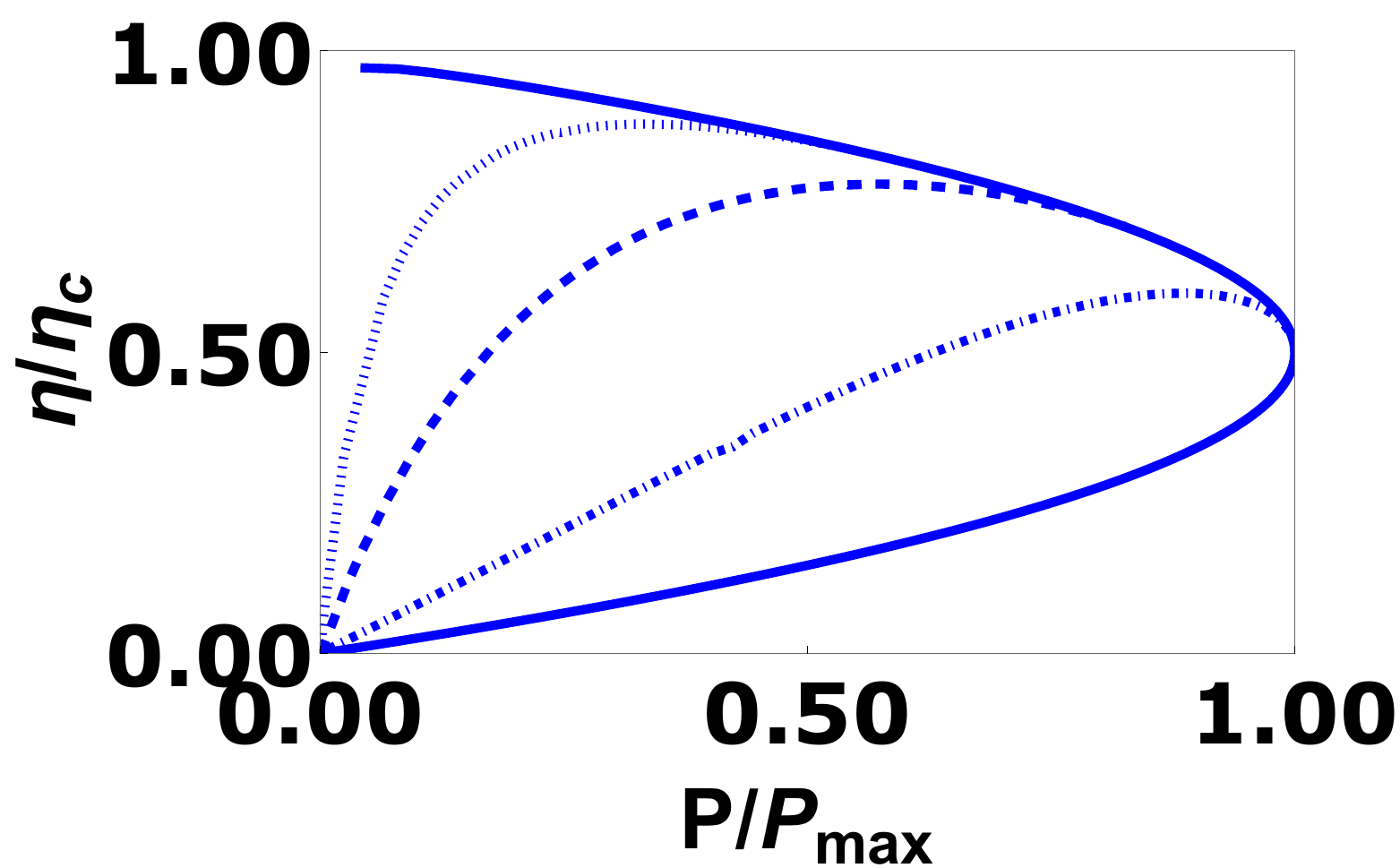}
         \caption{QH heat engine (linear)}
     \end{subfigure}
     \hspace{0.05cm}
     \begin{subfigure}[b]{0.45\textwidth}
         \centering
         \includegraphics[width=\textwidth]{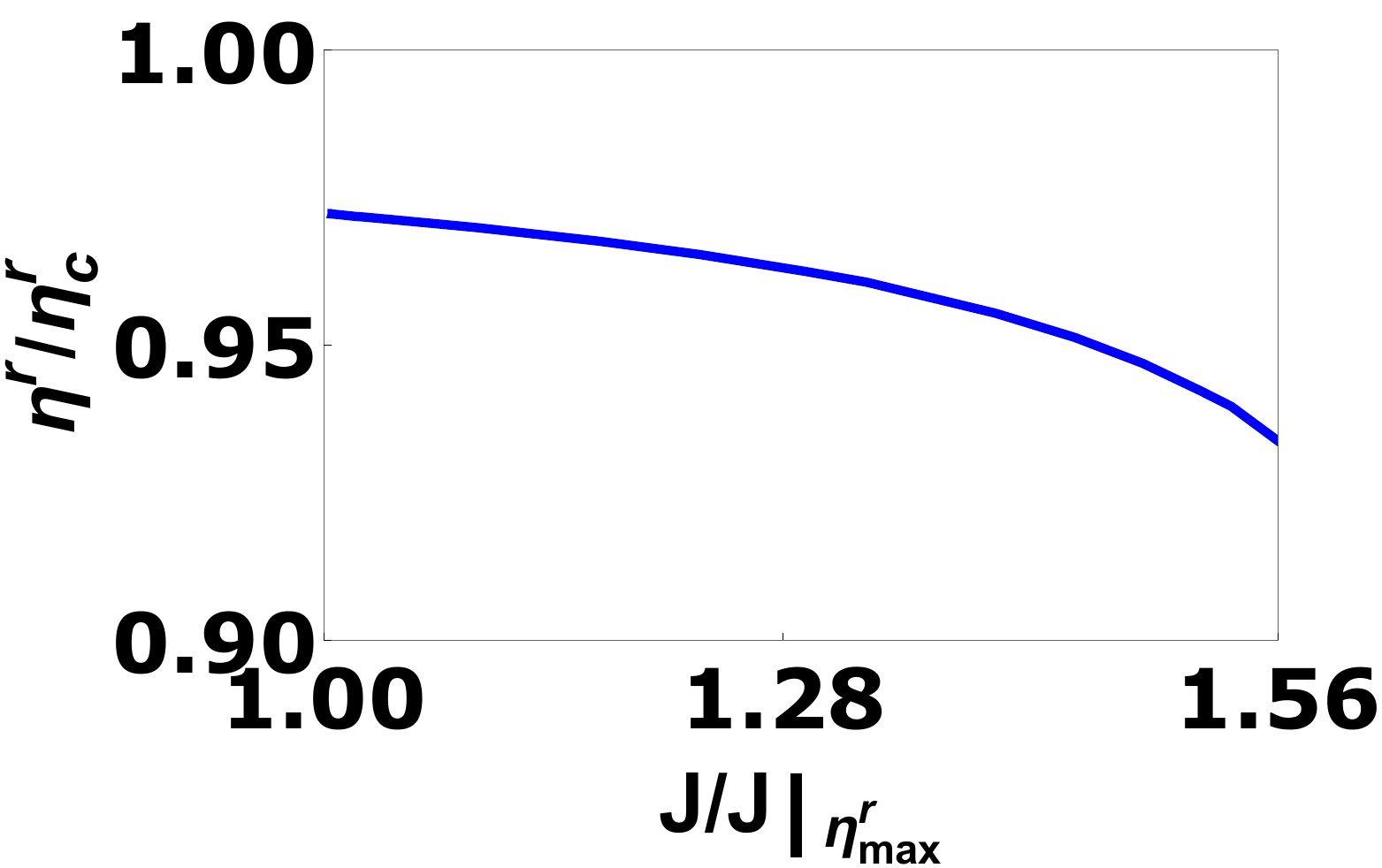}
         \caption{QH refrigerator (linear)}
     \end{subfigure}
        \caption{Parametric plot of $\frac{\eta}{\eta_c}$ vs $\frac{P}{P_{max}}$ (a) QH QHE. Parametric plot of $\frac{\eta^r}{\eta^r_c}$ vs $\frac{\textbf{J}}{\textbf{J}|_{\eta^r_{max}}}$ for (b) QH QR. Parameters are $eV_g = 83.8 k_B T$, $\omega = 0.1 k_B T / \hbar$, $T_1 = 1.01K, T_2 = 1.0K$, $\mu = 0$, $2E_1 = E_2 = 2k_B T$.}
         \label{fig:5}
       \end{figure}

In the main text, we have considered the case, where $E_1' = E_1 + eV_g$ and $E_2' = E_2 + e V_g$, with $E_1 = E_2 = k_B T$. One can also consider the case $E_1 \ne E_2$, see plots in Fig. \ref{fig:5}, where we take $E_1 = k_B T$, $E_2 = 2 k_B T$, and we achieve identical results in Fig. 2(a) and (c) of the main text.  

\subsection{Transport in the nonlinear response regime in the QH setup}\label{Sec:II(B)}
In this section, we turn to the non-linear regime of transport \cite{PhysRevLett.112.130601, PhysRevB.91.115425} in the same QH setup as shown in Fig. \ref{fig:3}. From Landaeur-Buttiker formalism, the charge and heat current in terminal $\alpha$ as derived in Eqs. (\ref{eq:9}) and (\ref{eq:12}), at finite temperature are given as, 
\begin{align}
\begin{split}\label{eq:39}
    I_{\alpha} &= \frac{2e}{h} \int_{-\infty}^{\infty} dE \sum_{\beta} \mathcal{T}_{\alpha \beta} (f_{\alpha} - f_{\beta}),\quad
    J_{\alpha} = \frac{2}{h} \int_{-\infty}^{\infty} dE (E - \mu_{\alpha}) \sum_{\beta}  \mathcal{T}_{\alpha \beta} (f_{\alpha} - f_{\beta}).
\end{split}
\end{align}

As discussed in Sec. \ref{Sec:II(A)}, when applied biases are small compared to the relevant thermal energy scale ($eV \ll k_B T$, $\tau \ll T$), then we are in the linear transport regime and when it is not that small i.e., $eV \approx k_B T$, $\tau \approx T$, then we are in the non-linear transport regime. In case of linear thermoelectrics, electron transport is only dependent on its kinetic energy, whereas, in non-linear regime, electron transport is dependent on both its kinetic energy as well as the potential ($U$) of the sample where it flows \cite{buttiker1993capacitance, christen1996gauge}. When the voltage bias and temperature bias are small compared to the relevant thermal energy scale ($k_B T$) i.e., in the linear transport regime, then $U$ is much smaller compared to the kinetic energy of the electron and does not play much role in electron transport, and therefore we ignore it in the linear response regime \cite{buttiker1993capacitance, christen1996gauge}. However, in the nonlinear transport regime, the applied biases are not small and are comparable to $k_B T$, and therefore in this case, $U$ cannot be ignored. When the electrons are injected from a terminal into a sample, it perturbs the charge distribution inside the sample, eventually affecting the behavior of electrons via electron-electron interaction. This means the potential energy of the whole system is changed by the voltage and temperature biases applied, which ultimately tells us that the transmission probability $\mathcal{T}_{\alpha \beta}$ is not only a function of kinetic energy here, but a function of potential ($U$) as well, which in turn is dependent on voltage as well as temperature biases applied across the sample. This has to be calculated in a self-consistent manner \cite{buttiker1993capacitance, christen1996gauge, PhysRevLett.110.026804, PhysRevB.88.045129, Meair_2013}. This is why $\mathcal{T}_{\alpha \beta}$ is dependent only on the kinetic energy of the electron for linear transport, while it is dependent on both kinetic energy as well as the potential of the sample in the nonlinear transport regime \cite{buttiker1993capacitance, christen1996gauge, PhysRevLett.110.026804, PhysRevB.88.045129, Meair_2013}.

In the subsequent subsections, we discuss the two-terminal (2T) QH setup as a nonlinear quantum heat engine and a quantum refrigerator in Sec. \ref{Sec:II(B1)} using Box-car type constriction. In Sec. \ref{Sec:II(B2)}, we discuss the two-terminal QH set up both as a heat engine and refrigerator using QPC-type constriction, which is an amalgamation of the approaches of Sanchez, Lopez and B\"uttiker \cite{buttiker1993capacitance, christen1996gauge, PhysRevLett.110.026804, PhysRevB.88.045129, Meair_2013} and Haack, Giazotto \cite{Haack_2021}. We then extend the same calculation to a three-terminal (3T) QH setup in Sec. \ref{Sec:II(B3)}. Finally, in Sec. \ref{Sec:II(B4)}, we discuss the thermoelectric properties of a 3T QH setup with a voltage-temperature probe.

\subsubsection{Nonlinear thermoelectrics in 2T QH setup with Box-car type constriction}\label{Sec:II(B1)}
\begin{figure} 
\centering
\includegraphics[width=0.60\linewidth]{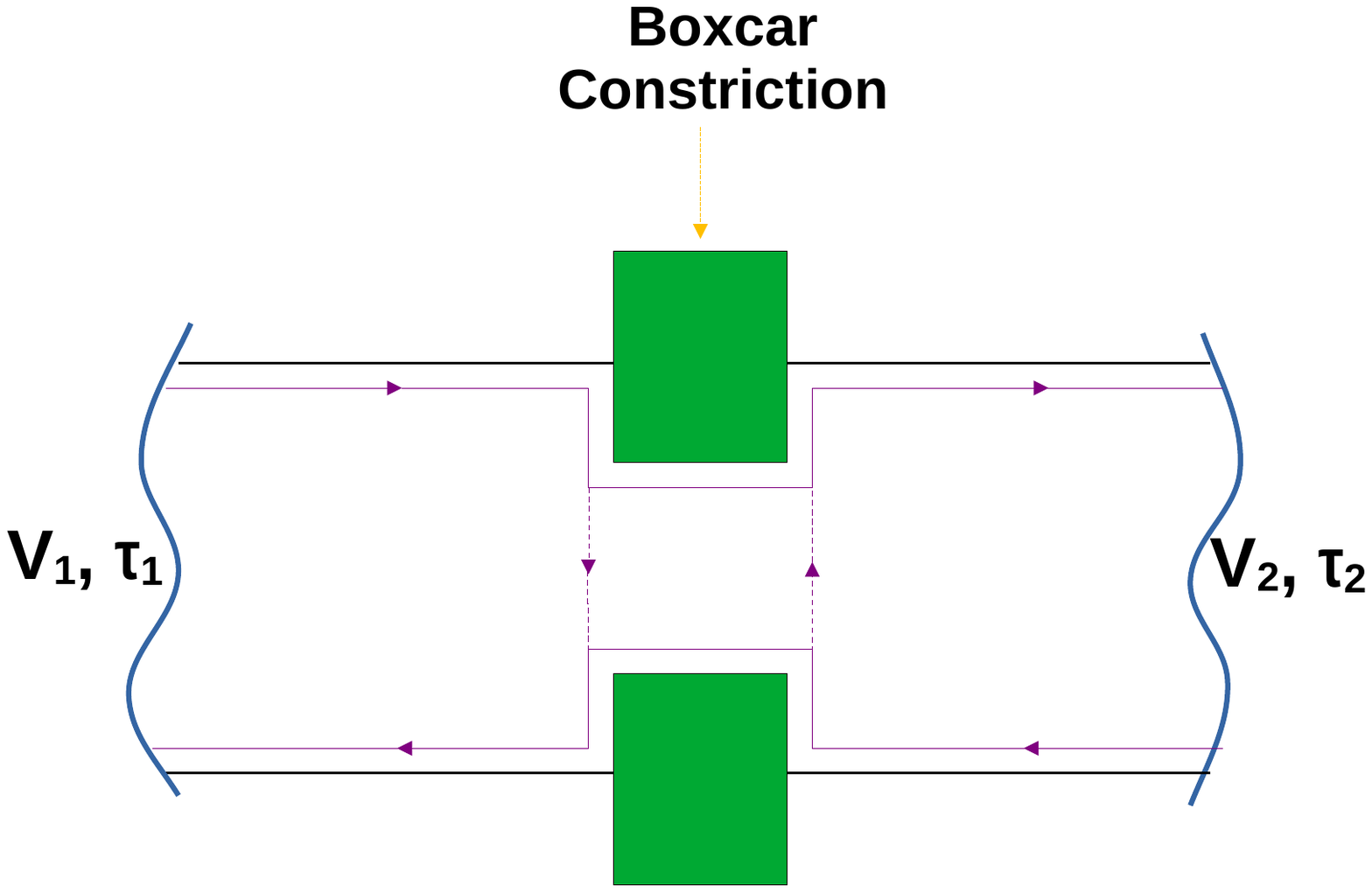}
\caption{Two-terminal QH setup with box-car type constriction (in nonlinear response), where the constriction in between has boxcar-type of transmission. }
\label{fig:6}
\end{figure}

Here, we discuss the nonlinear thermoelectricity of a 2T QH setup as shown in Fig. \ref{fig:6} using box car type constriction following the Refs. \cite{PhysRevLett.112.130601, PhysRevB.91.115425}.
 We consider the voltage biases to be $V_1 = -V, V_2 = 0$ and $T_1$ and $T_2$ being the temperatures of terminals 1 and 2 respectively. Here, the power generated ($P$) as a quantum heat engine is given as
\begin{equation}\label{eq:40}
    P = (V_2 - V_1)I_1 = \frac{2eV}{h} \int_{-\infty}^{\infty}dE \quad \mathcal{T}_{12}(f_1 - f_2),
\end{equation}
where, $\mathcal{T}_{12}$ is the transmission probability for an electron to scatter from terminal 2 to terminal 1 and the $f_1$ and $f_2$ are the Fermi-Dirac distributions of reservoirs 1 and 2 respectively. Similarly, from Eq. (\ref{eq:12}), the heat current out of terminal 1 is
\begin{equation}\label{eq:41}
    J_1 = \frac{2}{h}\int_{-\infty}^{\infty}dE (E + eV) \mathcal{T}_{12}(f_1 - f_2),
\end{equation}
where we have taken $\mu_1 = -eV$. Using Eqs. (\ref{eq:40}) and (\ref{eq:41}), one can evaluate the efficiency i.e.,  $\eta = P/J_1$.

In \cite{PhysRevLett.112.130601, PhysRevB.91.115425}, Whitney proved that $\mathcal{T}_{12}$ needs to be a boxcar-type transmission as shown in Fig. \ref{fig:1}(b) to extract the best possible performance as a heat engine.
Considering a single edge mode, the boxcar-type transmission probability $\mathcal{T}_{12}$ = 1 for $E_1 < E < E_{1}^{'}$, otherwise it is zero. When $E_{1}^{'} \rightarrow \infty$, $\mathcal{T}_{12}$ is a QPC type, where it is one above $E_1$ and zero otherwise as shown in Fig. \ref{fig:1}. Here, $E_1 = eV/\left(\frac{T_1}{T_2}-1\right)^{-1}$ and $E_{1}^{'} = eV J_1^{'}/P^{'}$ \cite{PhysRevLett.112.130601, PhysRevB.91.115425}, where the prime in $J_1$ and $P$ denotes the first derivatives with respect to V. The maximum possible power ($P_{max}$) is generated when $P^{'} = 0$, which makes $E_{1}^{'} = \infty$, and consequently one can also derive the efficiency at the maximum power ($\eta|_{P_{max}}$). Similarly, the maximum efficiency ($\eta_{max}$) can be achieved when $E_{1}^{'}$ is closer to $E_1$ so that the finite power ($P$) generated won't be only much lesser than $P_{max}$, but the maximum efficiency will also be close to Carnot efficiency ($\eta_c$). All these quantities of interest such as $P_{max}$, $\eta|_{P_{max}}$ and $\eta_{max}$ as derived by Whitney are given as \cite{PhysRevLett.112.130601, PhysRevB.91.115425},
\begin{align}\label{eq:42}
\begin{split}
    P_{max} &= P_{max}^{Wh} = 0.0642 \frac{\pi^2 k_B^2 \tau^2}{h},\quad 
    \eta|_{P_{max}} = \eta|_{P_{max}^{Wh}} = \frac{\eta_c}{1+ 0.936(1 + T_2/T_1)},\quad
    \eta_{max} = \eta_{max}^{Wh} = \eta_c \left(1-0.478\sqrt{\frac{T_2 P}{T_1 P_{max}^{Wh}}}\right),
    \end{split}
\end{align}
where, $\tau = T_1 - T_2$ = temperature bias applied across the sample, $\eta_c$ = Carnot efficiency, $T_1$ = temperature in terminal 1, $T_2$ = temperature in terminal 2, $P$ = Finite output power with $P \ll P_{max}^{Wh}$ and we have taken spin degeneracy into account.

 Similarly, for a quantum refrigerator, the cooling power ($J = -J_2$) of this 2T QH setup is given as
\begin{equation}\label{eq:43}
    J = J_2 = \frac{2}{h}\int_{-\infty}^{\infty}dE E \mathcal{T}_{21}(f_2-f_1).
\end{equation}
Using the cooling power $J$ and the power absorbed $P = I_1 V$ (see, Eq. (\ref{eq:40})), one can determine the coefficient of performance i.e., $\eta^r = J/P$.
Similar to the quantum heat engine also, $\mathcal{T}_{21}$ should be a Boxcar-type of transmission probability, which provides the best possible performance as a refrigerator in a two-terminal QH nonlinear setup as proved in Refs. \cite{PhysRevLett.112.130601, PhysRevB.91.115425}, except $E_{1}^{'}$ is $eV J_2^{'}/P^{'}$ and with the form of $E_1$ remaining same. 
Now the maximum cooling power ($J_{max}$) can only be extracted when $J_2^{'} = 0$ and the applied voltage bias is large making $E_{1}^{'} = 0$ and $E_{x} \rightarrow \infty$ \cite{PhysRevLett.112.130601, PhysRevB.91.115425}. When the maximum cooling power is reached, then due to the large voltage bias, the power absorbed $P$ by the system will be large, which makes the coefficient of performance ($\eta^r$) vanish. Similarly, the maximum coefficient of performance ($\eta^r_{max}$) can be achieved when $E_x$ is closer to $E_{1}^{'}$ and the extracted cooling power ($J$) is much lesser than $J_{max}$. So, the quantities of interest such as $J_{max}$ and $\eta^r_{max}$ as derived by Whitney are given as \cite{PhysRevLett.112.130601, PhysRevB.91.115425}  
\begin{equation}\label{eq:44}
    J_{max} = J_{max}^{Wh} = \frac{\pi^2}{6h}N k_B^2 T_2^2, \quad \text{and} \quad \eta^r_{max} = \eta^{r, Wh}_{max} = \eta^r_c\left(1-1.09 \sqrt{\frac{T_2}{T_1 - T_2}\frac{J}{J_{max}^{Wh}}}\right), 
\end{equation}
where $\eta_c^r$ = Carnot coefficient of performance, $T_1$ = temperature in terminal 1, $T_2$ = temperature in terminal 2, $J$ = Finite cooling output power and $J \ll J_{max}^{Wh}$. \\

\subsubsection{Nonlinear thermoelectrics in 2T QH setup with QPC type constriction}\label{Sec:II(B2)}

\begin{figure} 
\centering
\includegraphics[width=0.60\linewidth]{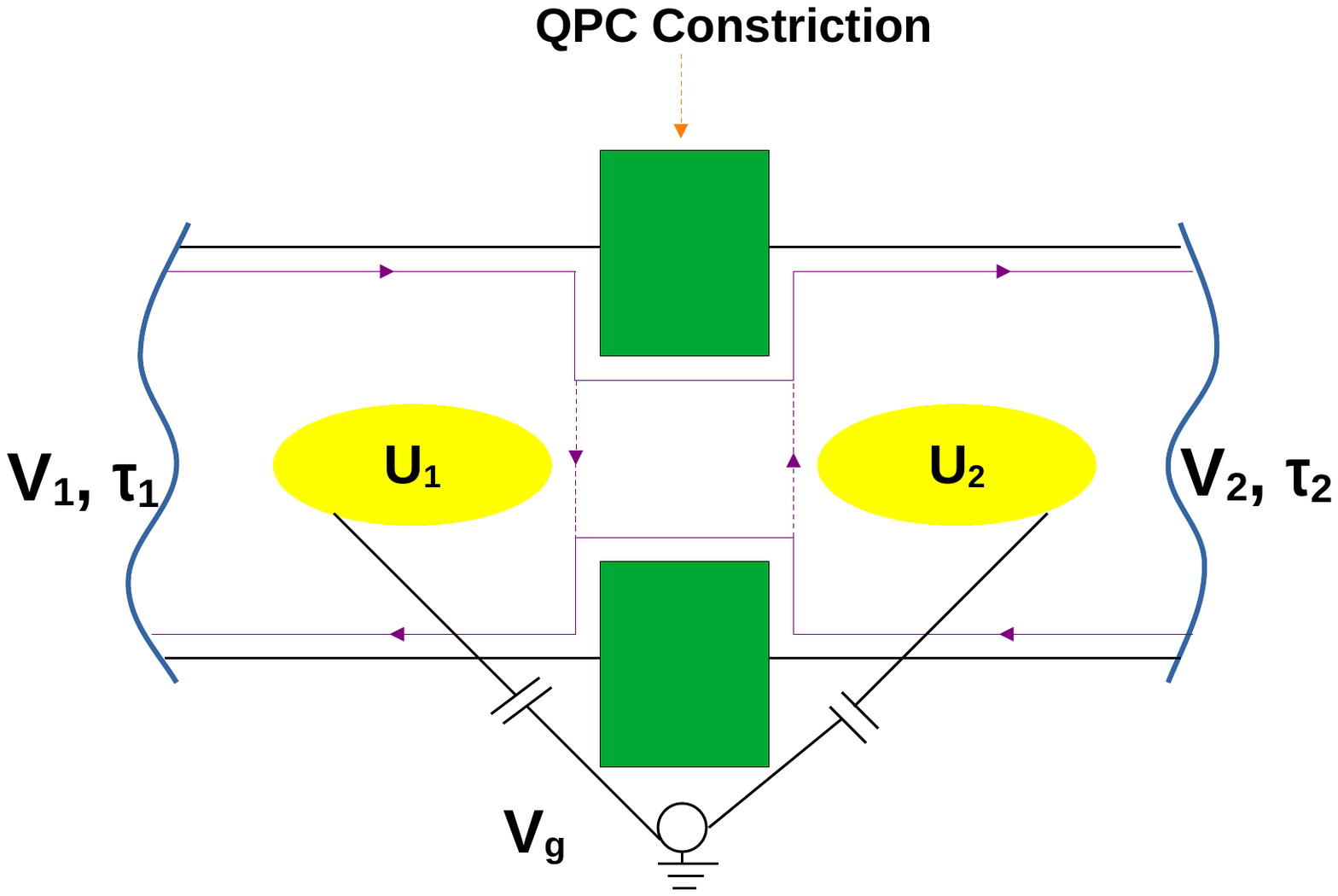}
\caption{{Two-terminal QH setup in nonlinear response (using our approach) discretized into two regions $r = 1, 2$ (yellow shaded regions) to the left and right of the QPC-type constriction. Each region is connected capacitively to a common external gate voltage $V_g$.} }
\label{fig:7}
\end{figure}

Now, we discuss the nonlinear thermoelectrics in a 2T QH setup as shown in Fig. \ref{fig:7} by considering a QPC-like constriction between the two terminals. As discussed in the introduction (Sec. \ref{Sec:II(B)}), in the nonlinear response regime, the interaction potential ($U$) comes into the picture because the applied biases are finite, with $eV \simeq k_B T$ \cite{BENENTI20171}. Thus, the interaction potential will be a function of the voltage bias ($V$) as well as the temperature bias ($\tau$). {The interaction potential as a function of the biases can be calculated both numerically and analytically \cite{PhysRevLett.110.026804, PhysRevB.88.045129, Meair_2013} for 2T setups. We discuss the numerical calculation method, which is based on the extra charge injected into the conductor when it deviates from equilibrium in the presence of an interaction potential, \( U \). When the biases are applied across the sample, the total charge injected into the conductor is given as \cite{PhysRevLett.110.026804, PhysRevB.88.045129, Meair_2013},}

\begin{equation}\label{eq:45}
    {q = e \int dE \left ( \sum_{\alpha} \nu_{\alpha}^p(E, U) f_\alpha(E) \right)}, 
\end{equation}

{where, $\nu_{\alpha}^p(E, U) = \frac{1}{2\pi i }\sum_{\beta}\text{Tr}\left[s_{\beta \alpha}^{\dagger}\frac{d s_{\beta \alpha}}{dE}\right]$ is the particle injectivity or density of states of terminal $\alpha$, and $s_{\beta \alpha}$ is the scattering amplitude for an electron to scatter from terminal $\alpha$ to $\beta$, and $f_{\alpha}$ is the Fermi-Dirac distribution for terminal $\alpha \in \{1,2\}$. $U$, the interaction potential arises due to applied bias ($V$) and thermal biases ($\tau$), i.e., $U = U(V, \tau)$. In an equilibrium situation, i.e., when applied biases are zero ($V_1 = V_2 = \tau_1 = \tau_2 = 0$), the Fermi-Dirac distributions of terminal 1 and 2 are same, i.e., $f_1 = f_2 = f$. In this case, the total equilibrium charge injected into the conductor is $q_{eq} = e \int \nu(E) f $, where $\nu(E) = \sum_{\alpha} \nu_{\alpha}(E)$ is the total particle injectivity or density of states \cite{PhysRevLett.110.026804, PhysRevB.88.045129, Meair_2013, buttiker1993capacitance, christen1996gauge}.} {The additional charge ($q_{ad}$) injected due to applied biases $V$ and $\tau$ into the conductor is $q_{ad} = q - q_{eq}$.}

{One can capacitively connect the conductor to an external gate voltage and  $q_{ad}$ can also be written as from Eq. (\ref{eq:45}) and Refs. \cite{PhysRevLett.110.026804, PhysRevB.88.045129, Meair_2013, buttiker1993capacitance, christen1996gauge}},
\begin{equation}\label{eq:46}
    {q_{ad} = e \int dE  \left ( \sum_{\alpha} \nu_{\alpha}^p(E, U) f_{\alpha}(E) \right) - e \int  \nu(E) f = C(U - V_g)}
\end{equation}

{For the proof of the equivalence of the Eq. (\ref{eq:46}), wherein $q_{ad}$ is derived from injectivities $\nu_{\alpha}^p(E, U)$ and $q_{ad}$ derived from capacitively connected gate voltage $V_g$ and equated is provided in Refs. \cite{PhysRevLett.110.026804, PhysRevB.88.045129, Meair_2013}. One can numerically estimate the interaction potential ($U$), see Refs. \cite{PhysRevLett.110.026804, PhysRevB.88.045129, Meair_2013} for some fixed parameters such as applied voltage bias ($V_{\alpha}$), temperature bias ($\tau_{\alpha}$), gate voltage ($V_g$) and capacitance ($C$).}

{Finding the exact local potential inside a particular region of the mesoscopic conductor is difficult to compute and to make the calculation simple, we discretize the conductor into different regions $r = 1, 2, ...$ each with interaction potential $U_r$. We then numerically estimate $U$ by finding $U_r$ for each region $r$. We assume that the interaction potentials for each region $r$ are independent of each other \cite{PhysRevLett.110.026804, PhysRevB.88.045129, Meair_2013, buttiker1993capacitance, christen1996gauge}. One can calculate $U_i$ in a region $i$ by assuming other potentials $U_r$ for $r \ne i$ to be zero. For example, if one wishes to find $U_1$ in region 1, then one can assume $U_2 = U_3.... = 0$ and find $U_1$. Below, we discuss this procedure in detail and later utilize this method to estimate $U$ in the QH setups in the presence of QPC-type constrictions.} 

{For every region $r$, an analogous equation to Eq. (\ref{eq:46}) for the addtional injected charge $q_{ad}^r$ into region $r$ can be derived,}
\begin{equation}\label{eq:47}
    {q_{ad}^r = e \int dE  \left ( \sum_{\alpha} \nu_{r \alpha}^p(E, U_r) f_{\alpha}(E) \right) - e \int  \nu_r(E) f = C (U_r - V_g).}
\end{equation}

{where $\nu_{r \alpha}^p$ is the particle injectivity into region $r$ from terminal $\alpha$, $C$ and $V_g$ are the capacitor and applied gate voltage in region $r$ \cite{PhysRevLett.110.026804, PhysRevB.88.045129, Meair_2013}.}

In \cite{PhysRevLett.110.026804, PhysRevB.88.045129}, the derivation of the interaction potential in a two-terminal QH setup with a resonant-tunneling-type of constriction is shown {where the discretization method is not used}, whereas in \cite{PhysRevLett.77.143, Meair_2013}, the same derivation has been done in presence of QPC-like constriction {by discretizing the conductor in two regions, i.e., $r = 1,2$, as shown in Fig. \ref{fig:7}. Here, $r = 1$ and $r = 2$ represent the left and right side of the QPC-type constriction. We find the interaction potential $U_r$ in this setup in the presence of QPC-type constrictions only. The electron propagating in the QH setup will encounter an average potential $U = \frac{U_1 + U_2}{2}$. First, we calculate $U_1$, assuming $U_2 = 0$. For region 1, the additional injected charge $q_{ad}^1$ is,}
\begin{align}\label{eq:48}
\begin{split}
    {q_{ad}^1} & {= e \int dE \left ( \sum_{\alpha \in \{1 ,2 \}} \nu_{1 \alpha}^p(E, U_1) f_{\alpha}(E) \right) - e \int dE  \nu_1^p(E) f}\\
    & {= e \int dE \left (  \nu_{11}^p(E, U_1) f_{1}(E) + \nu_{12}^p(E, U_1) f_{2}(E)\right) - e \int dE  \nu_1^p(E) f.}
    \end{split}
\end{align}
{where $\nu_1^p = \nu_{11}^p + \nu_{12}^p$ and $f_{\alpha}(E) = \left(1 + e^{(E + e V_{\alpha})/k_B T_{\alpha}} \right)^{-1}$ is the Fermi-Dirac distribution with $V_{\alpha}$ and $T_{\alpha}$ being the applied voltage bias and temperature at terminal $\alpha$. The temperature in terminal $\alpha$, $T_{\alpha} = T + \tau_{\alpha}$, where $T$ and $\tau_{\alpha}$ are the equilibrium temperature and temperature bias at terminal $\alpha$. Following the argument of Christen and Buttiker in \cite{christen1996gauge, PhysRevLett.77.143}, one can determine both $\nu_{11}^p$ and $\nu_{12}^p$. $\nu_{11}^p$ can be derived from the partial density of states $\nu_{ \beta 1 1}^p$ which is the partial density of state for electrons scattered from contact 1 to contact $\beta$, $\beta \in \{1,2 \}$. Therefore, $\nu_{\beta 1 1}^{p}$ is defined as the product of $\mathcal{T}_{\beta 1 }$ and $\nu_1^p$, where $\mathcal{T}_{\beta 1}$ is the probability for an electron to scatter from terminal 1 to terminal $\beta$. Similarly, $\nu_{12}^p = \nu_{112}^p + \nu_{212}^{p}$, where $\nu_{\beta 1 2}^p$ is the partial density of state for electrons scattered from contact 2 to contact $\beta$ for $\beta \in \{1,2 \}$. The partial density of states $\nu_{1 1 1}^p = \mathcal{T}_{11}\nu_1^p =  (1-\mathcal{T}_{QPC})\nu_1^p $ and $\nu_{2 1 1}^p = \mathcal{T}_{21}\nu_1^p =\frac{1}{2}\mathcal{T}_{QPC} \nu_1^p$, where $\mathcal{T}_{11}$ is the probability for an electron to scatter from terminal 1 to itself, whereas $\mathcal{T}_{21}$ is the probability to scatter from terminal 1 to terminal 2. Therefore, $\nu_{11}^p = \left(1 - \frac{\mathcal{T}_{QPC}}{2}\right) \nu_1^p$, where $\mathcal{T}_{QPC}$ is function of energy and $U_1$ and it is given as,}

\begin{equation}\label{eq:49}
    {\mathcal{T}_{QPC} = \frac{1}{1 + e^{-\frac{E - E_1 - e U_1}{\hbar \omega}}}}
\end{equation}
{where $E_1$ is defined as $V_0 + \hbar \omega_y \left(n_y + 1/2\right)$ as shown below Eq. (\ref{eq8}). There are no states in region 1 associated with the reflection of electrons from terminal 2 to itself, which is why, $\nu_{212}^p = \mathcal{T}_{22} \nu_1^p$ vanishes, even if $\mathcal{T}_{22}$ and $\nu_1^p$ are nonzero and only $\nu_{112}^p = \mathcal{T}_{12} \nu_1^p$ survives and is equal to $\nu_{12}^p = \frac{1}{2}\mathcal{T}_{QPC} \nu_1^p$. The prefactor of $\frac{1}{2}$ in front of $\mathcal{T}_{QPC}$ in $\nu_{211}^p$ and $\nu_{112}^p$ has been taken in order to ensure $\nu_{11}^p + \nu_{12}^p = \nu_1^p$ \cite{christen1996gauge, PhysRevLett.77.143}. Therefore, the additional injected charge into region 1 is given as,}
\begin{equation}\label{eq:50}
    {q^1_{ad} = e \int dE \left(\left(1 - \frac{\mathcal{T}_{QPC}}{2}\right)\nu_1^p(E, U_1) f_1 + \frac{\mathcal{T}_{QPC}}{2}\nu_1^p(E, U_1) f_2\right) - e \int dE \, \, \,  \nu_1^p(E) f}
\end{equation}

{The density of states $\nu_1^p (E, U_1)$ in region 1 in presence of QPC like constriction has been calculated in Ref. \cite{PhysRevB.57.1838} and is given as, } 
\begin{align}\label{eq:51}
    \begin{split}
       {\nu_1^p(E, U)} & {= \frac{4}{ 2\pi  \hbar \omega} \sinh^{-1} \sqrt{\left(\frac{1}{2} \frac{m \omega^2 \lambda^2}{E  - E_1 - eU_1}\right)}} , \quad {\text{for}} \quad {E > E_1 + eU_1,} \\
            & {= \frac{4}{ 2\pi  \hbar \omega} \cosh^{-1} \sqrt{\left(\frac{1}{2} \frac{m \omega^2 \lambda^2}{E_1 + eU_1 - E}\right)} ,} \quad {\text{for}} \quad {E_1 + eU_1 - \frac{1}{2} m \omega^2 \lambda^2 < E < E_1.}
    \end{split}
\end{align}
{where, $m$ and $\omega_x$ are defined in Eq. (\ref{eq7}) and $\lambda$ is the screening length defined in Ref. \cite{PhysRevB.57.1838}. One can capacitively connect the conductor region 1 of Fig. \ref{fig:7} to an external gate voltage $V_g$ and the additional injected charge then is given as,}
\begin{equation}\label{eq:52}
    {q_{ad}^1 = C(U_1 - V_g),}
\end{equation}
{utilizing Eqs. (\ref{eq:48}-\ref{eq:52}), one can numerically estimate $U_1$, which is a function of $V_{1}, V_{2}, \tau_{1}, \tau_{2}$ and $V_g$. As shown in Refs. \cite{PhysRevLett.110.026804, PhysRevB.88.045129, Meair_2013}, the interaction potential $U_1 (V_1, V_2, \tau_1, \tau_2, V_g)$ can also be written as}
\begin{equation}\label{eq:53}
    {U_1(V_1, V_2, \tau_1, \tau_2, V_g) = U_1'(V_1, V_2, \tau_1, \tau_2) + u_{g1}V_g}
\end{equation}
{where $U_1'(V_1, V_2, \tau_1, \tau_2)$ is the interaction potential induced solely by the voltage bias ($V_1, V_2$) and temperature bias ($\tau_1, \tau_2$) at zero gate voltage, i.e., $U_1(V_g = 0) = U_1'$ and $u_{g1} = \frac{ U_1 - U_1'}{V_g}$ determines the response of $U_1$ to an applied gate voltage ($V_g$), and is a unitless ratio.}  

{Similarly, one can again find $U_2$ in region 2 of Fig. \ref{fig:7}, assuming $U_1 = 0$. The additional injected charge $q_{ad}^2$ into region 2 is,}
\begin{equation}\label{eq:54}
\begin{split}
    {q_{ad}^2} & {= e \int dE \left ( \sum_{\alpha} \nu_{2 \alpha}^p(E, U_2) f_{\alpha}(E) \right) - e \int dE  \nu_2^p(E) f = e \int dE \left (  \nu_{21}^p(E, U_2) f_{1}(E) + \nu_{22}^p(E, U_2) f_{2}(E)\right) - e \int dE  \nu_2^p(E) f}\\
    & {= e \int dE  \left(\frac{\mathcal{T}_{QPC}}{2}\nu_{2}^p(E, U) f_1 + \left(1 - \frac{\mathcal{T}_{QPC}}{2}\right)\right) \nu_{2}^p(E, U) f_2 - e \int dE \, \, \, \nu_{2}^p(E) f}.
    \end{split}
\end{equation}

{Similar to the derivation in region 1 of $\nu_{11}^p$, for region 2, $\nu_{21}^p = \frac{\mathcal{T}_{QPC}}{2}\nu_2^p$ and $\nu_{22}^p = \left(1 - \frac{\mathcal{T}_{QPC}}{2}\right)\nu_2^p$, where $\mathcal{T}_{QPC}$ and $\nu_2^p$ are given in Eqs. (\ref{eq:49}) and (\ref{eq:51}) with $U_1$ replaced by $U_2$. Similar to region 1, one can also capacitively connect region 2 to an external gate voltage $V_g$ and therefore $q_{ad}^2$ is,}
\begin{equation}\label{eq:55}
    {q_{ad}^2 = C(U_2 - V_g)}.
\end{equation}
{One can estimate $U_2$ numerically as a function of $V_1, V_2, \tau_1, \tau_2$ and $V_g$ and similar to $U_1$, $U_2 (V_1, V_2, \tau_1, \tau_2, V_g)$ can be written as}
\begin{equation}\label{eq:56}
    {U_2(V_1, V_2, \tau_1, \tau_2, V_g) = U_2'(V_1, V_2, \tau_1, \tau_2 ) + u_{g2}V_g}
\end{equation}
{where $U_2'(V_1, V_2, \tau_1, \tau_2)$ is the interaction potential induced solely by the voltage bias ($V_1, V_2$) and temperature bias ($\tau_1, \tau_2$) at zero gate voltage ($V_g = 0$) and $u_{g2} = \frac{U_2 - U_2'}{ V_g}$ determines the response of $U_2$ with the applied gate voltage ($V_g$).}  
{Now, the average interaction potential $U = \frac{U_1 + U_2}{2}$, where $U_1$ and $U_2$ can be estimated numerically using Eqs. (\ref{eq:48}-\ref{eq:51}) and Eqs. (\ref{eq:54}-\ref{eq:55}). } 
{Thus, average interaction potential is,}
\begin{equation}\label{eq:57}
    {U = \frac{U_1 + U_2}{2} = U' + u_g V_g, \quad \text{for} \quad U' = \frac{U_1' + U_2'}{2} \quad \text{and} \quad u_g = \frac{u_{g1} + u_{g2}}{2} }
\end{equation}

{In our setup, we consider $V_1 = -V, V_2 = 0, \tau_1 = \tau, \tau_2 = 0$}. If the setup is kept at a constant voltage bias $V$ and temperature bias $\tau$, then the interaction potential $U'$ will also be constant. One can apply $V_g$ to nullify $U'$ as discussed quantum refrigerator, in Ref. \cite{Haack_2021} depending on the value of bias $V$ and $\tau$. At a fixed voltage bias $V$ and $\tau$, for the quantum heat engine, we consider the gate voltage to be
\begin{equation}\label{eq:58}
    V_g = (-U' + V_g')/u_g, 
\end{equation}
  The first term inside the parenthesis in Eq. (\ref{eq:58}) is used to nullify the interaction potential $U'$ and the second term $V_g'$ is used to lift the energy level of QPC, and the threshold energy of the QPC is thus,
\begin{align}\label{eq:59}
\begin{split}
    E_{1}' &= E_{1} + eU = E_{1} + eU' + eu_g V_g = E_{1} + eV_g'.
    \end{split}
\end{align}
The transmission probability of a QPC is, then, 
\begin{equation}\label{eq:60}
    \mathcal{T}_{QPC} = \frac{1}{1+e^{-2\pi(E - E_{1}')/\hbar \omega}} = \frac{1}{1+e^{-2\pi(E - E_1 - eV_g')/\hbar \omega}}.
\end{equation}
 {One can estimate the value of interaction potential $U'$ numerically and accordingly apply a gate voltage $V_g$ which can help nullify $U'$. We consider a simple example, let $V = 1.14 k_B T/e,$ as at this value, the QHE gives maximum power ($P_{max}^{Wh}$) and is optimal. We also take $ \tau = 1K, C = 0.01 F, \hbar \omega = 0.1 k_B T$ and $\frac{1}{2}m \omega^2 \lambda^2 = 20 k_B T$ as these values lie in the range where the setup of Fig. \ref{fig:7} can work as a QHE. With this information, we calculate $U_1$ and then $U_2$. First, we calculate $U_1$ at $V_g = 0$, i.e., $U_1'$ from Eqs. (\ref{eq:48}-\ref{eq:51}) and this is 10.88229$k_B T/e$ and later we calculate $U_1$ at an arbitrary gate voltage (for example $V_g = k_B T/e$) and got $U_1 = 11.04889 k_B T/e$. Using Eq. (\ref{eq:53}), we find $u_{g1} = (U_1 - U_1')/V_g = 0.1666$. Similarly, one can calculate $U_2'$ at $V_g = 0$ using Eqs. (\ref{eq:56}) and (\ref{eq:57}). We also calculate $U_2$ at any arbitrary gate voltage (for example $V_g = k_B T/e$). By this procedure, we get $U_2 = 5.64910 k_B T/e$ and $U_2' = 5.358072 k_B T/e$ with $u_{g2} = 0.291028$. Therefore, the average interaction potential for an arbitrary gate voltage $V_g$ is $U = U' + u_g V_g = 8.2657 k_B T/e  + u_g V_g$, where $u_g = 0.228815$. Thus, one applies a gate voltage $V_g = \frac{-8.2657}{0.228815} + \frac{V_g'}{0.228815} = -36.1239 + \frac{V_g'}{0.228815}$ in units of $k_B T/e$, with the first term of $V_g$ nullifying the interaction potential $U' = 8.2657 k_B T/e$ and the second term with $V_g'$ lifting the energy level of a QPC-type constriction. Thus, the threshold energy of the QPC is $E_1' = E_1 + e V_g' $. In Table \ref{Table1} above, we provide a list of 4 different bias voltages and the resulting interaction potentials along with the applied the gate voltage needed to nullify the interaction potential. The reason taking these values of biases is because at $V = 1.14 k_B T/e, 1.77 k_B T/e, 3.7 k_B T/e$, the 2T QH setup (Fig. \ref{fig:7}) acts as QHE and at $V = 20 k_B T/e$, it acts as QR.}

\begin{figure} 
\centering
\includegraphics[width=1.00\linewidth]{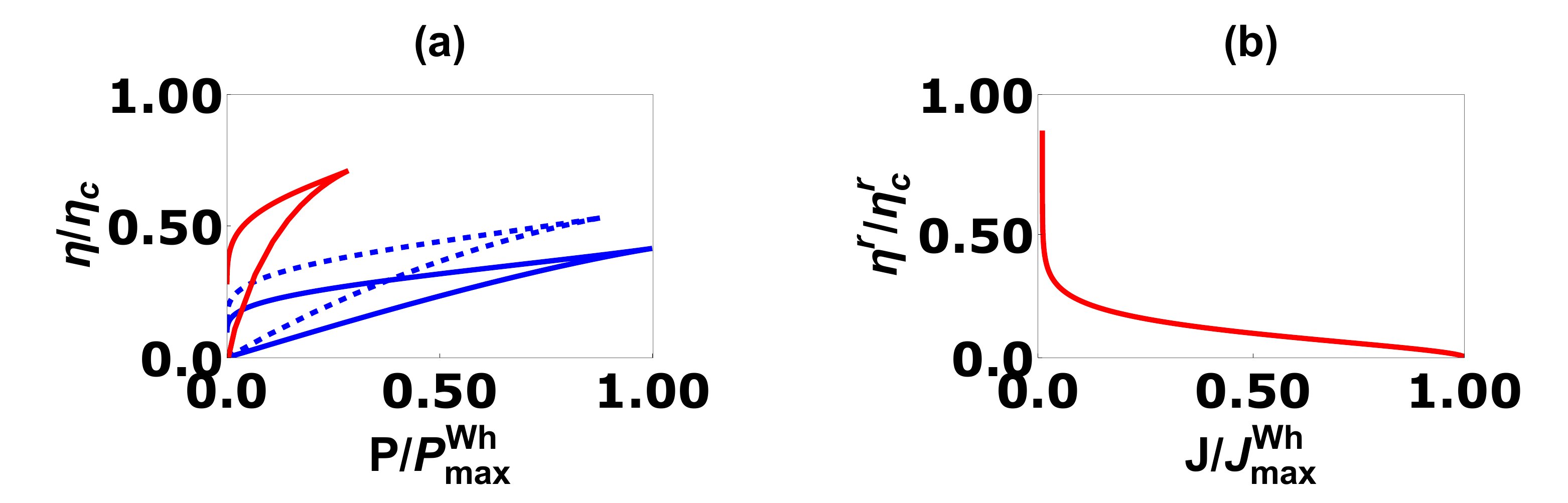}
\caption{(a) Parametric plot of the power ($P/P_{max}^{Wh}$) vs. efficiency ($\eta/\eta_c$) as a heat engine. The blue thick line corresponds to $V = 1.14 k_B T/e$, blue dotted line corresponds to $V = 1.77 k_B T/e$, red thick line corresponds to $V = 3.7 k_B T/e$, (b) Parametric plot of the cooling power ($J/J_{max}^{Wh}$) and coefficient of performance ($\eta^r/\eta^r_c$) for QR in two-terminal QH (chiral) setup at $V = 20 k_B T/e$. Here, we have taken $E_{QPC}' = k_B T$ and $T_1 = 2T_2 = 2K$, while $\tau = T_1 - T_2  = 1K$. For the QPC, we have taken $\omega = 0.1k_B T/e$. These results are similar to the three-terminal QH setup without any voltage-temperature probe with $V_1 = -V, V_2 = V_3 = 0$ and with $T_1 = 2K, T_2 = T_3 = 1K$. }
\label{fig:8}
\end{figure}

With this, the power delivered ($P = I_1 V$) and efficiency ($\eta = P / J_1$) can be found using the formulae given in Eq. (\ref{eq:35}). We fix the values of $V$ and $\tau$ and evaluate the power and efficiency with different values of $V_g'$. We plot them parametrically in Fig. \ref{fig:8}(a) fixing $V = 1.14 k_B T/e$ and $\tau = 1K$, we find that at a particular $V_g'$, the maximum power approaches $P_{max}^{Wh}$ as shown in Fig. \ref{fig:8}(a).

\begin{table}[]
\renewcommand{\arraystretch}{1.5} 
\centering
\caption{Numerical estimation of interaction potential and the applied gate voltage in 2T QH setup. Parameters taken are $\tau = 1K, C = 0.01 F, \hbar \omega = 0.1 k_B T$ and $\frac{1}{2}m \omega^2 \lambda^2 = 20 k_B T$.}
\begin{tabular}{|c|c|c|}
\hline
Applied Voltage bias ($k_B T/e$) & Interaction potential ($k_B T/e$) & Applied gate voltage ($k_B T/e$)\\ \hline
1.14 & 8.2657 & $-36.1239 + \frac{V_g'}{0.228815}$ \\ \hline
1.77 & 7.59397 & $-30.76910 + \frac{V_g'}{0.246805}$ \\ \hline
3.7 & 3.85578 & $73.17859 - \frac{V_g'}{0.05269}$ \\ \hline
20 & 6.506805 & $-764.15795 + \frac{V_g'}{0.008515}$ \\ \hline
\end{tabular}
\label{Table1}
\end{table}

Similarly, for the quantum refrigerator, we consider $V_g = -U + V_g'$ and calculate the cooling power $ J = J_2$ and coefficient of performance ($\eta^r = J/P$), with $P = I_1 V$ for different values of $V$. The parametric plot of the cooling power $J$ and $\eta^r$ is shown in Fig. \ref{fig:8}(b). The maximum cooling power approaches $J_{max}^{Wh}$, and the coefficient of performance reaches its highest, i.e., 86$\%$ of $\eta_c^r$.

\subsubsection{Nonlinear thermoelectrics in 3T QH setup (Our approach)}\label{Sec:II(B3)}

 Here, we discuss the power and efficiency of a simple three-terminal QH setup without voltage-temperature probes and with two QPC-like constrictions as shown in Fig. \ref{fig:3} by controlling the transmission of QPC 1 and QPC 2 by an applied gate voltage using the conditions: $E_1^{'} = E_1 + e V_g$ and $E_2^{'} = E_2 + e V_{g}$, and $E_1 = E_2$. We consider a simple case with $V_1 = -V, V_2 = V_3 =0$, and $T_1 = T + \tau, T_2 = T_3 = T$. Similar to the 2T QH case as in Sec. \ref{Sec:II(B2)}, the interaction potential in the presence of a gate voltage can be nullified by considering the gate voltage $V_g$ as in Eq. (\ref{eq:58}). For the heat engine, the power delivered ($P = I_1 V$) and efficiency ($\eta = P/J_1$) can be found using the Eqs. (\ref{eq:40}) and (\ref{eq:41}) for different values of voltage bias ($V$), the performance exactly matches with the result of two-terminal heat engine as shown in Fig. \ref{fig:8}(a). Similarly, for the refrigerator, we take $V_g = -U + V_g'$ and cooling power ($J = -(J_2 + J_3)$) and coefficient of performance ($\eta^r = J/P$) exactly matches the result of the two-terminal refrigerator as in Fig. \ref{fig:8}(b).

\subsubsection{Nonlinear thermoelectrics in 3T QH setup with voltage-temperature probe (with QPC type constriction)}\label{Sec:II(B4)}

\begin{figure} 
\centering
\includegraphics[width=0.60\linewidth]{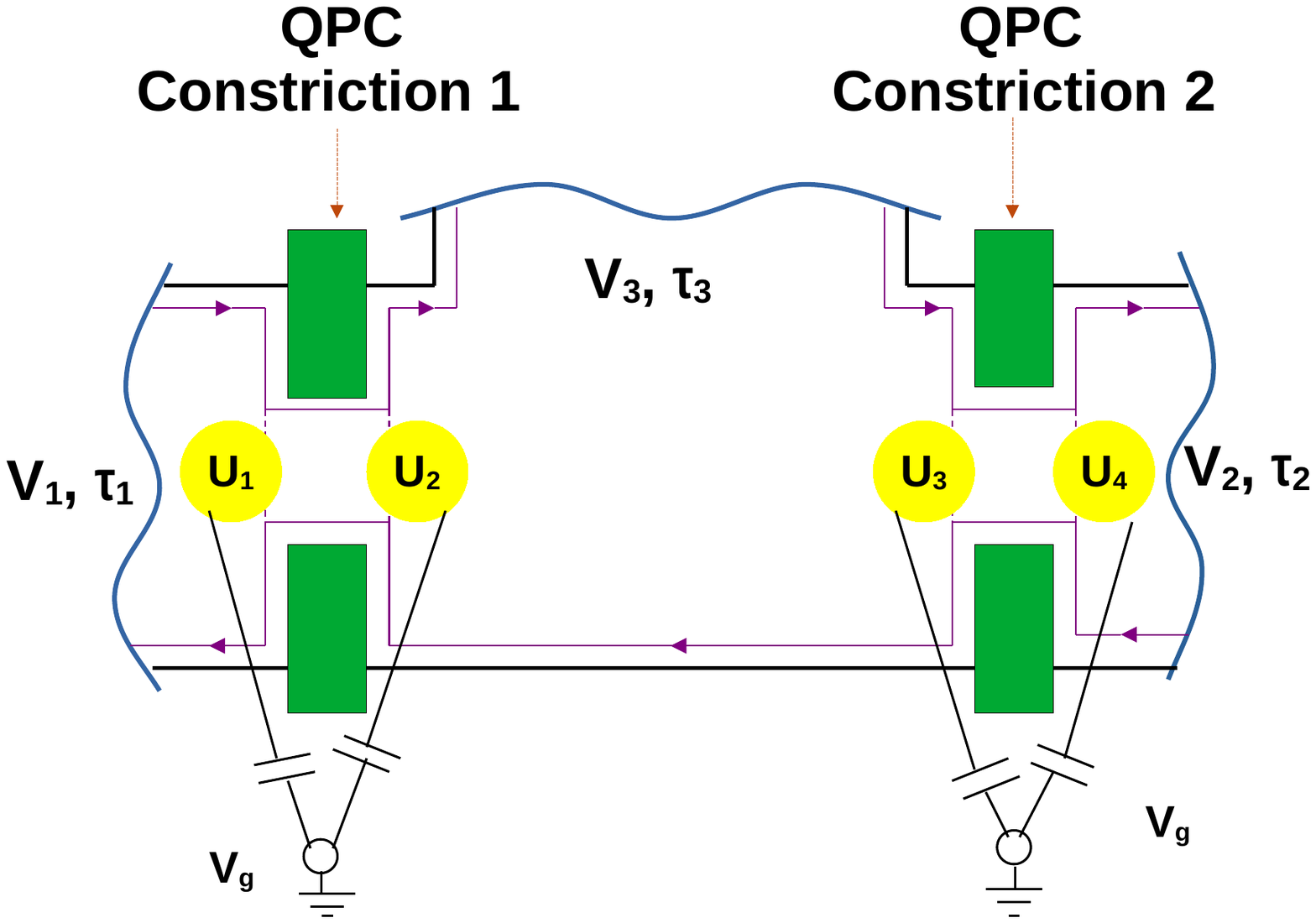}
\caption{{Three-terminal QH setup with voltage-temperature probe in nonlinear response, which is discretized into four regions $r = 1, 2, 3, 4$ (yellow shaded regions) to the left and right of the QPC-1 and QPC-2 type constrictions. Regions 1 and 2 are connected capacitively to a common external gate voltage $V_{g1}$ and $3$ and 4 are connected capacitively to another common external gate voltage $V_{g2}$.} }
\label{fig:9}
\end{figure}

 Next, we extend our approach to a 3T nonlinear QH setup as shown in Fig. \ref{fig:9} with a voltage-temperature probe by controlling the transmission of QPCs 1 and QPC 2 via a gate voltage. {We consider the threshold energies of either the QPC-type constrictions to be $E_1 = E_2 = k_B T$. We extend the calculation of a two-terminal QH setup with one QPC-type constriction to a three-terminal QH setup with two QPC-type constrictions. We consider two regions each on the left and right sides of QPC-1 and QPC-2. The regions to the left and right sides of QPC-1 are denoted as 1 and 2, and the QPC-2 regions are denoted as 3 and 4. One can find the interaction potentials in each region independently by assuming the interaction potentials of the other region to be zero. For finding $U_1$, we assume $U_2 = U_3 = U_4 = 0$. Using Eq. (48), the additional injected charge $q_{ad}^1$  in region 1 is \cite{PhysRevLett.110.026804, PhysRevB.88.045129, Meair_2013}} 
 \begin{equation}\label{eq:61}
 \begin{split}
    {q_{ad}^1} & {= e \int dE \left ( \sum_{\alpha} \nu_{1 \alpha}^p(E, U_1) f_{\alpha}(E) \right) - e \int dE  \nu_1^p(E) f,}\\
    & {= e \int dE \left(\nu_{1 1}^p(E, U_1)f_1(E) +  \nu_{1 2}^p(E, U_1)f_2(E) + \nu_{1 3}^p(E, U_1)f_3(E) \right) - e \int dE  \nu_1^p(E) f.} 
    \end{split}
\end{equation}
{Here, $\nu_{1\alpha}^p$ are the particle injectivities into region 1 from terminal $\alpha \in \{1, 2, 3 \}$ and it is given as $\nu_{1 \alpha}^p = \sum_{\beta \in \{1, 2, 3 \}} \nu_{\beta 1 \alpha}^p$, where $\nu_{\beta 1 \alpha}$ is the partial density of states of electrons scattered from contact $\alpha$
to contact $\beta$. One can calculate the partial density of states as follows. The partial density of state $\nu_{111}^p$ is $\mathcal{T}_{11}$ times $\nu_1^p$, i.e.,  $ \nu_{111}^p = (1 - \mathcal{T}_{QPC1})\nu_1^p$, where $\mathcal{T}_{QPC1}$ is the transmission probability of the QPC as shown in Fig. \ref{fig:9}. Similarly, $\nu_{211}^p$ is 0 as $\mathcal{T}_{12} = 0$ (because of chiral edge mode property) and $\nu_{311}^p$ is $\mathcal{T}_{31}$ times $\nu_1^p$, which is equal to $\nu_{311}^p = \frac{\mathcal{T}_{QPC1}}{2} \nu_1^p$. Therefore, $\nu_{11}^p = \nu_{111}^p + \nu_{2 11}^p + \nu_{311}^p = \left(1 - \frac{\mathcal{T}_{QPC1}}{2}\right) \nu_1^p$. Similarly, one can also calculate $\nu_{12}^p = \nu_{112}^p + \nu_{212}^p + \nu_{312}^p$ and $\nu_{13}^p = \nu_{113}^p + \nu_{213}^p + \nu_{313}^p$ as well. $\nu_{112}^p$ is given as $ \mathcal{T}_{12} \nu_{1}^p = \frac{\mathcal{T}_{QPC1} \mathcal{T}_{QPC2}}{2} \nu_1^p$, whereas $\nu_{213}^p$ vanishes as there are no states in region 1 associated with electrons scattered from terminal 3 to terminal 2. Similarly, $\nu_{313}^p = \mathcal{T}_{33} \nu_1^p = 0$ even if $\mathcal{T}_{33} \ne 0$ and $\nu_1^p \ne 0$ because there are no states in region 1 for electrons scattered from terminal 3 to itself. Therefore, $\nu_{12}^p = \frac{\mathcal{T}_{QPC1} \mathcal{T}_{QPC2}}{2} \nu_1^p$. For $\nu_{13}^p$, the contribution comes only from $\nu_{113}^p$, which is equal to $\frac{\mathcal{T}_{QPC1}(1- \mathcal{T}_{QPC2})}{2} \nu_1^p$, whereas $\nu_{213}^p$ and $\nu_{313}^p$ vanishes as there are no states in region 1 associated with the scattering of electron from terminal 3 to terminal 2 and therefore, $\nu_{13}^p= \nu_{113}^p + \nu_{213}^p + \nu_{313}^p = \frac{\mathcal{T}_{QPC1}(1 - \mathcal{T}_{QPC2})}{2} \nu_1^p$, where $\mathcal{T}_{QPC1} = \frac{1}{1 + e^{-\frac{E - E_1 - eU_1}{\hbar \omega}}}$ and $\mathcal{T}_{QPC2} = \frac{1}{1 + e^{-\frac{E - E_2}{\hbar \omega}}}$. Now, one can capacitively connect the region 1 with an external gate voltage and the additional injected charge is given as,}
\begin{equation}\label{eq:62}
    {q_{ad}^1 = C(U_1 - V_g)}.
\end{equation}
{From here, one can calculate $U_1$ numerically. First, we find the interaction potential $U_1'$, which is solely dependent upon the applied biases. In the presence of gate voltage $V_g$, the interaction potential $U_1$ is $U_1' + u_{g1} V_g$, where $u_{g1}$ determines the response of $U_1$ to the applied gate voltage $V_g$. One needs the value of $u_{g1}$, which will help in nullifying the interaction potential $U_1'$, and it is given as $u_{g1} = \frac{U_1 - U_1'}{V_g}$. One can thus calculate $u_{g1}$ at any arbitrary value of $V_g$, implying numerically one can calculate the interaction potential $U_1$ at any arbitrary $V_g$, say $1 k_B T/e$, giving $u_{1g}$ is $U_1 - U_1'$. For our purpose, we consider $V_1 = -V, V_2 = 0, \tau_1 = \tau, \tau_2  = 0$. Then, using the condition $I_3 = J_3  = 0$, we can calculate the interaction potential $U_1$ and $U_1'$ in region 1. The Python code for this calculation can be found in Ref. \cite{My_mathematica_notebook}.}

{Using a similar procedure, we can calculate $U_2$, $U_3$, $U_4$. The partial density of states or particle injectivities for each region are,}
\begin{equation}\label{eq:63}
\begin{split}
    {\nu_{21}^p} & {= \sum_{\beta \in \{1, 2, 3 \}} \nu_{\beta 21}^p = \frac{\mathcal{T}_{QPC1} \nu_2^p}{2}, \quad \nu_{22}^p  = \sum_{\beta \in \{1, 2, 3 \}} \nu_{\beta22}^p = \left(1 - \frac{\mathcal{T}_{QPC1}}{2}\right)\mathcal{T}_{QPC2}\nu_2^p,} \\ {\nu_{23}^p} & {=  \sum_{\beta \in \{1, 2, 3 \}} \nu_{\beta23}^p = \left(1 - \frac{\mathcal{T}_{QPC1}}{2}\right)(1 - \mathcal{T}_{QPC2})\nu_2^p, \quad \nu_{31}^p = \sum_{\beta \in \{1, 2, 3 \}} \nu_{\beta31}^p = 0,}\\
    {\nu_{32}^p} & {= \sum_{\beta \in \{1, 2, 3 \}} \nu_{\beta32}^p = \frac{\mathcal{T}_{QPC2}\nu_3^p}{2},\quad \nu_{33}^p = \sum_{\beta \in \{1, 2, 3 \}} \nu_{\beta33}^p \left(1 - \frac{\mathcal{T}_{QPC2}}{2}\right)\nu_3^p, \quad \nu_{41}^p = \sum_{\beta \in \{1, 2, 3 \}} \nu_{\beta41}^p = 0,}\\
    {\nu_{42}^p} & {= \sum_{\beta \in \{1, 2, 3 \}} \nu_{\beta42}^p = \left(1 - \frac{\mathcal{T}_{QPC2}}{2}\right)\nu_4^p,\quad \text{and} \quad \nu_{43}^p = \sum_{\beta \in \{1, 2, 3 \}} \nu_{\beta43}^p = \frac{\mathcal{T}_{QPC2}}{2}.}
    \end{split}
    \end{equation}

\begin{table}[h]
\renewcommand{\arraystretch}{1.5}
\centering
\caption{Estimation of interaction potential and the applied gate voltages at QPC-1 and QPC-2 in a 3T QH setup with the voltage-temperature probe. Parameters taken are $\tau = 1K, C = 0.01 F, \hbar \omega = 0.1 k_B T$ and $\frac{1}{2}m \omega^2 \lambda^2 = 20 k_B T$.}
\renewcommand{\arraystretch}{1.5} 
\scalebox{0.90}{
\begin{tabular}{|c|c|c|c|c|}
\hline
Applied Voltage Bias ($k_B T/e$) & \begin{tabular}[c]{@{}c@{}}Interaction Potential\\ for QPC-1 ($k_B T/e$)\end{tabular} & \begin{tabular}[c]{@{}c@{}}Applied Gate Voltage\\ at QPC-1 ($k_B T/e$)\end{tabular} & \begin{tabular}[c]{@{}c@{}}Interaction Potential\\ at QPC-2 ($k_B T/e$)\end{tabular} & \begin{tabular}[c]{@{}c@{}}Applied Gate Voltage\\ at QPC-2\end{tabular} \\ \hline
1.14                 &           4.55493                        &    $-32.71279 + \frac{V_g'}{0.13924}$                             &        4.71373                         &   $-65.26452 + \frac{V_g'}{0.072225}$                              \\ \hline
1.77                 &       4.80969                            &        $69.24701 - \frac{V_g'}{0.069457}$                         &          -1.46130                        &      $137.92 - \frac{V_g'}{0.010595}$                           \\ \hline
3.7                  &              6.39887                     &      $-44310.1379 + \frac{V_g'}{0.000145}$                           &         -0.20151                        &         $0.26547 + \frac{V_g'}{0.75906}$                        \\ \hline
20                   &               4.95627                    &   $-10434.2526 + \frac{V_g'}{0.000475}$                              &      6.85963                            &                              $-1294.26 + \frac{V_g'}{0.0053}$   \\ \hline
\end{tabular}}
\label{Table2}
\end{table}

{For QPC-1, the average interaction potential:} 

\begin{equation}\label{eq:64}
    {U_{QPC1} = \frac{U_1 + U_2}{2} = \frac{U_1' + U_2'}{2} + u_{g1} V_{g1} = U_{QPC}' + u_{g1} V_{g1}.}
\end{equation}    
    {We consider the same values of $C, \hbar \omega$ and $\frac{1}{2}m \omega^2 \lambda^2$, which we took during the calculation of interaction potential in two-terminal QH setup, see below Eq. (\ref{eq:60}). For the calculation, we again consider $V = 1.14 k_B T/e$ and $\tau = 1K$. With this information, we got $u_{g1} = 0.13924$, $U_{QPC1}' = 4.55493 k_B T/e$ and the gate voltage is $V_{g1} = \frac{-U_{QPC}'}{u_{g1}} + \frac{V_g'}{u_{g1}} = -32.71279 + \frac{V_g'}{0.13924}$. Therefore, the threshold energy of QPC-1 becomes $E_1 + e V_g'$. For QPC-2, the average interaction potential is
    \begin{equation}\label{eq:65}
    U_{QPC2} = \frac{U_3  + U_4}{2} = U_{QPC2} = \frac{U_3' + U_4'}{2} + u_{g2} V_{g2}. 
    \end{equation}
    Considering identical values of $C, \hbar \omega$ and $\frac{1}{2}m \omega^2 \lambda^2$ for the QPC-2, we get the interaction potential $U_{QPC2}' = 4.71373 k_B T/e$ and the gate voltage $V_{g2} = -66.26452 k_B T/e + \frac{V_g'}{0.072225}$ and the threshold energy of the QPC-2 is $E_2  + e V_g'$. We tabulate the value of interaction potential and gate voltage required to nullify it for different voltage biases such as $V = 1.77 k_B T/e$, $3.7 k_B T/e$ and $20 k_B T/e$ in Table \ref{Table2}. The voltage biases $V = 1.14 k_B T/e$ (blue dot-dashed curve in Fig. 3(a) of the main text), $1.77 k_B T/e$ (blue dashed curve in Fig. 3(a) of the main text) and $V = 3.7 k_B T/e$ (blue dotted curve in Fig. 3(a) of the main text) are considered to study the performance of the setup as QHE as these values fall in the regime, where the setup can work as QHE. On the other hand, $V= 20 k_B T/e$ (blue curve in Fig. 3(c) of the main text) is considered to study the setup's performance as a QR.}


  \begin{figure}
     \centering
     \begin{subfigure}[b]{0.45\textwidth}
         \centering
         \includegraphics[width=\textwidth]{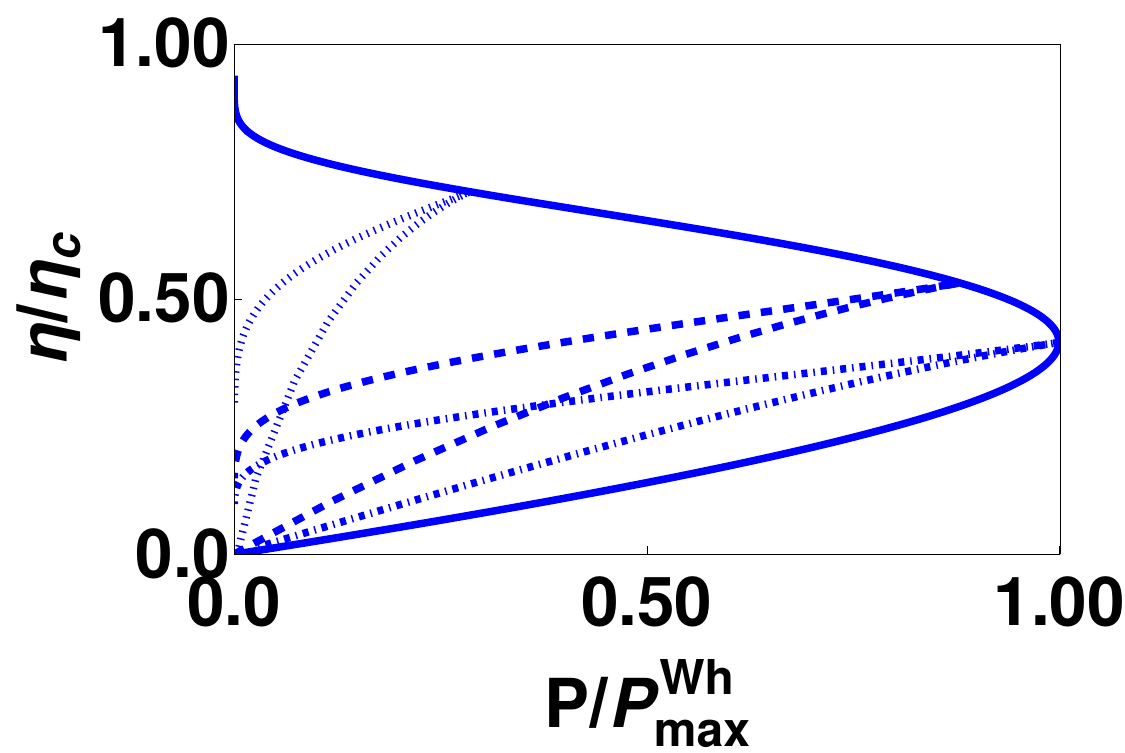}
         \caption{QH heat engine (nonlinear)}
     \end{subfigure}
     \hspace{0.05cm}
     \begin{subfigure}[b]{0.45\textwidth}
         \centering
         \includegraphics[width=\textwidth]{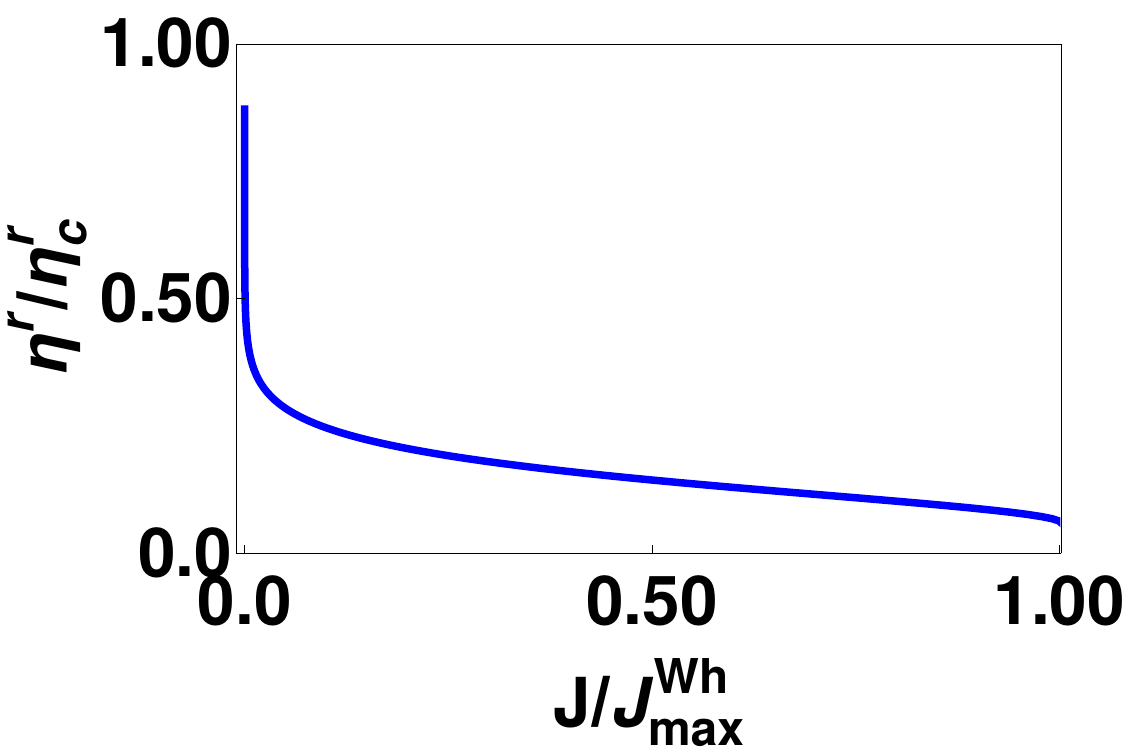}
         \caption{QH refrigerator (nonlinear)}
     \end{subfigure}
        \caption{Parametric plot of $\frac{\eta}{\eta_c}$ vs $\frac{P}{P_{max}^{Wh}}$ for (a) QH QHE. The blue dot-dashed curve is for $V = 1.1 k_B T/e$, the blue dashed curve for $V = 1.77 k_B T/e$, and the blue dotted curve for $V = 3.7 k_B T/e$. The blue line shows the bound to the power and efficiency. Parametric plot for $\frac{\eta^r}{\eta^r_c}$ vs $\frac{\textbf{J}}{\textbf{J}^{Wh}_{max}}$ for (c) QH QR.
         For QH  QR, the blue line is for $V = 20k_B T/e$. Parameters taken are $\omega = \frac{0.1 k_B T}{h}$, $T_1 = 2K, T_2 = 1K$, $\mu = 0$, $2E_1 = E_2 = 2k_B T$.}
         \label{fig:10}
       \end{figure}
       
 
 Considering this, one can calculate power and efficiency as a quantum heat engine by changing $V_g^{'}$ at fixed values of $V, \tau$ and we observe that the maximum power approaches the Whitney bound, whereas the maximum efficiency is 0.93 $\eta_c$ as shown in Fig. 3(a) of the main text.

  Similar to the quantum heat engine, we investigate the three-terminal QH setup as a quantum refrigerator with voltage-temperature probe by our approach and considering identical values of $E_1^{'}$ and $E_2^{'}$ same as used for quantum heat engine. The coefficient of performance $\eta^r$ $= \frac{J}{P}$ and the cooling power $J$ is $-(J_2 + J_3)$ are evaluated with either probe and the parametric plot are shown in Fig. 3(c) of the main text.
 With these conditions, in the three-terminal QH setup for a voltage-temperature probe, we observe that $J_{max}$ reaches $J^{Wh}_{max}$, while $\eta^r$ reaches about 86$\%$ of $\eta^r_c$ as shown in Fig. 3(c) of the main text. In the main text, we considered $E_1 = E_2 = k_B T$ and achieved the Whitney limits for both the quantum heat engine and refrigerator. When they are not the same, then also the Whitney limits are achieved, see Fig. \ref{fig:10}, with $2E_1 = E_2 = 2k_B T$, which are identical Fig. 3(a) and (c) in main text with $E_1 = E_2 = k_B T$.

\section{Charge and heat current in QSH setup}\label{Sec:III}
For a multiterminal QSH setup, the charge and heat current in terminal $\alpha$ can be derived using Landauer-Buttiker formalism \cite{PhysRevB.91.155407, PhysRevE.97.022114}. The total current in terminal $\alpha$ is a sum of currents due to spin-up as well as spin-down electrons. The charge current due to an electron with spin $\sigma = \uparrow \text{or} \downarrow$ is,
\begin{equation} \label{eq:66}
    I_{\alpha}^{\sigma} = \frac{e}{h}\int_{-\infty}^{\infty}dE \sum_{\beta}f_{\beta}(E)[N_{\alpha}\delta_{\alpha \beta} - \mathcal{T}_{\alpha \beta}^{\sigma}],
\end{equation} 
where, $\mathcal{T}_{\alpha \beta}^{\sigma} = \sum_{\rho = \uparrow/\downarrow} \mathcal{T}_{\alpha \beta}^{\sigma \rho}$ with $\mathcal{T}_{\alpha \beta}^{\sigma \rho}$ being the transmission probability for an electron to transmit from terminal $\beta$ with initial spin $\rho$ to terminal $\alpha$ with final spin $\sigma$. Similarly, the heat current is given as,
\begin{equation} \label{eq:67}
    J_{\alpha}^{\sigma} = \frac{1}{h}\int_{-\infty}^{\infty}dE (E - \mu_{\alpha}) \sum_{\beta} f_{\beta}(E)[N_{\alpha} \delta_{\alpha \beta} - \mathcal{T}_{\alpha \beta}^{\sigma}].
\end{equation}

\subsection{Transport in the linear response regime in the QSH setup}\label{Sec:III(A)}
Similar to the chiral case, in the linear response regime, both charge and heat currents can be derived as,
\begin{align} \label{eq:68}
    I_{\alpha}^{\sigma} &= \sum_{\beta}(G_{\alpha \beta}^{\sigma} V_{\beta} + L_{\alpha \beta}^{\sigma}\tau_{\beta}),\quad
    J_{\alpha}^{\sigma} = \sum_{\beta}(\Pi_{\alpha \beta}^{\sigma} V_{\beta} + K_{\alpha \beta}^{\sigma}\tau_{\beta}),
\end{align}
where,
\begin{align} \label{eq:69}
     \begin{split}
         G_{\alpha \beta}^{\sigma} &= \frac{e^2}{h} \int_{-\infty}^{\infty}dE \left(- \frac{\partial f}{\partial E}\right)[N_{\alpha} \delta_{\alpha \beta} - \mathcal{T}_{\alpha \beta}^{\sigma}],\quad
         L_{\alpha \beta}^{\sigma} = \frac{e}{h T} \int_{-\infty}^{\infty} dE (E - \mu) \left(-\frac{\partial f}{\partial E}\right) [N_{\alpha}\delta_{\alpha \beta} - \mathcal{T}_{\alpha \beta}^{\sigma}],\\
         \Pi_{\alpha \beta}^{\sigma} &= \frac{e}{h} \int_{-\infty}^{\infty} dE (E - \mu) \left(-\frac{\partial f}{\partial E}\right) [N_{\alpha}\delta_{\alpha \beta} - \mathcal{T}_{\alpha \beta}^{\sigma}],\quad
         K_{\alpha \beta}^{\sigma} = \frac{1}{h T} \int_{-\infty}^{\infty}dE (E - \mu)^2 \left(-\frac{\partial f}{\partial E}\right)[N_{\alpha}\delta_{\alpha \beta} - \mathcal{T}_{\alpha \beta}^{\sigma}].
     \end{split}
 \end{align}\\
The quantities as written in Eq. (\ref{eq:69}) are the spin-polarized Onsager coefficients for helical edge mode transport.

In the subsequent subsections, we derive Onsager coefficients for our QH setup (See, Sec. \ref{Sec:III(A1)}) with a three-terminal voltage-temperature probe. In Sec. \ref{Sec:III(A2)}, we derive the general formulae for power and efficiency as a quantum heat engine. Further in Sec. \ref{Sec:III(A3)}, we discuss the cooling power and coefficient of performance as a quantum refrigerator.
 
\subsubsection{Derivation of Onsager coefficients in a three-terminal QSH setup in the linear response regime}\label{Sec:III(A1)}
\begin{figure} 
\centering
\includegraphics[width=0.60\linewidth]{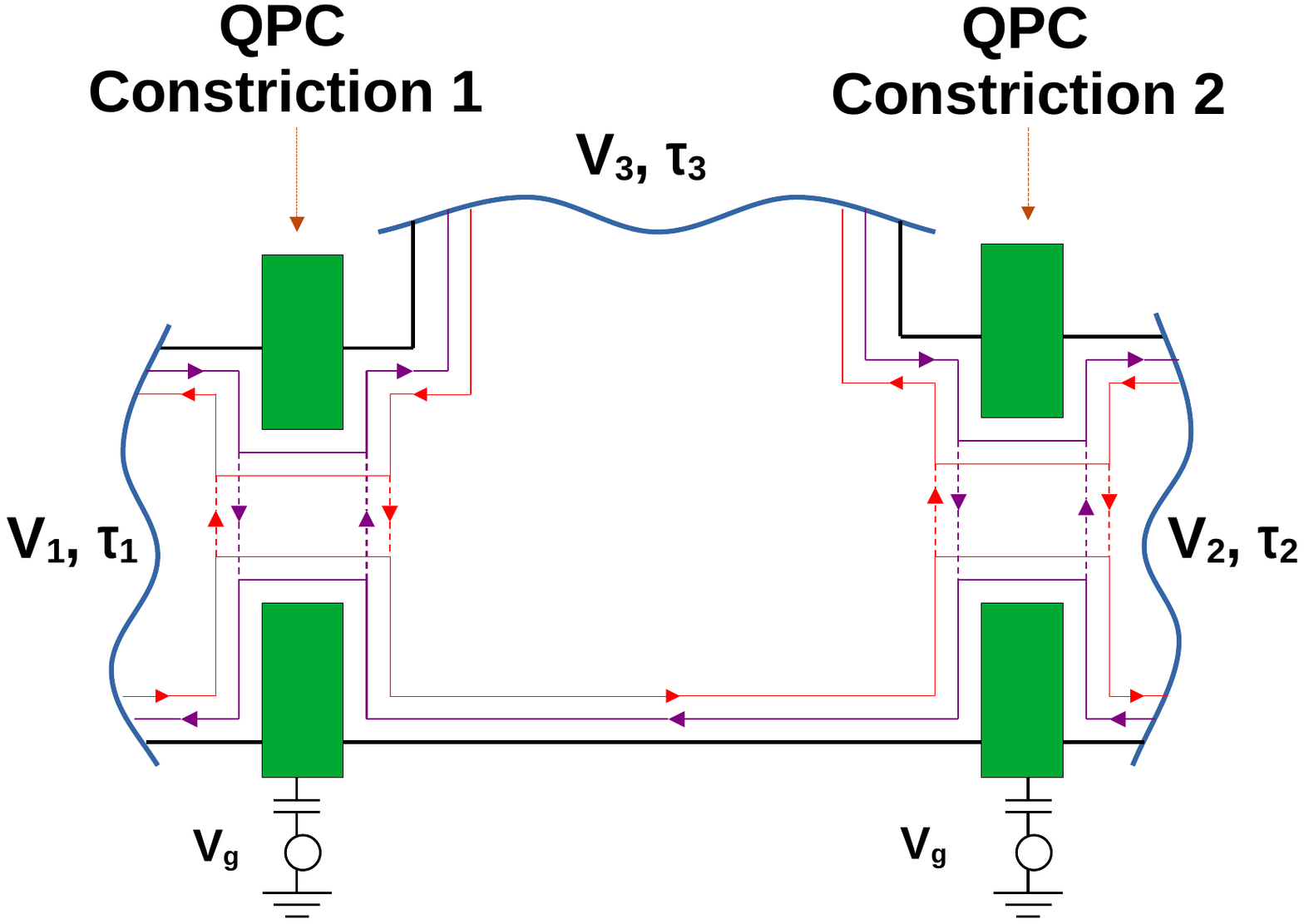}
\caption{Three terminal QSH sample with helical edge modes and two constrictions. The red solid (dashed) line represents the edge mode for spin-down electrons, whereas the purple solid (dashed) line represents the edge mode for spin-up electrons, which can scatter from the constrictions. Terminal 3 is a voltage-temperature probe.}
\label{fig:11}
\end{figure}

In this section, we discuss the thermoelectric properties of a three-terminal QSH setup as shown in Fig. \ref{fig:11} with terminal 3 acting as either a voltage probe or a voltage-temperature probe. For this setup, we consider $\tau_{2} = \tau_3 = 0$, i.e., $T_2 = T_3 = T$ and $T_1 = T + \tau_1$, wherein $\tau_1 = \tau$. Similarly for the QSH setup as shown in Fig. \ref{fig:11}, due to the presence of helical edge modes, there will be electron motion via both spin-up and spin-down electrons in the opposite directions. Therefore, the transmission probabilities ($\mathcal{T}_{\alpha \beta}^{s\sigma}$) for an electron to scatter from terminal $\beta$ with initial spin $\sigma$ to enter terminal $\alpha$ with final spin $s$ for $s = \uparrow/ \downarrow$ and $\sigma = \uparrow / \downarrow$ are given as,
\begin{align}\label{eq:70}
\begin{split}
    \mathcal{T}_{11}^{\uparrow \uparrow} &= \mathcal{T}_{11}^{\downarrow \downarrow} = (1-\mathcal{T}_1),\quad \mathcal{T}_{12}^{\uparrow \uparrow} = \mathcal{T}_{21}^{\downarrow \downarrow} = \mathcal{T}_1 \mathcal{T}_2, \quad
    \mathcal{T}_{12}^{\downarrow \downarrow} = \mathcal{T}_{21}^{\uparrow \uparrow} = 0, \quad \mathcal{T}_{13}^{\uparrow \uparrow} = \mathcal{T}_{31}^{\downarrow \downarrow} = \mathcal{T}_1(1 - \mathcal{T}_2),\quad \mathcal{T}_{13}^{\downarrow \downarrow} = \mathcal{T}_{31}^{\uparrow \uparrow} = \mathcal{T}_1\\
    \mathcal{T}_{22}^{\uparrow \uparrow} &= \mathcal{T}_{22}^{\downarrow \downarrow} = (1-\mathcal{T}_2),\quad
    \mathcal{T}_{23}^{\uparrow \uparrow} = \mathcal{T}_{32}^{\downarrow \downarrow} = \mathcal{T}_2, \quad \mathcal{T}_{23}^{\downarrow \downarrow} = \mathcal{T}_{32}^{\uparrow \uparrow} =  (1-\mathcal{T}_1)\mathcal{T}_2, \quad 
    \mathcal{T}_{33}^{\uparrow \uparrow} = \mathcal{T}_{33}^{\downarrow \downarrow} = (1-\mathcal{T}_1)(1-\mathcal{T}_2).
    \end{split}
    \end{align}

 Similar to the previous case, we consider $\tau_1 = \tau$, $\tau_2 = 0$, but we consider terminal 3 to be a voltage-temperature probe, i.e., both the charge current $I_3$ and the heat current $J_{3}$ through terminals are zero, and terminal 2 is a current probe, with, $V_2 = 0$.  Here, chemical potential $\mu$ is taken to be zero. We derive the Onsager coefficients, as follows from Eq. (\ref{eq:68}), the charge and heat currents in terminal 3 is given as,
 \begin{align} \label{eq:71}
\begin{split}
    I_3 &= \sum_{\sigma \in \{\uparrow, \downarrow\}} I_3^{\sigma} = \sum_{\sigma \in \{\uparrow, \downarrow\}} \left(\sum_{\beta} G_{3 \beta}^{\sigma}V_{\beta} + \sum_{\beta} L_{3 \beta}^{\sigma} \tau_{\beta}\right),\quad
    J_3 = \sum_{\sigma \in \{\uparrow, \downarrow\}} J_3^{\sigma} = \sum_{\sigma \in \{\uparrow, \downarrow\}} \left(\sum_{\beta} \Pi_{3 \beta}^{\sigma} V_{\beta} + \sum_{\beta}K_{3 \beta}^{\sigma} \tau_{\beta}\right).
    \end{split}
\end{align} 
Substituting $I_3 = J_3 = 0$, we get solutions for $V_3$ and $\tau_3$, as,
\begin{align} \label{eq:72}
\begin{split}
    V_3 &= \frac{(-K_{33} L_{31} + L_{33}K_{31}) \tau - (-K_{33}G_{31}+L_{33}\Pi_{31})V}{X},\quad
    \tau_3 = \frac{(-G_{33} K_{31} + \Pi_{33}L_{31}) \tau - (-\Pi_{33}G_{31}+G_{33}\Pi_{31})V}{X}.
    \end{split}
\end{align}
where, $X = (G_{33}K_{33} - L_{33} \Pi_{33})$ and  $G_{ij} = \sum_{\sigma \in \{\uparrow, \downarrow\}} G_{ij}^{\sigma}, L_{ij} = \sum_{\sigma \in \{\uparrow, \downarrow\}} L_{ij}^{\sigma}, \Pi_{ij} = \sum_{\sigma \in \{\uparrow, \downarrow\}} \Pi_{ij}^{\sigma}$ and $K_{ij} = \sum_{\sigma \in \{\uparrow, \downarrow\}} K_{ij}^{\sigma}$ for $i, j = 1, 2, 3$. The spin-polarized charge and heat currents in terminal 1 is given as,
 \begin{align} \label{eq:73}
\begin{split}
    I_1^{\sigma} &= \sum_{\beta} G_{1 \beta}^{\sigma} V_{\beta} + \sum_{\beta} L_{1 \beta}^{\sigma} \tau_{\beta},\quad
 \text{and}   \quad J_1^{\sigma} = \sum_{\beta} \Pi_{1 \beta}^{\sigma} V_{\beta} + \sum_{\beta}K_{1 \beta}^{\sigma} \tau_{\beta}.
    \end{split}
\end{align}
Using $V_3$ and $\tau_3$ from Eq. (\ref{eq:72}), $I_1^{\sigma}$ and $J_1^{\sigma}$ can be written as,
\begin{equation} \label{eq:74}
    \begin{pmatrix}
        I_1^{\sigma} \\
        J_1^{\sigma}
    \end{pmatrix} = \begin{pmatrix}
        L_{eV}^{\sigma} && L_{e T}^{\sigma}\\
        L_{hV}^{\sigma} && L_{h T}^{\sigma}
    \end{pmatrix} \begin{pmatrix}
        -V\\
        \tau
    \end{pmatrix},
\end{equation}
where,
    \begin{align} \label{eq:75}
        \begin{split}
            L_{eV}^{\sigma} &= G_{11}^{\sigma} + \frac{G_{13}^{\sigma}(L_{33} \Pi_{31} - G_{31}K_{33})}{X} - \frac{L_{13}^{\sigma}(G_{33}\Pi_{31}-G_{31}\Pi_{33})}{X},\\
            L_{e T}^{\sigma} &= L_{11}^{\sigma} + \frac{G_{13}^{\sigma}(L_{33} K_{31} - L_{31}K_{33})}{X} - \frac{L_{13}^{\sigma}(G_{33}K_{31}-L_{31}\Pi_{33})}{X},\\
            L_{hV}^{\sigma} &= \Pi_{11}^{\sigma} + \frac{\Pi_{13}^{\sigma}(L_{33} \Pi_{31} - G_{31}K_{33})}{X} - \frac{K_{13}^{\sigma}(G_{33}\Pi_{31}-G_{31}\Pi_{33})}{X},\\
              L_{h T}^{\sigma} &= K_{11}^{\sigma} + \frac{\Pi_{13}^{\sigma}(L_{33} K_{31} - L_{31}K_{33})}{X} - \frac{K_{13}^{\sigma}(G_{33}K_{31}-L_{31}\Pi_{33})}{X}.
             \end{split}
    \end{align}

Eq. (\ref{eq:75}) are the Onsager coefficients for 3T QSH system with voltage-temperature probe. The total conductance $G$, Seebeck coefficient $S$, Peltier coefficient $\Pi$ and thermal conductance $K$ are given as,
\begin{align} \label{eq:76}
\begin{split}
    G = \sum_{\sigma \in \{\uparrow, \downarrow\}}G^{\sigma},\quad
    S = \frac{1}{2}\sum_{\sigma \in \{\uparrow, \downarrow\}} S^{\sigma},\quad
    \Pi = \frac{1}{2}\sum_{\sigma \in \{\uparrow, \downarrow\}} \Pi^{\sigma},\quad
    K = \sum_{\sigma \in \{\uparrow, \downarrow\}} K^{\sigma}.
    \end{split}
\end{align}
where
\begin{align} \label{eq:77}
    G^{\sigma} &= L_{eV}^{\sigma},\quad S^{\sigma} = \frac{L_{eT}^{\sigma}}{L_{eV}^{\sigma}},\quad \Pi^{\sigma} = \frac{L_{hV}^{\sigma}}{L_{eV}^{\sigma}},\quad K^{\sigma} = L_{hT}^{\sigma} - \frac{L_{hV}^{\sigma}L_{eT}^{\sigma}}{L_{eV}^{\sigma}}.
\end{align}

\subsubsection{Power and efficiency in 3T QSH setup as a quantum heat engine}\label{Sec:III(A2)}
Now, the charge power generated in terminal 1 is given as,
\begin{align} \label{eq:78}
\begin{split}
    P &= VI_1 = V\sum_{\sigma \in \{\uparrow, \downarrow\}} I_1^{\sigma} 
    = V \sum_{\sigma \in \{\uparrow, \downarrow\}}(-L_{eV}^{\sigma} V + L_{e T}^{\sigma} \tau) = V (-L_{eV} V + L_{e T} \tau)
    \end{split}
\end{align}

The maximum charge power can be obtained from $\frac{\partial P}{\partial V} = 0$, i.e., $V = \frac{ L_{e T} \tau}{2 L_{eV}}$. Thus maximum power is $P_{max} = \frac{L_{e T}^2 \tau^2}{4 L_{eV}}.$
Similarly, the efficiency at maximum power is given as,
\begin{equation} \label{eq:79}
    \eta|_{P_{max}} = \frac{P_{max}}{J} = \frac{ T \eta_c}{2}\frac{L_{e T}^2}{2 L_{h T} L_{eV} - L_{hV} L_{e T}} = \frac{\eta_c}{2}\frac{ZT}{ZT + 2}.
\end{equation}
 Here, $Z T$ is the figure of merit, which is defined as
\begin{equation}\label{eq:80}
    Z T = \frac{L_{hV}L_{eT}}{L_{eV}L_{hT}-L_{eT}L_{hV}} = \frac{G S^2 T}{K}.
\end{equation}
For $Z T \rightarrow \infty$, $\eta|_{P_{max}} = \frac{\eta_c}{2}$. The maximum value of efficiency at maximum power ($\eta|_{P_{max}}$) one can attain is half of Carnot efficiency, which is defined as Curzon-Ahlborn efficiency \cite{BENENTI20171}.

Similarly, one can find maximum efficiency. The general expression for efficiency is given as
\begin{equation} \label{eq:81}
    \eta = \frac{V I_1}{J} = \frac{V(-L_{eV}V + L_{e T}\tau)}{L_{hV} V + L_{h T} \tau}.
\end{equation}
The voltage bias $V_{max}$ is given as,
\begin{equation} \label{eq:82}
    V_{max} = \frac{L_{h T}}{L_{h V}}\left(1 - \sqrt{\frac{L_{eV}L_{h T} - L_{eT} L_{h V}}{L_{eV} L_{h T}}}\right)\tau
\end{equation}
which can be derived from the condition $\frac{d \eta}{d V} = 0$. Thus, the maximum efficiency is given as
\begin{equation} \label{eq:83}
    \eta_{max} = \eta_c \frac{\sqrt{Z T + 1} - 1}{\sqrt{Z T + 1} + 1},\quad \text{where} \quad Z T = \frac{G S^2 T}{K} = \quad \text{Figure of merit}.
\end{equation}

\begin{figure}
     \centering
     \begin{subfigure}[b]{0.45\textwidth}
         \centering
         \includegraphics[width=\textwidth]{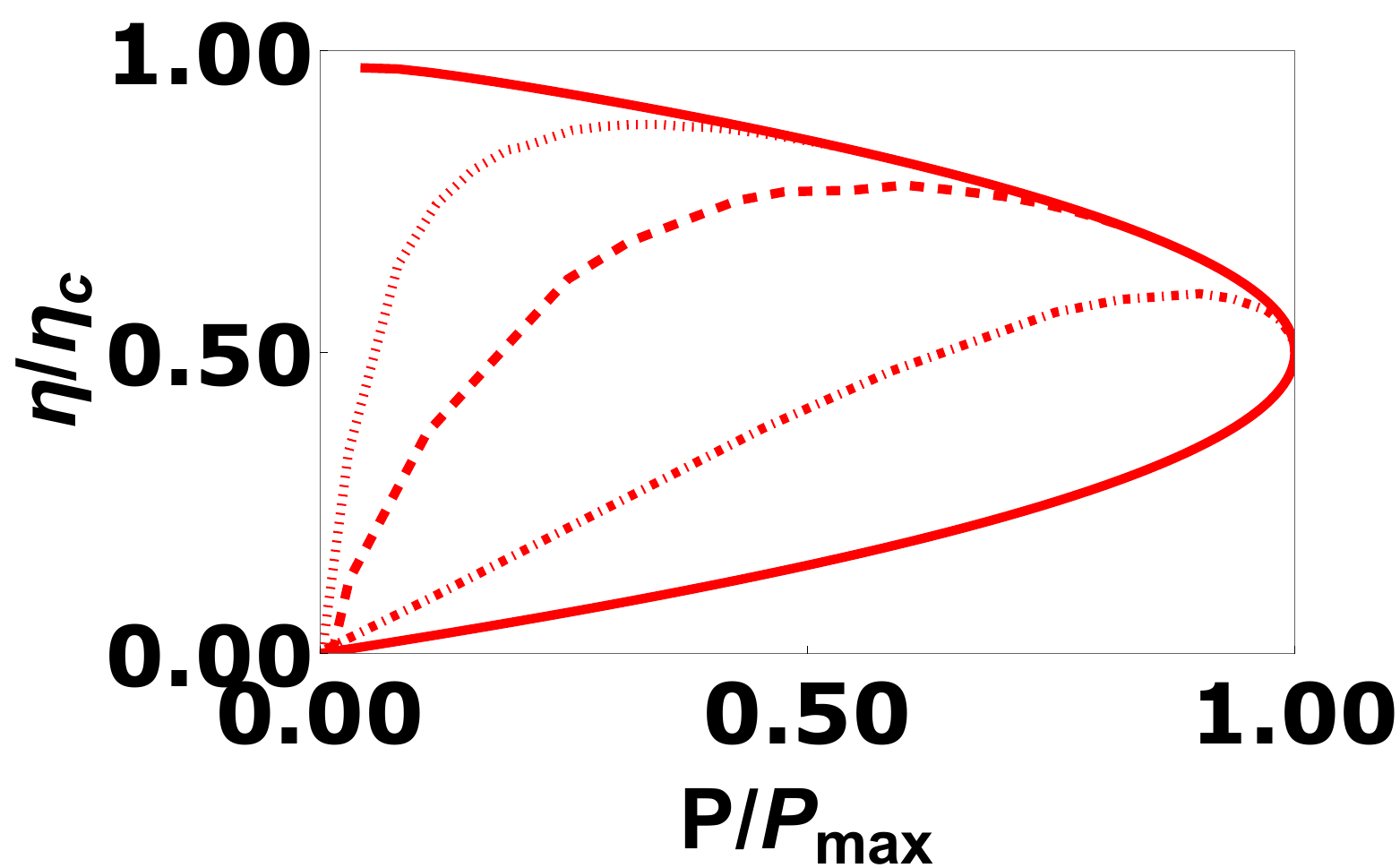}
         \caption{QH heat engine (linear)}
     \end{subfigure}
     \hspace{0.05cm}
     \begin{subfigure}[b]{0.45\textwidth}
         \centering
         \includegraphics[width=\textwidth]{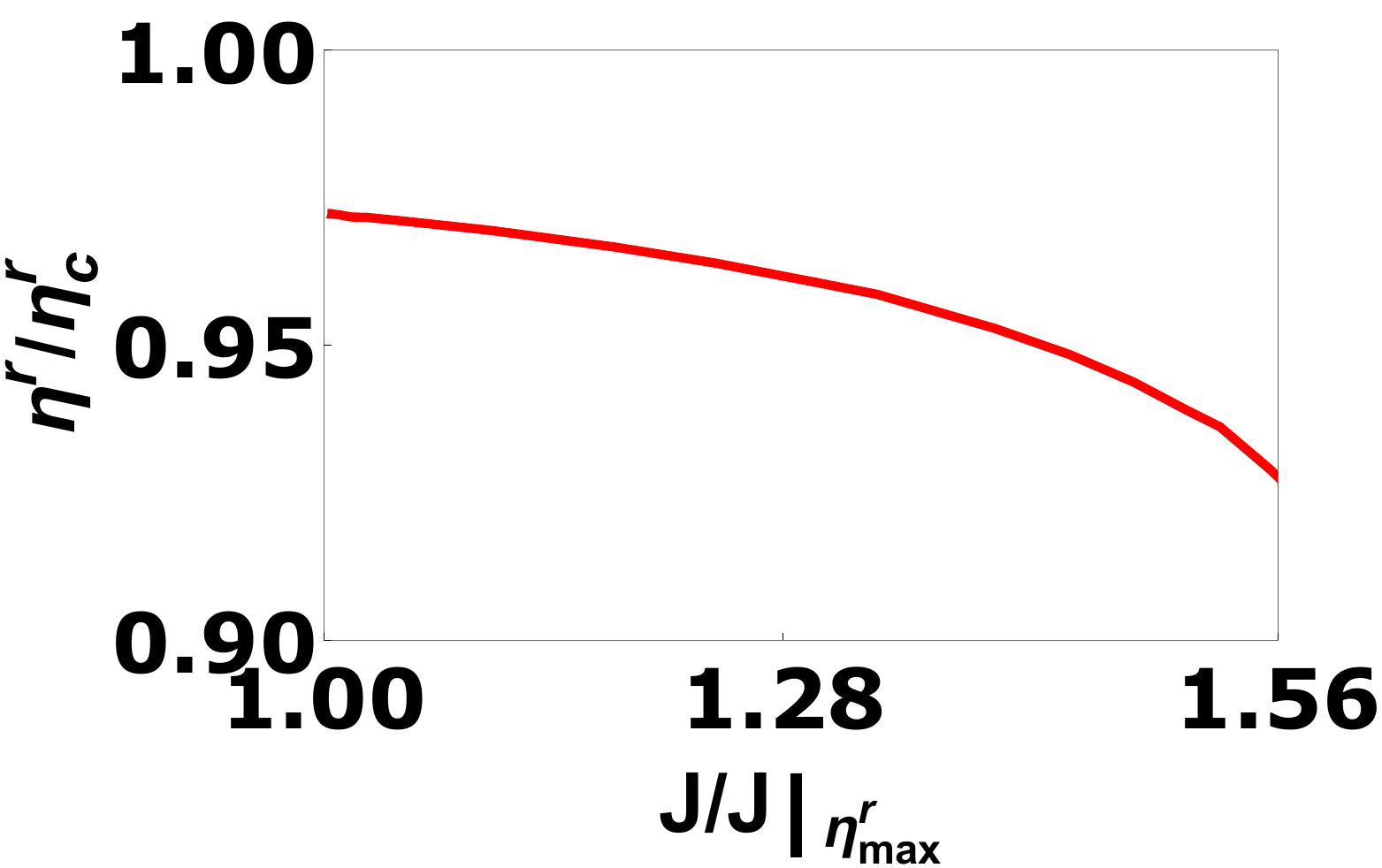}
         \caption{QH refrigerator (linear)}
     \end{subfigure}
        \caption{Parametric plot of $\frac{\eta}{\eta_c}$ vs $\frac{P}{P_{max}}$ (a) QSH QHE. Parametric plot of $\frac{\eta^r}{\eta^r_c}$ vs $\frac{\textbf{J}}{\textbf{J}|_{\eta^r_{max}}}$ for (b) QSH QR. Parameters are $eV_g = 83.8 k_B T$, $\omega = 0.1 k_B T / \hbar$, $T_1 = 1.01K, T_2 = 1.0K$, $\mu = 0$, $2E_1 = E_2 = 2k_B T$.}
         \label{fig:12}
       \end{figure}

\subsubsection{Cooling Power and coefficient of performance in 3T QSH setup as a quantum refrigerator}\label{Sec:III(A3)}

One can similarly derive the maximum coefficient of performance ($\eta^r_{max}$) and cooling power at maximum coefficient of performance ($J|_{\eta^r_{max}}$) for a QSH setup to act as a quantum refrigerator. Here, the heat will be absorbed from the cooler terminals (terminals 2 and 3 with voltage probe condition, and only terminal 2 with voltage-temperature probe) and dumped into the hotter terminal (terminal 1 in either voltage probe or voltage-temperature probe). For a quantum refrigerator, the coefficient of performance ($\eta^r$) is defined as the ratio of heat taken from the cooler terminal ($J^Q$) and the power absorbed ($P$) by the setup, i.e.,   $\eta^r = \frac{J^Q}{P}$. Here, $J^Q = -(J_2 + J_3)$ for voltage probe and -$J_2$ for voltage-temperature probe and can be derived from Eq. (\ref{eq:68}). Similarly, $P = I_1 V$ can also be derived from Eq. (\ref{eq:68}). Now, $\eta^r_{max}$ can be found using the condition $\frac{d \eta^r}{dV}$ = 0, which gives
\begin{equation}\label{eq:84}
    V|_{\eta^r_{max}} = \frac{L_{hT}}{L_{hV}}\left(1 + \sqrt{\frac{L_{eV} L_{h T} - L_{hV}L_{e T}}{L_{eV}L_{h T}}}\right)
\end{equation}

where, $V|_{\eta^r_{max}}$ is the voltage required to achieve $\eta^r_{max}$. From here onwards one can derive $\eta^r_{max}$ by using Eq. (\ref{eq:65}) in $\eta^r = \frac{J^Q}{P}$, which yields
\begin{equation}\label{eq:85}
    \eta_{max}^r = \eta_c^r \frac{\sqrt{Z T + 1} - 1}{\sqrt{Z T + 1} + 1} 
\end{equation}
where $\eta_c^r$ = Carnot COP = $\frac{T}{\Delta T} = \eta_c^{-1}$. Similarly, the cooling power at maximum coefficient of performance ($J|_{\eta^r_{max}}$) is given as,
\begin{equation}\label{eq:86}
    J|_{\eta^r_{max}} = L_{hV} V_{max} + L_{hT} \tau = L_{hT} \left(\sqrt{\frac{L_{eV} L_{h T} - L_{hV}L_{e T}}{L_{eV}L_{h T}}}\right)\tau.
\end{equation}

For our QSH setup, we have considered $E_1 = E_2 = k_B T$ and achieved the best possible results both as a quantum heat engine and refrigerator. Here, for the case, $E_1 \ne E_2$, i.e., if $2E_1 = E_2 = 2k_B T$, then results don't change, see Fig. \ref{fig:12}, which is identical to Fig. 2(b) and (d) of main text with $E_1 = E_2$.

\subsection{Transport in the Nonlinear response regime in the QSH setup}\label{Sec:III(B)}
The charge and heat current in a QSH setup is given as
\begin{equation} \label{eq:87}
    I_{\alpha}^{\sigma} = \frac{e}{h}\int_{-\infty}^{\infty}dE \sum_{\beta}f_{\beta}(E)[N_{\alpha}\delta_{\alpha \beta} - \mathcal{T}_{\alpha \beta}^{\sigma}],\quad J_{\alpha}^{\sigma} = \frac{1}{h}\int_{-\infty}^{\infty}dE (E - \mu_{\alpha}) \sum_{\beta} f_{\beta}(E)[N_{\alpha} \delta_{\alpha \beta} - \mathcal{T}_{\alpha \beta}^{\sigma}].
\end{equation}

Here too, the theory of nonlinear transport is similar as discussed in Sec. \ref{Sec:II(B)}. In the subsequent subsections, we follow the same pattern as QH, where we first discuss 2T nonlinear QSH setup as a heat engine and refrigerator using Whitney's approach in Sec. \ref{Sec:III(B1)}, then do the same thing with our approach in 2T setup in Sec. \ref{Sec:III(B2)} and in 3T setup without any probes in Sec. \ref{Sec:III(B3)}. Finally, we conclude this section by addressing the nonlinear transport property of QSH setup with both voltage and voltage-temperature probe in Sec. \ref{Sec:III(B4)}. 

\begin{figure} 
\centering
\includegraphics[width=0.60\linewidth]{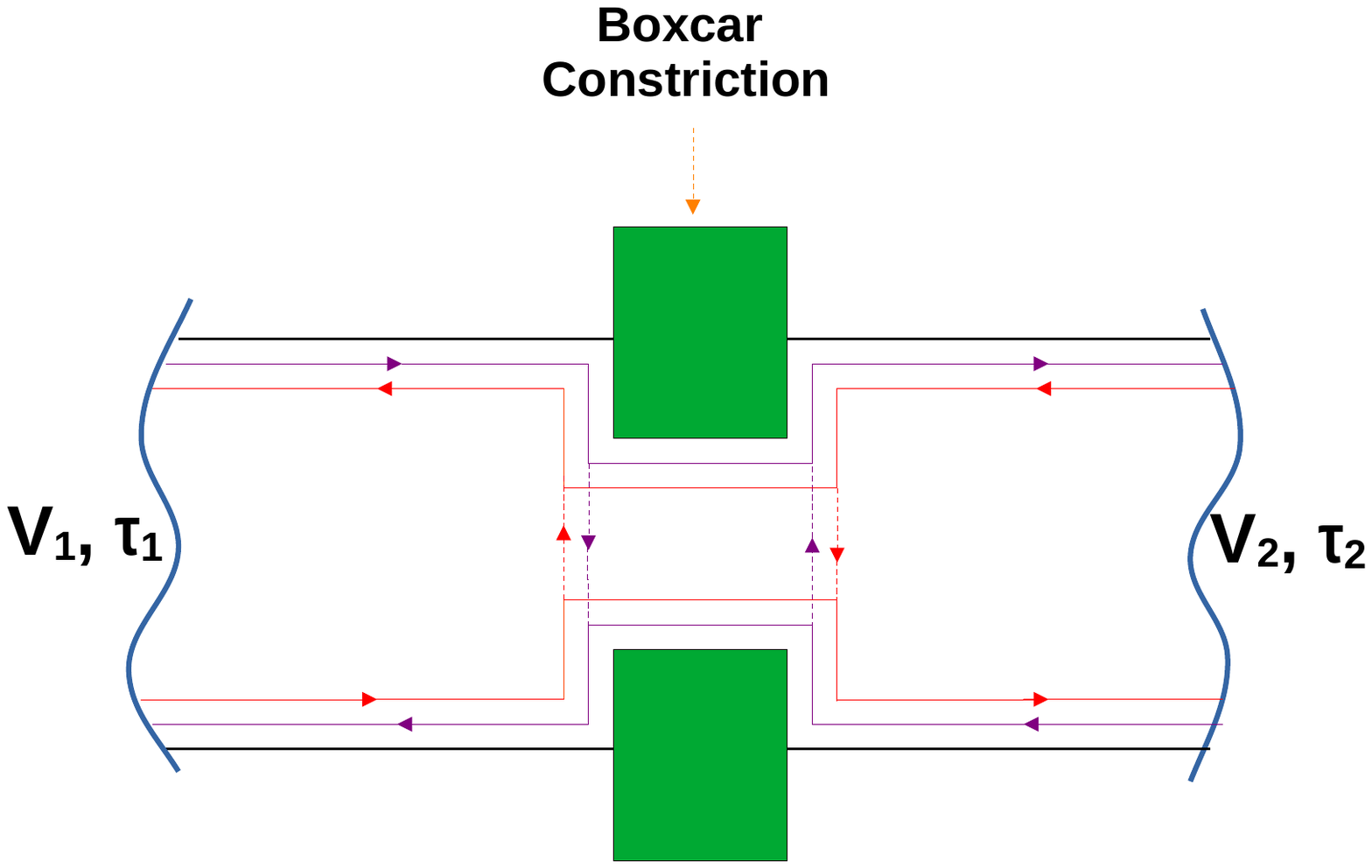}
\caption{(a) Two-terminal QSH setup with box-car type constriction (in nonlinear response regime), wherein the constriction has a boxcar-type transmission.}
\label{fig:13}
\end{figure}

\subsubsection{Nonlinear thermoelectrics in 2T QSH setup using Box-car type constriction}\label{Sec:III(B1)}
Here, we discuss the nonlinear thermoelectricity of a 2T QSH setup as shown in Fig. \ref{fig:13}(a) following the approach of Whitney \cite{PhysRevLett.112.130601, PhysRevB.91.115425} both as a quantum heat engine as well as a quantum refrigerator. This analysis is novel as Whitney only used his approach for a 2T QH setup. Similar to the QH case, we consider $V_1 = -V, V_2 = 0$, and $T_1$ and $T_2$ are the temperatures of terminals 1 and 2 respectively.
First, we discuss the quantum heat engine and later, we will discuss the quantum refrigerator.

\begin{figure} 
\centering
\includegraphics[width=0.50\linewidth]{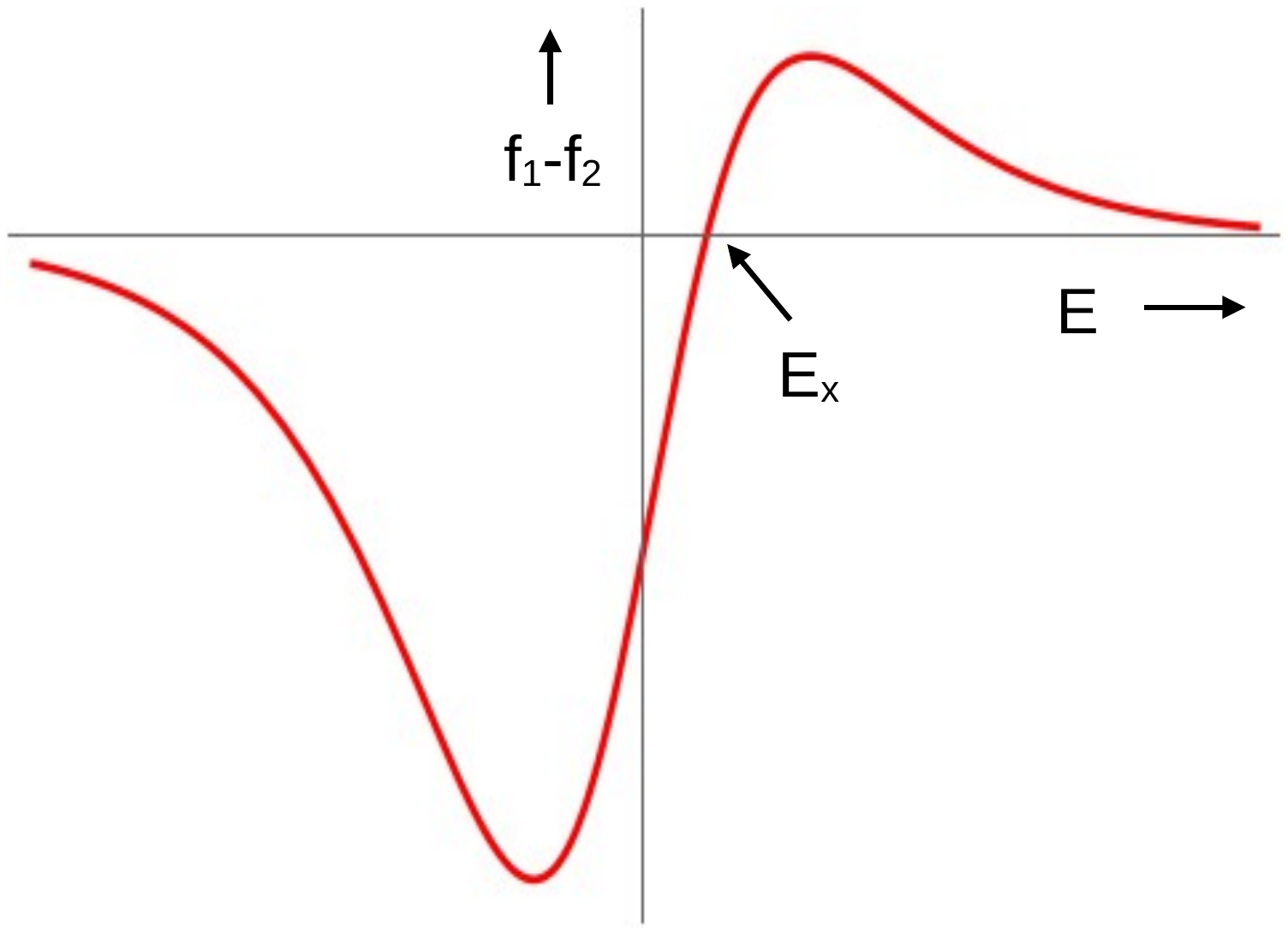}
\caption{Plot of $f_1 - f_2$ in the 2T QSH setup. }
\label{fig:14}
\end{figure}

Using the formula given in Eq. (\ref{eq:87}), the power generated ($P$) for a quantum heat engine is
\begin{equation}\label{eq:88}
    P = V(I_1^{\uparrow} + I_1^{\downarrow}) = \frac{eV}{h}\int_{-\infty}^{\infty}dE (\mathcal{T}_{12}^{\uparrow} + \mathcal{T}_{12}^{\downarrow})(f_1 - f_2),
\end{equation}
where, $\mathcal{T}_{12}^{\uparrow}$ is the transmission probability for a spin-up electron to scatter from terminal 2 to terminal 1 and $f_1$ and $f_2$ are the Fermi-Dirac distribution for terminals 1 and 2 respectively. Similarly, $\mathcal{T}_{12}^{\downarrow}$ is the transmission probability for the spin-down electron to scatter from terminal 2 to terminal 1. Similarly, the heat current out of terminal 1 using Eq. (\ref{eq:87}) is given as,
\begin{equation}\label{eq:89}
    J_1 = J_1^{\uparrow} + J_1^{\downarrow} = \frac{1}{h}\int_{-\infty}^{\infty}dE (E + eV)(\mathcal{T}_{12}^{\uparrow} + \mathcal{T}_{12}^{\downarrow})(f_1 - f_2) = \frac{1}{h}\int_{-\infty}^{\infty}dE (E )(\mathcal{T}_{12}^{\uparrow} + \mathcal{T}_{12}^{\downarrow})(f_1 - f_2) + P,
\end{equation}
wherein we have taken $\mu_1 = -eV$. Using Eqs. (\ref{eq:88}) and (\ref{eq:89}), one can evaluate the efficiency, i.e., $\eta = P/J_1$. For a heat engine, the power generated is always positive. From Eq. (\ref{eq:88}), one can see that the positive contribution to $P$ comes, when $f_1 - f_2 > 0$. Now, from Fig. \ref{fig:14}, one sees that $f_1 - f_2$ is positive above an energy value $E_x$, which is given as,
\begin{equation}\label{eq:90}
    E_x = \frac{eV}{\left(\frac{T_1}{T_2} - 1\right)},\quad \text{(Derived from the condition $f_1 = f_2$)}.
\end{equation}

As proposed by Whitney \cite{PhysRevLett.112.130601, PhysRevB.91.115425}, one can maximize the power generated by a heat engine by blocking the transmission
of the electrons with energy $E$ below $E_x$, which means only the electrons above energy $E_x$ will contribute to the
power $P$. Now, by comparing Eqs. (\ref{eq:74}) and (\ref{eq:75}), one can see that for higher energy values i.e., for $E > E_x$, the rate
of increase of $J_1$ is higher than that of $P$ as $J_1$ is always more than $P$, See Eq. (\ref{eq:89}). This means that high-energy electrons contribute to power ($P$) less efficiently than it does for the heat current ($J_1$). So, one can guess to have a
transmission probability in the energy range $E_x < E < E_{x}^{'}$, where $E_{x}^{'}$ is the voltage-dependent energy just like $E_x$,
above which the transmission is zero.
Here the analysis is exactly similar to the QH case as in Sec. \ref{Sec:II(B1)} as the helical edge modes is a composition of two chiral edge modes each for spin-up and spin-down electrons, then both $\mathcal{T}_{12}^{\uparrow}$ and $\mathcal{T}_{12}^{\downarrow}$ are boxcar type as shown in Fig. in some energy range $E_x < E < E_{x}^{'}$ to have the best possible performance as a quantum heat engine, where $E_{x}^{'} = eV J_1^{'}/P^{'}$. The expression for $E_{x}^{'}$ is derived in \cite{PhysRevB.91.115425}. Considering single edge mode for spin-up electron, $\mathcal{T}_{12}^{\uparrow}$ will be 1 for $E_x < E < E_{x}^{'}$, otherwise zero. Similarly, in the same energy range $E_X < E < E_{x}^{'}$, $\mathcal{T}_{12}^{\downarrow}$ will be 1, otherwise zero. At some voltage bias, when $E_{x}^{'} \rightarrow \infty$, then both $\mathcal{T}_{12}^{\uparrow}$ as well as $\mathcal{T}_{12}^{\downarrow}$ will become QPC-like transmission as shown in Fig. \ref{fig:1}. Here, $E_{x}^{'} = eV J_1^{'}/P^{'}$ \cite{PhysRevLett.112.130601, PhysRevB.91.115425}, where the prime in $J_1$ and $P$ denotes the first derivatives of $J_1$ and $P$ with respect to the applied voltage bias $V$. Now, the maximum possible power ($P_{max}$) is generated, when $P^{'} = 0$, which makes $E_{x}^{'} = \infty$, and consequently one can derive the efficiency at the maximum power ($\eta|_{P_{max}}$). The maximum power is given as
\begin{equation}\label{eq:91}
    P_{max} = \frac{2e}{h}\int_{E_x}^{\infty}dE (f_1 - f_2),
\end{equation}
and the heat current from the terminal 1, when $P^{'} = 0$ is
\begin{equation}\label{eq:92}
    J|_{P_{max}} = \frac{2e}{h}\int_{E_x}^{\infty}dE (f_1 - f_2).
\end{equation}
One can derive the efficiency at maximum power by using the formula $\eta|_{P_{max}} = P_{max}/J|_{P_{max}}$. Following the Ref. \cite{PhysRevB.91.115425}, one can similarly derive the expressions for the maximum power ($P_{max}$) and efficiency at maximum power ($\eta|_{P_{max}}$) and they are given as
\begin{equation}\label{eq:93}
    P_{max} = P_{max}^{Wh} = 0.0642 \frac{\pi^2 k_B^2 \tau^2}{h},\quad 
    \eta|_{P_{max}} = \eta|_{P_{max}^{Wh}} = \frac{\eta_c}{1+ 0.936(1 + T_2/T_1)}.
\end{equation}
Similarly, when $E_{x}$ and $E_{x}^{'}$ are close to each other the finite power generated will be low, and then the expression for maximum efficiency $\eta_{max}$ can be derived for generic values of $E_x$ and $E_{x}^{'}$. Now, following Whitney's result \cite{PhysRevLett.112.130601, PhysRevB.91.115425}, $\eta_{max}$ is given as
\begin{align}\label{eq:94}
\begin{split}
    \eta_{max} = \eta_{max}^{Wh} = \eta_c \left(1-0.478\sqrt{\frac{T_2 P}{T_1 P_{max}^{Wh}}}\right),
    \end{split}
\end{align}
where, $\tau = T_1 - T_2$ = temperature bias applied across the sample, $\eta_c$ = Carnot efficiency, $T_1$ = temperature in terminal 1, $T_2$ = temperature in terminal 2, $P$ = Finite output power with $P \ll P_{max}^{Wh}$.

Similarly, for a quantum refrigerator, the cooling power ($J = J_2$) of this 2T QSH setup is given as
\begin{equation}\label{eq:95}
    J^Q = J_2 = (J_2^{\uparrow} + J_2^{\downarrow}) =\frac{1}{h}\int_{-\infty}^{\infty}dE E (\mathcal{T}_{21}^{\uparrow} + \mathcal{T}_{21}^{\downarrow})(f_2 - f_1). 
\end{equation}
From Eq. (\ref{eq:95}), the positive contribution to the cooling power $J$ comes when $(f_2 - f_1)>0$, means when $f_1 - f_2 < 0$. Now, from Fig. \ref{fig:14}, it is clear that, $f_1 - f_2$ is negative below the energy value $E_x$, which is given in Eq. (\ref{eq:90}). According to the argument of Whitney \cite{PhysRevLett.112.130601, PhysRevB.91.115425}, one needs to maximize the cooling power by allowing the transmission of electrons only below energy $E_x$
and blocking above them. Similarly, one can also look into maximizing the coefficient of performance $\eta^r = J/P_1$, where $P = I_1 V$ is the power delivered on the system, which is given in Eq. (\ref{eq:88}). Now, one can guess a transmission function in the energy range $E_{x2} < E < E_{x}$, where $E_{x2}$ is also a voltage-dependent energy, above which the transmission is zero. From now onwards, the analysis is similar to the QH case as in Sec. \ref{Sec:II(B2)} as the helical edge mode is a composition of two chiral edge modes each for spin-up and spin-down electrons. Then, $\mathcal{T}_{21}^{\uparrow}$ and $\mathcal{T}_{21}^{\downarrow}$ are boxcar-type transmission, which is 1 in the energy range $E_{x2}<E<E_{x}$, where $E_{x2} = eV J_2^{'}/P^{'}$, as derived by Whitney \cite{PhysRevB.91.115425}. So, the cooling power $J$ is given as
\begin{equation}\label{eq:96}
    J = J_2 = \frac{2}{h}\int_{E_{x2}}^{E_x}dE E (f_2 - f_1), 
\end{equation}
where the cooling power is maximum when $J_2^{'} = 0$, which makes the lower limit of integration $E_{x2} = 0$. Now, to have a large cooling power, the upper limit of the integration $E_x$ needs to be large, which means $E_x \rightarrow \infty$. Now, from Eq. (\ref{eq:95}), it is clear that $E_x \rightarrow \infty$, when $V \rightarrow \infty$, which will make $f_1 \rightarrow 0$. Then the maximum cooling power ($J_{max}^{Wh}$) following Whitney's approach is given as
\begin{equation}\label{eq:97}
    J_{max}^{Wh} = \frac{2}{h}\int_{0}^{\infty}dE E (f_2) = \frac{\pi^2}{6h}k_B^2T_2^2 
\end{equation}

Similarly, the power delivered on the system ($P = I_1 V$) is large, when the maximum cooling power ($J_{max}^{Wh}$) is reached. So, the coefficient of performance at $J_{max}^{Wh}$ is $J_{max}^{Wh}/P$ is zero as $P$ is very large. Similarly, one achieves the maximum coefficient of performance when the finite cooling power is much smaller than $J_{max}^{Wh}$, which can be achieved when $E_x$ and $E_{x2}$ are close. Then one can calculate the cooling power and power delivered at generic values of $E_x$ and $E_{x2}$, which are quite close as done by Whitney \cite{PhysRevLett.112.130601, PhysRevB.91.115425}. In that case, the coefficient of performance is large, which is given as 
\begin{equation}\label{eq:98}
    \eta_{max}^r = \eta_{max}^{r, Wh} =  \eta^r_c\left(1-1.09 \sqrt{\frac{T_2}{T_1 - T_2}\frac{J}{J_{max}^{Wh}}}\right).
\end{equation}
where $J \ll J_{max}^{Wh}$. 

One can conclude that the best possible performance of a 2T QSH setup is exactly similar to that of 2T QH setup as done by Whitney.

\begin{figure} 
\centering
\includegraphics[width=0.60\linewidth]{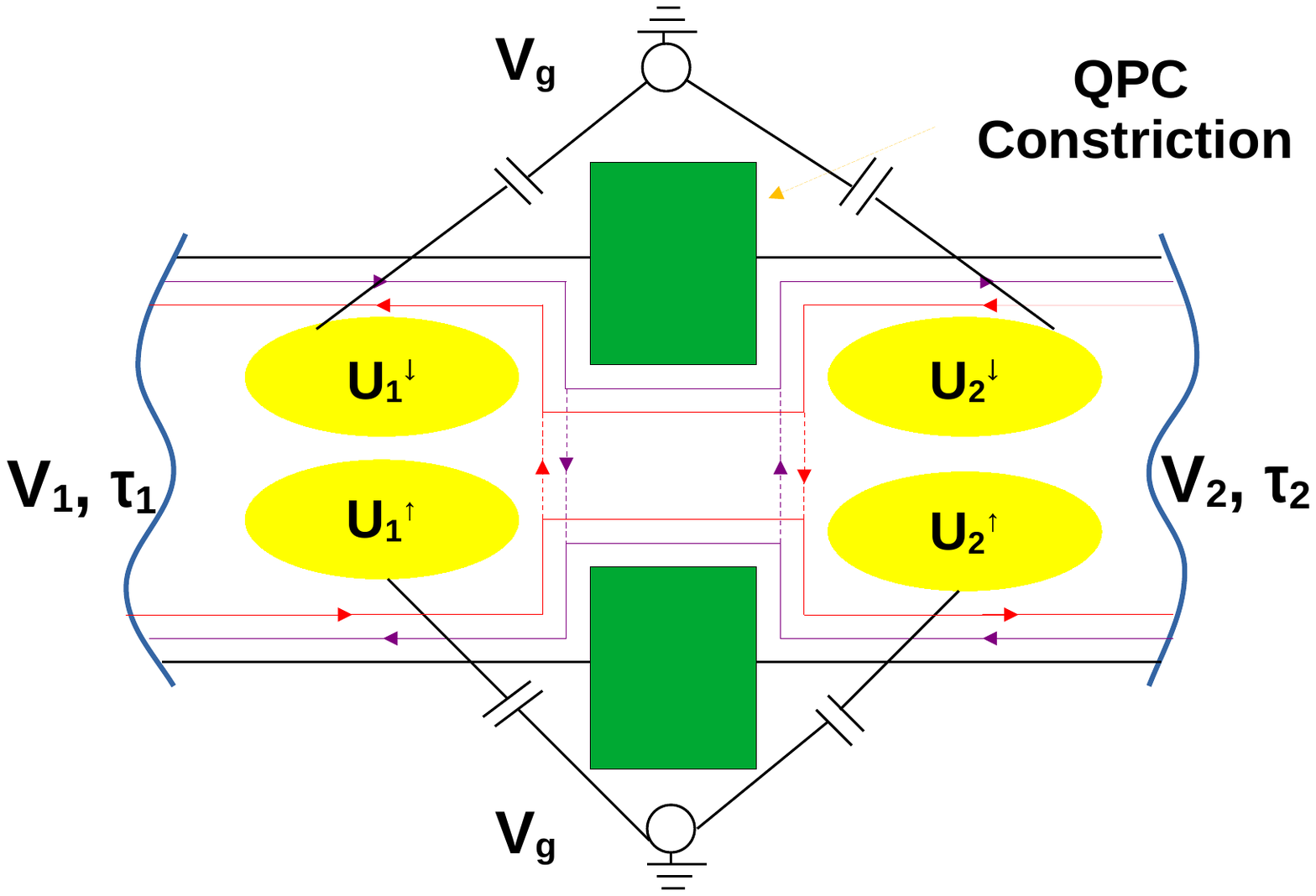}
\caption{{Two-terminal QSH setup partitioned into four regions defined by interaction potential $U_r^{\sigma}$ with $r \in \{ 1, 2 \}$ and $\sigma \in \{\uparrow, \downarrow \}$ (the yellow shaded regions) to the left and right of the QPC-type constriction. Each region is connected capacitively to a common external gate voltage $V_g$.}}
\label{fig:15}
\end{figure}

\subsubsection{Nonlinear thermoelectrics in 2T QSH setup using QPC type constriction}\label{Sec:III(B2)}

Here, we discuss nonlinear thermoelectrics for a 2T QSH setup by considering a QPC-like constriction in between, See Fig. \ref{fig:15}, where we have applied a gate voltage ($V_g$) to the constriction. Here, the gate voltage controls the energy level of QPC i.e., $E_1^{'} = E_1 + e U$, where $U$ is the interaction potential. For this case, we consider $V_1 = -V, V_2 = 0, \tau_1 = \tau$ and $\tau_2 = 0$. Here, also we follow the same argument used for the 2T QH setup as in Sec. \ref{Sec:II(B2)}. {For this, we consider 4 regions with spin-polarized interaction potentials $U_1^{\uparrow}$, $U_1^{\downarrow}$, $U_2^{\uparrow}$ and $U_2^{\downarrow}$ following Ref. \cite{PhysRevB.90.115301}, see Fig. \ref{fig:15}. Again, we find $U_r^{\sigma}$ in a particular region $r$ for spin $\sigma$ assuming other potentials $U_{r'}^{\sigma'} = 0$ for $r' \ne r$ and $\sigma' \ne \sigma$. We extend the theory discussed in Sec. \ref{Sec:II(B2)} to helical edge modes. The additional injected charge with spin $\sigma \in \{\uparrow, \downarrow \}$ into regions $r \in \{1,2 \}$ is \cite{PhysRevLett.110.026804, PhysRevB.88.045129, Meair_2013, PhysRevB.90.115301},}
\begin{equation}\label{eq:99}
\begin{split}
    {q_{ad}^{r\sigma}} & {= e \int dE \left ( \sum_{\alpha} \nu_{r \alpha}^{p\sigma}(E, U_r^{\sigma}) f_{\alpha}(E) \right) - e \int dE  \nu_r^{p\sigma}(E) f}\\
    & {= e \int dE \left (  \nu_{r1}^{p\sigma}(E, U_r^{\sigma}) f_{1}(E) + \nu_{r2}^p(E, U_r^{\sigma}) f_{2}(E)\right) - e \int dE  \nu_r^{p\sigma}(E) f.}
    \end{split}
\end{equation}
{We capacitively connect region $r$ with spin $\sigma$ to an external gate voltage $V_g$ giving $q_{ad}^{r \sigma}$ = $C (U_r^{\sigma}   - V_g)$. Equating it with Eq. (\ref{eq:99}), one can numerically determine $U_r^{\sigma}$.} {The average interaction potential the electron experiences is thus}
\begin{equation}\label{eq:100}
    {U = \sum_{\sigma \in \{\uparrow, \downarrow\}} U^{\sigma} = \sum_{r \in \{1, 2\}, \sigma \in \{\uparrow, \downarrow \}} \frac{U_r^{\sigma}}{2}} 
\end{equation}

{First, we discuss the evaluation of $U_1^{\uparrow}$ in an analogous way to the two-terminal QH case as done in Sec. II B 2. The additional injected charge with up spin into region 1 is}
\begin{equation}\label{eq:102}
\begin{split}
    {q_{ad}^{1\uparrow}} & {= e \int dE \left ( \sum_{\alpha} \nu_{1 \alpha}^{p\uparrow}(E, U_1^{\uparrow}) f_{\alpha}(E) \right) - e \int dE  \nu_1^{p\uparrow}(E) f}, \\
    &{= e \int dE \left (  \nu_{11}^{p\uparrow}(E, U_1^{\uparrow}) f_{1}(E) + \nu_{12}^p(E, U_1^{\uparrow}) f_{2}(E)\right) - e \int dE  \nu_1^{p\uparrow}(E) f.}
    \end{split}
\end{equation}

{The injectivity $\nu_{11}^{p\uparrow}$ can be derived from the partial density of states $\nu_{ \beta 1 1}^{p\uparrow}$ for up spin electrons scattered from contact 1 to contact $\beta$ for $\beta \in \{1,2 \}$. Thus, $\nu_{\beta 1 1}^{p \uparrow}$ is $\mathcal{T}_{\beta 1 }^{\uparrow}$  $\nu_1^p$, where $\mathcal{T}_{\beta 1}^{\uparrow}$ is the probability for an up spin electron to scatter from terminal 1 to terminal $\beta$. Similarly, $\nu_{12}^{p \uparrow} = \nu_{112}^{p \uparrow} + \nu_{212}^{p \uparrow}$, where $\nu_{\beta 1 2}^{p\uparrow}$ is the partial density of states for up spin electrons scattered from contact 2 to contact $\beta$, for $\beta \in \{1,2 \}$. The partial density of states $\nu_{1 1 1}^{p \uparrow} = (1-\mathcal{T}_{QPC})\nu_1^{p\uparrow} $ and $\nu_{2 1 1}^{p \uparrow} = \frac{1}{2}\mathcal{T}_{QPC} \nu_1^{p \uparrow}$, therefore $\nu_{11}^{p \uparrow} = \left(1 - \frac{\mathcal{T}_{QPC}}{2}\right) \nu_1^{p \uparrow}$, where $\mathcal{T}_{QPC}$ is function of energy and $U_1^{\uparrow}$, and is given as, }

\begin{equation}\label{eq:103}
    {\mathcal{T}_{QPC} = \frac{1}{1 + e^{-\frac{E - E_1 - e U_1^{\uparrow}}{\hbar \omega}}}},
\end{equation}
{where $E_1$ = $V_0 + \hbar \omega_y \left(n_y + 1/2\right)$ as shown below Eq. (\ref{eq8}). Since there are zero states in region 1 associated with the reflection of up spin electrons from terminal 2 to itself, see Fig. \ref{fig:15}, therefore, $\nu_{212}^{p\uparrow} $ vanishes even if $\mathcal{T}_{22}^{\uparrow} \ne 0$ and $\nu_1^{p \uparrow} \ne 0$ and only $\nu_{112}^{p \uparrow}$ continues with $\nu_{12}^{p \uparrow} = \frac{1}{2}\mathcal{T}_{QPC} \nu_1^{p\uparrow}$. The prefactor of $\frac{1}{2}$ in front of $\mathcal{T}_{QPC}$ in both $\nu_{211}^{p\uparrow}$ and $\nu_{112}^{p\uparrow}$ has been taken in order to ensure $\nu_{11}^{p\uparrow } + \nu_{12}^{p\uparrow} = \nu_1^{p\uparrow}$ \cite{christen1996gauge, PhysRevLett.77.143}. Therefore, the additional up spin-polarized injected charge into region 1 is,}
\begin{equation}\label{eq:104}
    {q^{1 \uparrow}_{ad}= e \int dE \left(\left(1 - \frac{\mathcal{T}_{QPC}}{2}\right)\nu_1^{p \uparrow}(E, U_1^{\uparrow}) f_1 + \frac{\mathcal{T}_{QPC}}{2}\nu_1^{p\uparrow}(E, U_1^{\uparrow}) f_2\right) - e \int dE \, \, \,  \nu_1^{p \uparrow}(E) f}
\end{equation}

{The partial density of states for a QPC in 2DEG a $\nu_r^{p \sigma} (E, U_1)$ for any region $r \in \{1,2 \}$ can be calculated as in Ref. \cite{PhysRevB.57.1838}, } 
\begin{align}\label{eq:105}
    \begin{split}
       {\nu_1^{p \uparrow}(E, U)} & {= \frac{4}{ 2\pi  \hbar \omega} \sinh^{-1} \sqrt{\left(\frac{1}{2} \frac{m \omega^2 \lambda^2}{E  - E_1 - eU_1}\right)}} , \quad {\text{for}} \quad {E > E_1 + eU_1,} \\
            & {= \frac{4}{ 2\pi  \hbar \omega} \cosh^{-1} \sqrt{\left(\frac{1}{2} \frac{m \omega^2 \lambda^2}{E_1 + eU_1 - E}\right)} ,} \quad {\text{for}} \quad {E_1 + eU_1 - \frac{1}{2} m \omega^2 \lambda^2 < E < E_1.}
    \end{split}
\end{align}
{where, $m$ and $\omega$ are defined in Eq. (\ref{eq7}) and $\lambda$ is the screening length defined in Ref. \cite{PhysRevB.57.1838}. One can capacitively connect the conductor region 1 of Fig. \ref{fig:15} to an external gate voltage $V_g$ and the additional injected charge is,}
\begin{equation}\label{eq:106}
    {q_{ad}^{1 \uparrow} = C(U_1^{\uparrow} - V_g).}
\end{equation}
{Utilizing Eqs. (\ref{eq:104}-\ref{eq:106}), one can numerically estimate $U_1^{\uparrow}$, which is a function of $V_{1}, V_{2}, \tau_{1}, \tau_{2}$ and $V_g$. As shown in Refs. \cite{PhysRevLett.110.026804, PhysRevB.88.045129, Meair_2013}, the interaction potential $U_1^{\uparrow} (V_1, V_2, \tau_1, \tau_2, V_g)$ can also be written as}
\begin{equation}\label{eq:107}
    {U_1^{\uparrow}(V_1, V_2, \tau_1, \tau_2, V_g) = U_1^{\uparrow \prime}(V_1, V_2, \tau_1, \tau_2) + u_{g1}^{\uparrow}V_g,}
\end{equation}
{where $U_1^{\uparrow \prime}(V_1, V_2, \tau_1, \tau_2)$ is the interaction potential induced solely by the voltage bias ($V_1, V_2$) and temperature bias ($\tau_1, \tau_2$) at zero gate voltage, i.e., $U_1^{\uparrow}(V_g = 0) = U_1^{\uparrow \prime}$ and $u_{g1}^{\uparrow} = \frac{ U_1^{\uparrow} - U_1^{\uparrow \prime}}{V_g}$ determines the response of $U_1^{\uparrow}$ to an applied gate voltage ($V_g$). For our calculation, we consider $V_1 = -V, V_2  = 0, \tau_1 = \tau $ and $\tau_2 = 0$ and first calculate $U_1^{\uparrow \prime}$ from Eqs. (106-106) at zero gate voltage $V_g = 0$ when $C = 0.01 F, \hbar \omega = 0.1 k_B T$ and $\frac{1}{2}m \omega^2 \lambda^2 = 20 k_B T$ with voltage bias $V = 1.14 k_B T/e$ and $\tau = 1K$. At $V_g = 0$, we get $U_1^{\uparrow \prime} = 10.88229 k_B T/e$. We then calculate $U_1^{\uparrow}$ at an arbitrary gate voltage, say $V_g = k_B T/e$, which gives $U_1^{\uparrow} = 11.04889 k_B T/e$, which implies $u_{g1}^{\uparrow} = \frac{U_1^{\uparrow} - U_{1}^{\uparrow \prime}}{V_g} = 0.1666$. Similarly, one can also calculate other interaction potentials of other regions, i.e., $U_1^{\downarrow}, U_2^{\uparrow}$ and $U_2^{\downarrow}$ using Eqs. (\ref{eq:99}) and capacitively connecting each region with an external gate voltage $V_g$ in the same parameter regime as in $U_1^{\uparrow}$.}
{The partial density of states or particle injectivities for each region are}
\begin{equation}\label{eq:108}
\begin{split}
    {\nu_{11}^{p \downarrow}} & {= \sum_{\beta \in \{1, 2, 3 \}} \nu_{\beta 11}^{p \downarrow} = \left(1 - \frac{\mathcal{T}_{QPC}}{2} \right)\nu_{1}^{p \downarrow}, \quad \nu_{12}^{p \downarrow} = \sum_{\beta \in \{1, 2, 3 \}} \nu_{\beta 12}^{p \downarrow} = \frac{\mathcal{T}_{QPC}}{2}\nu_{1}^{p \downarrow}, \quad \nu_{21}^{p \uparrow} = \sum_{\beta \in \{1, 2, 3 \}} \nu_{\beta 21}^{p \uparrow} = \frac{\mathcal{T}_{QPC}}{2}\nu_{1}^{p \uparrow},}\\
     {\nu_{22}^{p \uparrow}} & {= \sum_{\beta \in \{1, 2, 3 \}} \nu_{\beta 22}^{p \uparrow} = \left(1 - \frac{\mathcal{T}_{QPC}}{2}\right)\nu_{2}^{p \uparrow}, \quad}
    {\nu_{21}^{p \downarrow}} {= \sum_{\beta \in \{1, 2, 3 \}} \nu_{\beta 21}^{p \downarrow} = \frac{\mathcal{T}_{QPC}}{2}\nu_{1}^{p \downarrow}, \quad \nu_{22}^{p \downarrow} = \sum_{\beta \in \{1, 2, 3 \}} \nu_{\beta 22}^{p \downarrow} = \left(1 - \frac{\mathcal{T}_{QPC}}{2}\right)\nu_{2}^{p \downarrow}}
\end{split}
\end{equation}

\begin{table}[]
\renewcommand{\arraystretch}{1.5} 
\centering
\caption{Numerical estimation of interaction potential ($U'$) and the applied gate voltage ($V_g$) in 2T QSH setup. Parameters taken are $\tau = 1K, C = 0.01 F, \hbar \omega = 0.1 k_B T$ and $\frac{1}{2}m \omega^2 \lambda^2 = 20 k_B T$.}
\begin{tabular}{|c|c|c|}
\hline
Applied voltage bias ($k_B T/e$) & Interaction potential ($k_B T/e$) & Applied gate voltage ($k_B T/e$)\\ \hline
1.14 & 16.5314 & $-36.1239 + \frac{V_g'}{0.45763}$ \\ \hline
1.77 & 15.18794 & $-30.76910 + \frac{V_g'}{0.49361}$ \\ \hline
3.7 & 7.71156 & $73.17859 - \frac{V_g'}{0.10538}$ \\ \hline
20 & 13.01361 & $-764.15795 + \frac{V_g'}{0.01703}$ \\ \hline
\end{tabular}
\label{Table3}
\end{table}

{The numerical calculation for $U_1^{\downarrow}, U_2^{\uparrow}$ and $U_2^{\downarrow}$ are done in Python \cite{My_mathematica_notebook}. Just like Eq. (\ref{eq:107}), the interaction potential in other regions can be written as}
\begin{equation}\label{eq:109}
\begin{split}
    {U_1^{\downarrow}(V_1, V_2, \tau_1, \tau_2, V_g)} & {= U_1^{\downarrow \prime}(V_1, V_2, \tau_1, \tau_2) + u_{g1}^{\downarrow}V_g,}\\
    {U_2^{\uparrow}(V_1, V_2, \tau_1, \tau_2, V_g)} & {= U_2^{\uparrow \prime}(V_1, V_2, \tau_1, \tau_2) + u_{g1}^{\uparrow}V_g,}\\
    {U_2^{\downarrow}(V_1, V_2, \tau_1, \tau_2, V_g)} & {= U_2^{\downarrow \prime}(V_1, V_2, \tau_1, \tau_2) + u_{g2}^{\downarrow}V_g.}
    \end{split}
\end{equation}

\begin{figure} 
\centering
\includegraphics[width=1.00\linewidth]{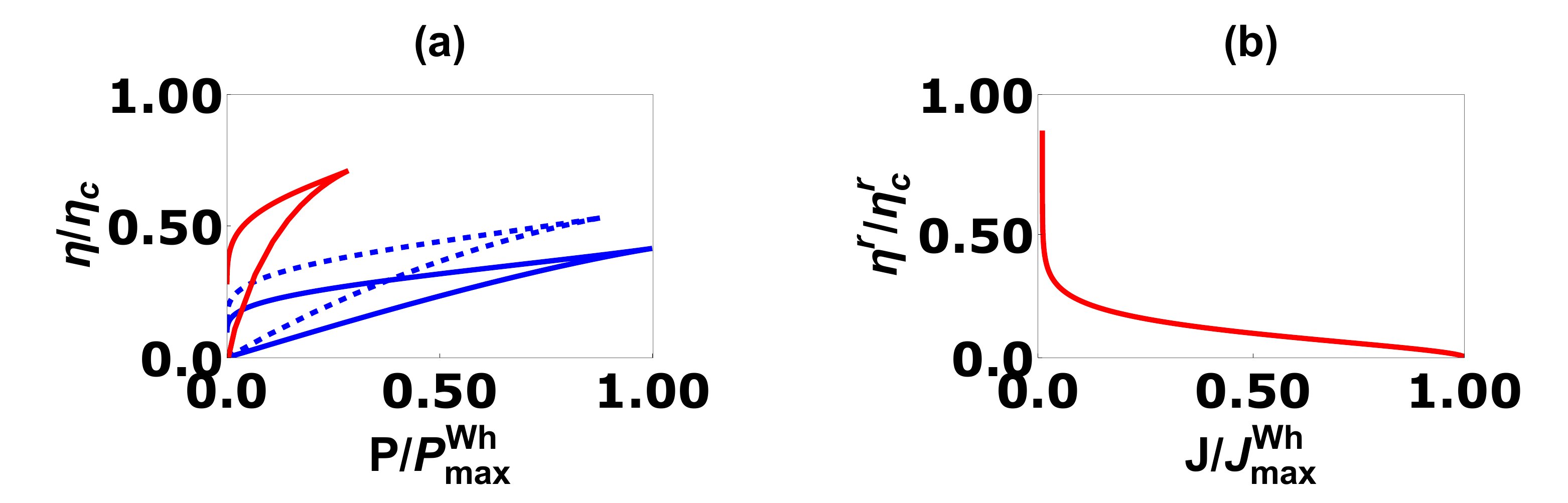}
\caption{(a) Parametric plot of the power ($P$) and efficiency as a heat engine. The blue thick line corresponds to $V = 0.57 k_B T/e$, blue dotted line corresponds to $V = 1.77 k_B T/e$, red thick line corresponds to $V = 2.52 k_B T/e$, (b) Parametric plot of the cooling power ($J$) and coefficient of performance ($\eta^r$) as a refrigerator of a two-terminal QSH (helical) setup at $V = 20 k_B T/e$. Here, we have taken $E_{QPC}' = k_B T$, where $T_1 = 2T_2 = 2K$ and $\tau = T_1 - T_2  = 1K$. For the QPC, we have taken $\omega = 0.1k_B T/e$. These results are similar to the three-terminal QH setup without any probe with $V_1 = -V, V_2 = V_3 = 0$ and $\tau_1 = \tau, \tau_2 = \tau_3 = 0$. }
\label{fig:16}
\end{figure}

{The total interaction potential at arbitrary $V_g$ is given as,}
\begin{equation}\label{eq:110}
    {U = \frac{U_1^{\uparrow} + U_1^{\downarrow} + U_2^{\uparrow} + U_2^{\downarrow}}{2} = U' + u_g V_g,\quad \text{where} \quad U' = \frac{U_1^{\uparrow \prime} + U_1^{\downarrow \prime} + U_2^{\uparrow \prime} + U_2^{\downarrow \prime}}{2}, \quad u_g = \frac{u_{g1}^{\uparrow} + u_{g1}^{\downarrow} + u_{g2}^{\uparrow} + u_{g2}^{\downarrow}}{2}}
\end{equation}
{We numerically calculate the interaction potentials with zero gate voltage ($U_1^{\downarrow \prime}, U_2^{\uparrow \prime}, U_2^{\uparrow \prime}$) and with finite gate voltage ($U_1^{\downarrow}, U_2^{\uparrow}, U_2^{\uparrow}$) at $V_g$. We thus get $U_1^{\downarrow \prime} = 10.88229 k_B T/e, U_2^{\uparrow \prime} = 5.358072 k_B T/e, U_2^{\downarrow \prime} = 5.358072 k_B T/e$ and $u_{g1}^{\downarrow} = 0.1666, u_{g2}^{\uparrow} = 0.291028, u_{g2}^{\downarrow} = 0.291028$ and the interaction potential without gate voltage ($U'$) is $16.5314 k_B T/e$ and $u_g = 0.45763$. One can apply a gate voltage $V_g = \frac{-16.5314}{0.45763} + \frac{V_g'}{0.45763} = -36.1239 + \frac{V_g'}{0.45763}$, where the first terms nullify $U'$ and the second term lifts the threshold energy value of QPC. Thus, the threshold energy of QPC is $E_1 + e V_g'$. One can also estimate the interaction potential at different voltage biases such as $V = 1.77 k_B T/e, V = 3.7 k_B T/e$ and $V = 20 k_B T/e$ and the applied gate voltage, and we mention their values in Table \ref{Table3}. The voltage biases $V = 1.14k_B T/e$ (blue solid curve in Fig. \ref{fig:16}(a)), $V = 1.77k_B T/e$ (blue dashed curve in Fig. \ref{fig:16}(a)), $V = 3.7k_B T/e$ (red curve in Fig. \ref{fig:16}(a)) are used to study the performance of the 2T QSH setup as a QHE. Similarly, at $V = 20 k_B T/e$ is used to study the setups performance as a QR.}
With this, the power delivered ($P = I_1 V$) and efficiency ($\eta = P / J_1$) can be found using the formulae given in Eqs. (\ref{eq:88}) and (\ref{eq:89}). In this work, we fix the values of $V$ and $\tau$ and evaluate the power and efficiency with different values of $V_g'$. We plot them parametrically in Fig. \ref{fig:16}(a). We fix $V = 1.14 k_B T/e$, $\tau = 1K$, we find that at a particular $V_g$, the maximum power approaches $P_{max}^{Wh}$ as shown in Fig. \ref{fig:16}(a) (solid blue line). 

Similarly, for the refrigerator, we consider $V_g = -U + V_g'$. The cooling power $ J = -J_2$ and coefficient of performance ($\eta^r = J/P$), where $P = I_1 V$ for different values of $V$. The parametric plot of the cooling power $J$ and $\eta^r$ is shown in Fig. \ref{fig:16}(b). The maximum cooling power approaches $J_{max}^{Wh}$ and the best coefficient of performance reach is 87$\%$ of $\eta_c^r$. 

\subsubsection{Nonlinear thermoelectrics in 3T QSH setup using QPC type constriction}\label{Sec:III(B3)}

The power and efficiency of a simple three-terminal QSH setup without any voltage-temperature probes with two QPC-like constrictions as shown in Fig. \ref{fig:11} controlling the transmission of QPC 1 and QPC 2 by an applied gate voltage using the conditions: $E_1^{'} = E_1 + e V_g'$ and $E_2^{'} = E_2 + e V_{g}'$, with $E_1 = E_2$. We consider a simple case with $V_1 = -V, V_2 = V_3 =0$, and $\tau_1 = \tau, \tau_2 = \tau_3 = 0$. We enable the 3T QSH setup to work as QHE and QR. Similar to a 2T QSH case as in Sec. \ref{Sec:III(B2)}, the interaction potential with and without gate voltage can be calculated. Then, the interaction potential can be nullified by considering the gate voltage $V_g$ of the form $\frac{-U'}{u_g} + \frac{V_g'}{u_g}$. For QHE, the power delivered ($P = I_1 V$) and efficiency ($\eta = P/J_1$) can be found by using the formula given in Eqs. (\ref{eq:74}) for different values of voltage bias ($V$), the performance exactly matches with the result of two-terminal heat engine as shown in Fig. \ref{fig:16}(a). Similarly, for the refrigerator, we take $V_g = \frac{-U'}{u_g} + \frac{V_g'}{u_g}$ and cooling power ($J = -(J_2 + J_3)$) and coefficient of performance ($\eta^r = J/P$) exactly matches the result of the two-terminal refrigerator as in Fig. \ref{fig:16}(b).
 
\subsubsection{Nonlinear thermoelectrics in 3T QSH setup with a voltage-temperature probe (Our approach)}\label{Sec:III(B4)}
\begin{figure} 
\centering
\includegraphics[width=0.60\linewidth]{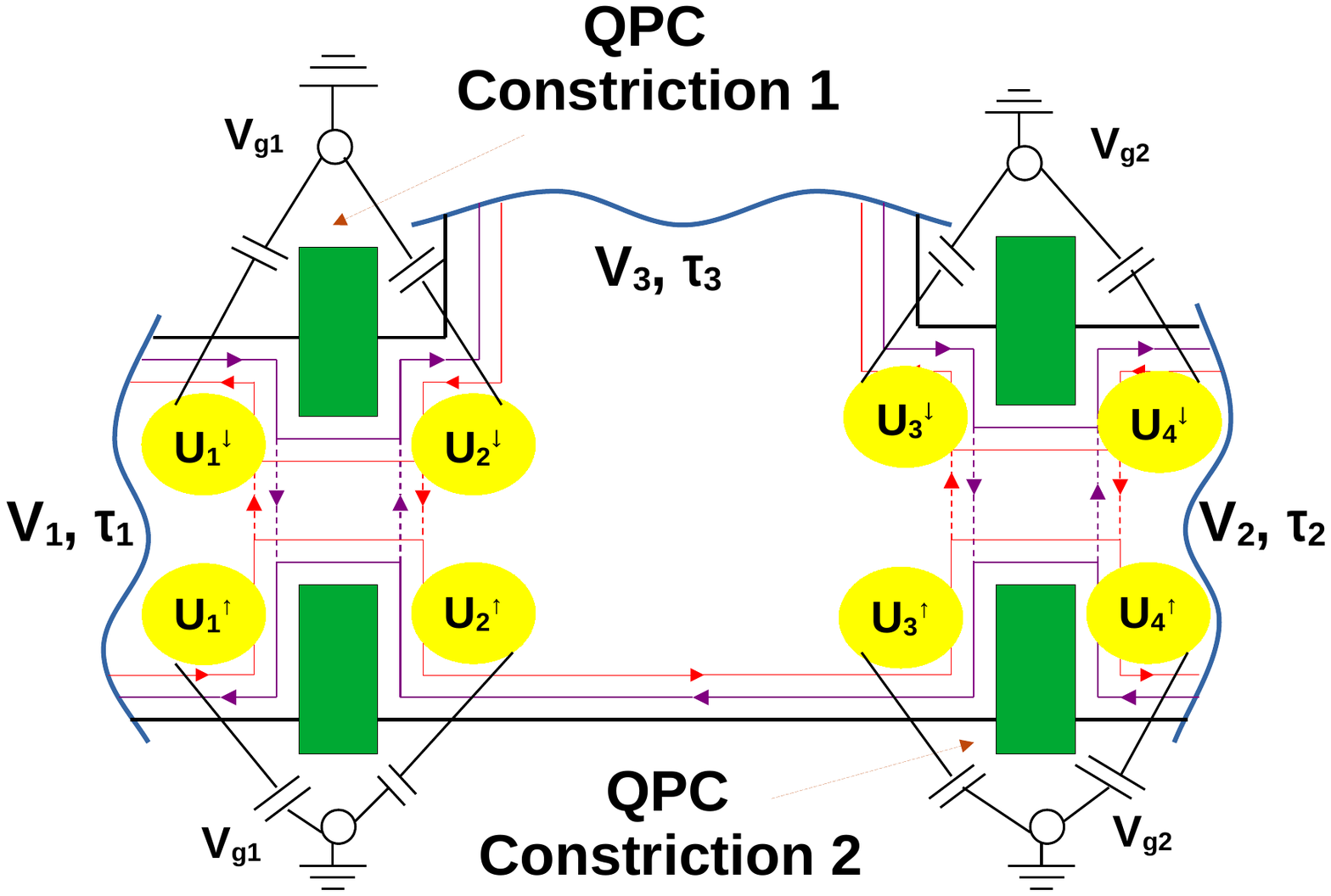}
\caption{{Three-terminal QSH setup, partitioned into 8 regions with interaction potentials $U_r^{\sigma}, r \in \{1, 2\}, \sigma \in \{\uparrow, \downarrow\}$ (yellow shaded regions) to the left and right of the QPC-1 and QPC-2 constrictions. Regions 1 and 2 are connected capacitively to a common external gate voltage $V_{g1}$ and regions $3$ and 4 are connected capacitively to another common external gate voltage $V_{g2}$.} }
\label{fig:17}
\end{figure}

 We next extend our approach to a 3T nonlinear QSH setup as shown in Fig. \ref{fig:17} with a voltage-temperature probe by controlling the transmission of QPC 1 and QPC 2 via a gate voltage. {Here, we consider four regions near the first QPC constriction with potentials $U_1^{\uparrow}, U_1^{\downarrow}, U_2^{\uparrow}$ and $U_2^{\downarrow}$ and then consider another four regions near QPC constriction 2 with potentials $U_3^{\uparrow}, U_3^{\downarrow}, U_4^{\uparrow}$ and $U_4^{\downarrow}$, see Fig. \ref{fig:17}. One can calculate the interaction potential $U_r^{\sigma}$ for regions $r$ with spin $\sigma$ by assuming all other potentials to be zero. The additional injected charge into the region $r$ with spin $\sigma$ ($q_{ad}^{r \sigma}$) can be calculated from Eq. (\ref{eq:99}). One can also connect the region capacitively to an external gate voltage $V_g$ and the injected charge, can be written as,}
 \begin{equation}\label{eq:111}
     {q_{ad}^{r \sigma} = C (U_r^{\sigma} - V_{g}).}
 \end{equation}
 {Equating the additional injected charge derived from Eqs. (\ref{eq:99}) and (\ref{eq:111}), one can numerically calculate $U_r^{\sigma}$.} 
{Using Eq. (\ref{eq:99}), the additional injected charge $q_{ad}^{1 \uparrow}$ in region 1 is, } 
 \begin{equation}\label{eq:112}
 \begin{split}
    {q_{ad}^{1\uparrow}} & {= e \int dE \left ( \sum_{\alpha} \nu_{1 \alpha}^{p \uparrow}(E, U_1^{\uparrow}) f_{\alpha}(E) \right) - e \int dE  \nu_1^{p \uparrow}(E) f,}\\
    & {= e \int dE \left(\nu_{1 1}^{p\uparrow}(E, U_1^{\uparrow})f_1(E) +  \nu_{1 2}^{p\uparrow}(E, U_1^{\uparrow})f_2(E) + \nu_{1 3}^{p \uparrow}(E, U_1^{\uparrow})f_3(E) \right) - e \int dE  \nu_1^{p \uparrow}(E) f,} 
    \end{split}
\end{equation}
{where, $\nu_{1\alpha}^{p \uparrow} = \sum_{\beta \in \{1, 2, 3 \}} \nu_{\beta 1 \alpha}^{p \uparrow}$ is the partial density of states of spin up electrons scattered from contact $\alpha$ to contact $\beta$. One can determine the partial density of states as follows: The partial density of state $\nu_{111}^{p \uparrow}$ is $\mathcal{T}_{11}^{\uparrow}$ times $\nu_1^{p\uparrow}$ = $(1 - \mathcal{T}_{QPC1})\nu_1^{p\uparrow}$, where $\mathcal{T}_{QPC1}$ is the transmission probability for the first QPC, see Fig. \ref{fig:17}. Similarly, $\nu_{211}^{p \uparrow}$ = 0, because $\mathcal{T}_{12}^{\uparrow} = 0$ and $\nu_{311}^{p \uparrow}$ is $\mathcal{T}_{31}^{\uparrow}$ times $\nu_1^{p\uparrow}$, i.e., $\nu_{311}^{p \uparrow} = \frac{\mathcal{T}_{QPC1}}{2} \nu_1^p$. Therefore, $\nu_{11}^{p\uparrow} = \nu_{111}^{p\uparrow} + \nu_{2 11}^{p \uparrow} + \nu_{311}^{p \uparrow} = \left(1 - \frac{\mathcal{T}_{QPC1}}{2}\right) \nu_1^{p\uparrow}$. In the similar way, we calculate $\nu_{12}^{p \uparrow} = \nu_{112}^{p \uparrow} + \nu_{212}^{p \uparrow} + \nu_{312}^{p \uparrow}$ and $\nu_{13}^{p \uparrow} = \nu_{113}^{p \uparrow} + \nu_{213}^{p \uparrow} + \nu_{313}^{p \uparrow}$ as well. $\nu_{112}^{p \uparrow}$ = $ \mathcal{T}_{12}^{\uparrow} \nu_{1}^{p \uparrow} = \frac{\mathcal{T}_{QPC1} \mathcal{T}_{QPC2}}{2} \nu_1^{p \uparrow}$, whereas $\nu_{213}^{p \uparrow}$ vanishes as there are no states in region 1 associated with electrons scattered from terminal 3 into terminal 2. Similarly, $\nu_{313}^{p \uparrow}$ = 0, as there are no states in region 1 for spin up electrons scattered from terminal 3 to itself. Therefore, $\nu_{12}^{p \uparrow} = \frac{\mathcal{T}_{QPC1} \mathcal{T}_{QPC2}}{2} \nu_1^{p \uparrow}$. For $\nu_{13}^{p \uparrow}$, the only contribution comes from $\nu_{113}^{p \uparrow}$ = $\frac{\mathcal{T}_{QPC1}(1- \mathcal{T}_{QPC2})}{2} \nu_1^{p \uparrow}$, whereas $\nu_{213}^{p \uparrow}$ and $\nu_{313}^{p \uparrow}$ vanish as there exist no states in region 1 associated with the scattering of spin up electron from terminal 3 to terminal 2 and therefore, $\nu_{13}^{p \uparrow}= \nu_{113}^{p \uparrow} + \nu_{213}^{p \uparrow} + \nu_{313}^{p \uparrow} = \frac{\mathcal{T}_{QPC1}(1 - \mathcal{T}_{QPC2})}{2} \nu_1^{p \uparrow}$, with $\mathcal{T}_{QPC1} = \frac{1}{1 + e^{-\frac{E - E_1 - eU_1}{\hbar \omega}}}$ and $\mathcal{T}_{QPC2} = \frac{1}{1 + e^{-\frac{E - E_2}{\hbar \omega}}}$. Now, one can capacitively connect the region 1 with spin up electron to an external gate voltage and the additional injected charge is given as,}
\begin{equation}\label{eq:113}
    {q_{ad}^{1 \uparrow} = C(U_1^{\uparrow} - V_g)}.
\end{equation}
{One can calculate $U_1^{\uparrow}$ numerically from Eqs. (\ref{eq:112}) and (\ref{eq:113}). In our setup, we consider $V_1 = -V, V_2 = 0, \tau_1 = \tau, \tau_2 = 0$ and $\tau = 1K, C = 0.01 F, \hbar \omega = 0.1 k_B T$ and $\frac{1}{2}m \omega^2 \lambda^2 = 20 k_B T$. We also impose that the terminal 3 is a voltage-temperature probe, i.e., $I_3 = J_3 = 0$. The interaction potential ($U_1^{\uparrow \prime}$), which solely depends upon the biases ($V$ and $\tau$) at zero gate voltage is found to be $3.75180 k_B T/e$ and with gate voltage ($V_g = 1 k_B T/e$) is $3.73925 k_B T/e$ at $V = 1.14 k_B T/e$. Therefore, $u_{g1}^{\uparrow} = \frac{U_{1}^{\uparrow} - U_1^{\uparrow \prime}}{V_g} = -0.01255$. The Python code for this calculation can be found in Ref. \cite{My_mathematica_notebook}. Similarly, we calculate the interaction potentials ($U_1^{\downarrow}, U_2^{\uparrow}$, $U_{2}^{\downarrow}, U_{3}^{\uparrow}, U_{3}^{\downarrow}, U_{4}^{\uparrow}, U_{4}^{\downarrow}$) of other regions around QPC-1 and QPC-2 numerically. The partial density of states or injectivities for $\sigma = \uparrow$ for each region are given by}
 
\begin{equation}\label{eq:114}
\begin{split}
    {\nu_{21}^{p\uparrow}} & {= \sum_{\beta \in \{1, 2, 3 \}} \nu_{21\beta}^{p\uparrow} = \frac{\mathcal{T}_{QPC1} \nu_2^{p\uparrow}}{2},\quad} {\nu_{22}^{p\uparrow}}   {= \sum_{\beta \in \{1, 2, 3 \}} \nu_{22\beta}^{p \uparrow} = \left(1 - \frac{\mathcal{T}_{QPC1}}{2}\right)\mathcal{T}_{QPC2}\nu_2^{p\uparrow},}\\
    {\nu_{23}^{p\uparrow}} & {=  \sum_{\beta \in \{1, 2, 3 \}} \nu_{23\beta}^{p \uparrow} = \left(1 - \frac{\mathcal{T}_{QPC1}}{2}\right)(1 - \mathcal{T}_{QPC2})\nu_2^{p\uparrow},\quad} {\nu_{31}^{p\uparrow}}  {= \sum_{\beta \in \{1, 2, 3 \}} \nu_{31\beta}^{p \uparrow} = 0,} \\ 
    {\nu_{32}^{p\uparrow}} & {= \sum_{\beta \in \{1, 2, 3 \}} \nu_{32\beta}^{p \uparrow} = \frac{\mathcal{T}_{QPC2}\nu_3^{p\uparrow}}{2},\quad \nu_{33}^{p\uparrow} = \sum_{\beta \in \{1, 2, 3 \}} \nu_{33\beta}^{p \uparrow} \left(1 - \frac{\mathcal{T}_{QPC2}}{2}\right)\nu_3^{p\uparrow},\quad} {\nu_{41}^{p\uparrow}} {= \sum_{\beta \in \{1, 2, 3 \}} \nu_{41\beta}^{p \uparrow} = 0,\quad} \\ 
    {\nu_{42}^{p\uparrow}} & {= \sum_{\beta \in \{1, 2, 3 \}} \nu_{42\beta}^{p \uparrow} = \left(1 - \frac{\mathcal{T}_{QPC2}}{2}\right)\nu_4^{p\uparrow},\quad \text{and} \quad \nu_{43}^{p\uparrow} = \sum_{\beta \in \{1, 2, 3 \}} \nu_{43\beta}^{p \uparrow} = \frac{\mathcal{T}_{QPC2}}{2}.}
    \end{split}
    \end{equation}

 {Similarly, the partial density of states or injectivities for $\sigma = \downarrow$ for each region are given by}
\begin{equation}\label{eq:115}
\begin{split} 
    {\nu_{11}^{p \downarrow}} & {= \sum_{\beta \in \{1, 2, 3 \}} \nu_{\beta 11}^{p \downarrow} = \left(1 - \frac{\mathcal{T}_{QPC1}}{2}\right)\nu_{1}^{p \downarrow}, \quad} {\nu_{12}^{p \downarrow} = = \sum_{\beta \in \{1, 2, 3 \}} \nu_{\beta 12}^{p \downarrow} = 0, \quad \nu_{13}^{p \downarrow} =  \sum_{\beta \in \{1, 2, 3 \}} \nu_{\beta 13}^{p \downarrow} = \frac{\mathcal{T}_{QPC1}}{2}\nu_{1}^{p \downarrow},}\\
    {\nu_{21}^{p\downarrow}} & {=\sum_{\beta \in \{1, 2, 3 \}} \nu_{21\beta}^{p\downarrow} = \frac{\mathcal{T}_{QPC1} \nu_2^{p\downarrow}}{2}, \quad} 
    {\nu_{22}^{p\downarrow}} = {\sum_{\beta \in \{1, 2, 3 \}} \nu_{22\beta}^{p \downarrow} = 0,}\\ 
    {\nu_{23}^{p\downarrow}} &= {\sum_{\beta \in \{1, 2, 3 \}} \nu_{23\beta}^{p \downarrow} = \left(1 - \frac{\mathcal{T}_{QPC1}}{2}\right)\nu_2^{p\downarrow},} \quad
    {\nu_{31}^{p\downarrow}} = {\sum_{\beta \in \{1, 2, 3 \}} \nu_{31\beta}^{p \downarrow} = \left(1 - \frac{\mathcal{T}_{QPC2}}{2}\right)\mathcal{T}_{QPC1}\nu_3^{p \downarrow},}\\ 
    {\nu_{32}^{p\downarrow}} &  {= \sum_{\beta \in \{1, 2, 3 \}} \nu_{32\beta}^{p \downarrow} = \frac{\mathcal{T}_{QPC2}\nu_3^{p\downarrow}}{2},\quad } 
    {\nu_{33}^{p\downarrow}} = {\sum_{\beta \in \{1, 2, 3 \}} \nu_{33\beta}^{p \downarrow} = \left(1 - \frac{\mathcal{T}_{QPC2}}{2}\right)(1-\mathcal{T}_{QPC1})\nu_3^{p \downarrow},}\\ 
    {\nu_{41}^{p\downarrow}} & {=\sum_{\beta \in \{1, 2, 3 \}} \nu_{41\beta}^{p \downarrow} = \frac{\mathcal{T}_{QPC1} \mathcal{T}_{QPC2}}{2}\nu_{4}^{p \downarrow},}\quad 
    {\nu_{42}^{p\downarrow}} = {\sum_{\beta \in \{1, 2, 3 \}} \nu_{42\beta}^{p \downarrow} = \left(1 - \frac{\mathcal{T}_{QPC2}}{2}\right)\nu_4^{p\downarrow}, \quad} 
    {\text{and}}\\ 
    {\nu_{43}^{p\downarrow}} & {=\sum_{\beta \in \{1, 2, 3 \}} \nu_{43\beta}^{p \downarrow} = \frac{(1-\mathcal{T}_{QPC1})\mathcal{T}_{QPC2}}{2}\nu_4^{p \downarrow}}.
\end{split}
\end{equation}

\begin{figure}
     \centering
     \begin{subfigure}[b]{0.45\textwidth}
         \centering
         \includegraphics[width=\textwidth]{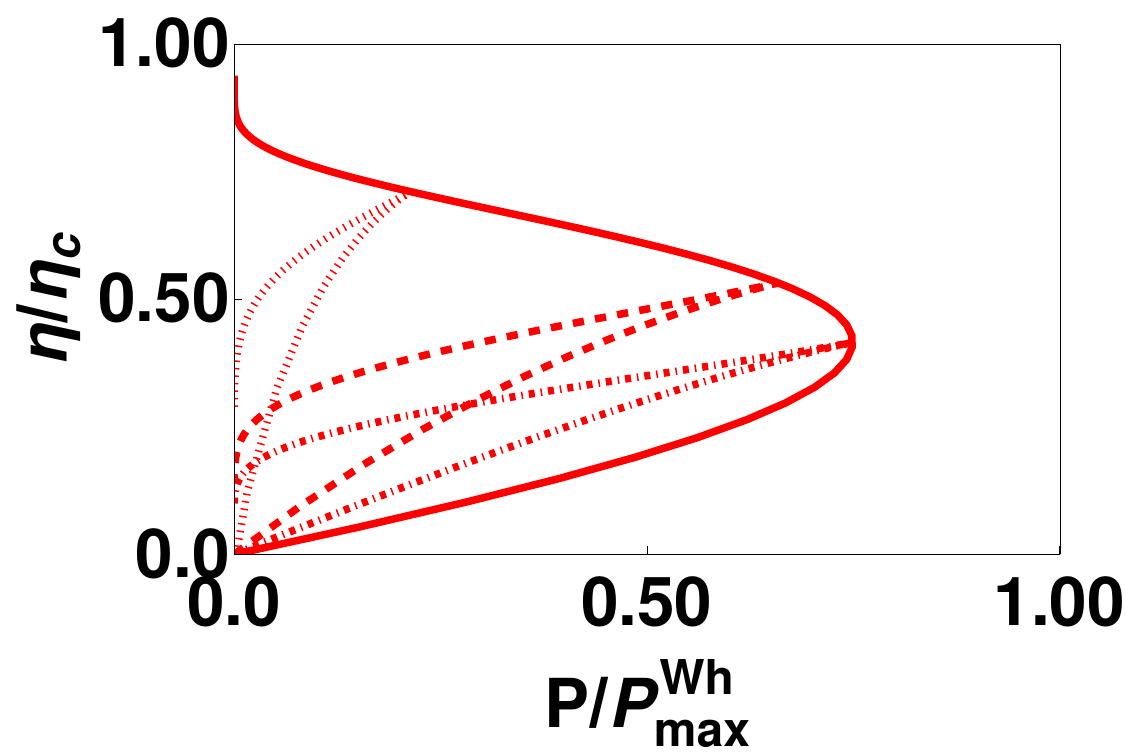}
         \caption{QH heat engine (linear)}
     \end{subfigure}
     \hspace{0.05cm}
     \begin{subfigure}[b]{0.45\textwidth}
         \centering
         \includegraphics[width=\textwidth]{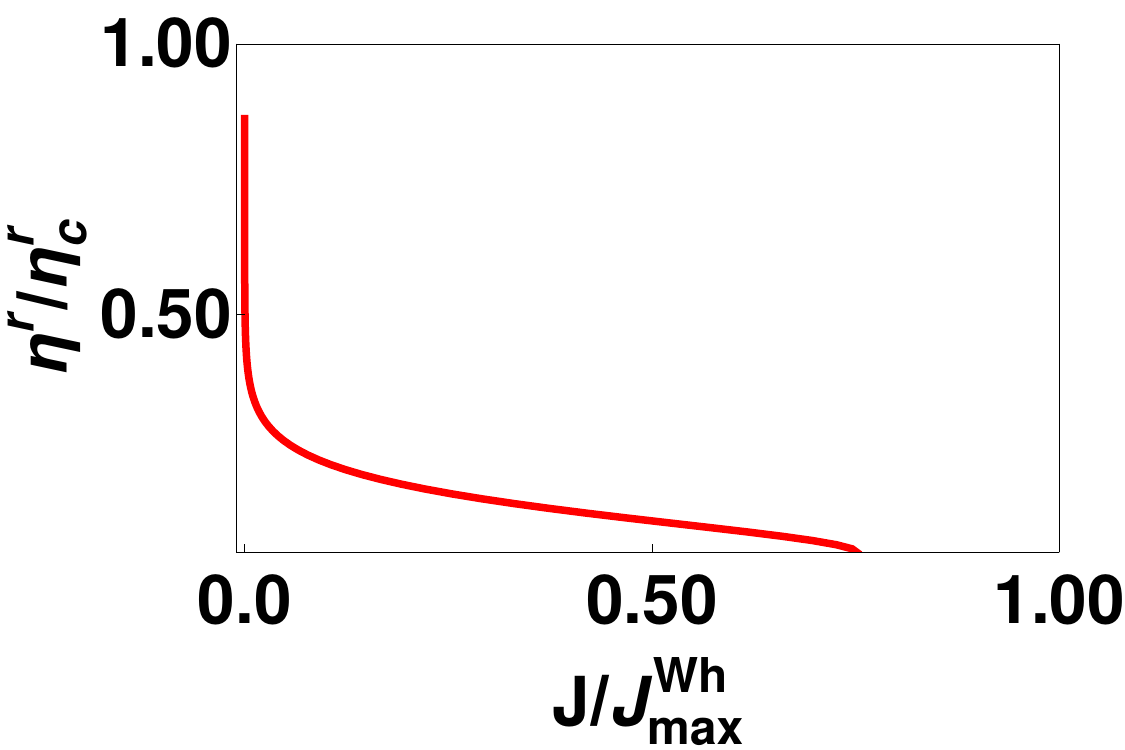}
         \caption{QH refrigerator (linear)}
     \end{subfigure}
        \caption{Parametric plot of $\frac{\eta}{\eta_c}$ vs $\frac{P}{P_{max}^{Wh}}$ for (a) QSH QHE. The red dot-dashed curve is for $V = 1.1 k_B T/e$, the red dashed curve for $V = 1.77 k_B T/e$, and the red dotted curve for $V = 3.7 k_B T/e$. The red line shows the bound to the power and efficiency. Parametric plot for $\frac{\eta^r}{\eta^r_c}$ vs $\frac{\textbf{J}}{\textbf{J}^{Wh}_{max}}$ for (b) QSH QR.
         For QSH  QR, the blue line is for $V = 20k_B T/e$. Parameters taken are $\omega = \frac{0.1 k_B T}{h}$, $T_1 = 2K, T_2 = 1K$, $\mu = 0$, $2E_1 = E_2 = 2k_B T$.}
         \label{fig:18}
       \end{figure}

 {The average interaction potential for QPC-1 is given as }
 \begin{equation}\label{eq:116}
     {U_{QPC1} = \frac{U_1^{\uparrow} + U_1^{\downarrow} + U_2^{\uparrow} + U_2^{\downarrow}}{2} = \frac{U_1^{\uparrow \prime} + U_1^{\downarrow \prime} + U_2^{\uparrow \prime} + U_2^{\downarrow \prime} }{2} + u_{g1} V_{g1} = U_{QPC1}' + u_{g1} V_{g1}}.
 \end{equation}

{As $\nu_1^{p \uparrow} = \nu_1^{p \downarrow} = \nu_2^{p \uparrow} = \nu_2^{p \downarrow}$, see Eq. (\ref{eq:105}). The interaction potentials $U_1^{\uparrow}, U_1^{\downarrow}, U_2^{\uparrow}, U_2^{\downarrow}$ can be evaluated both at zero gate voltage and finite gate voltage ($V_g = 1.0k_B T/e$) at applied bias $V = 1.14k_B T/e$. The average interaction potential ($U_{QPC1}'$) in absence of a gate voltage is 1.27055$k_B T/e$ and at finite gate voltage ($V_{g1} = 1.0k_B T/e$) is 1.25144 $k_B T/e$, which implies $u_{g1} = \frac{U_{QPC1} - U_{QPC1}'}{V_g} = -0.01911$. One then applies a gate voltage $V_{g1} = \frac{-U_{QPC1}'}{u_{g1}} + \frac{V_g'}{u_{g1}} = 66.48613 - \frac{V_g'}{0.01911}$, such that the first term in $V_{g1}$ nullifies the interaction potential, and the second term lifts the threshold energy of QPC1. The threshold energy of QPC1 is then $E_1 + eV_g'$. We tabulate the values of interaction potential and the applied gate voltages in Table IV for different applied voltages, i.e., $V = 1.77 k_B T/e, V = 3.7 k_B T/e $ and $V = 20 k_B T/e$. The voltages $V = 1.14 k_B T/e, 1.77 k_B T/e$ and $3.7 k_B T/e$ are used for studying the setup's performance as QHE, whereas $V = 20k_B T/e$ is used to study the setup's performance as QR.}

  {Similarly, the average interaction potential for QPC2 is given as, }
 \begin{equation}\label{eq:117}
     {U_{QPC2} = \frac{U_3^{\uparrow} + U_3^{\downarrow} + U_4^{\uparrow} + U_4^{\downarrow}}{2} = \frac{U_3^{\uparrow \prime} + U_3^{\downarrow \prime} + U_4^{\uparrow \prime} + U_4^{\downarrow \prime} }{2} + u_{g2} V_{g2} = U_{QPC2}' + u_{g2} V_{g2}},
 \end{equation}

{As $\nu_1^{p \uparrow} = \nu_3^{p \uparrow} = \nu_3^{p \downarrow} = \nu_4^{p \uparrow} = \nu_4^{p \downarrow}$, from Eq. (\ref{eq:105}), one can estimate the interaction potentials $U_3^{\uparrow}, U_3^{\downarrow}, U_4^{\uparrow}, U_4^{\downarrow}$ both at zero gate voltage and finite gate voltage ($V_{g2} = 1.0k_B T/e$) at applied bias $V = 1.14k_B T/e$. The average interaction potential ($U_{QPC2}'$) at zero gate voltage $V_{g2} (= 0)$ is 3.195815$k_B T/e$ and at finite gate voltage ($V_{g2} = 1.0k_B T/e$) is 3.54566 $k_B T/e$, which gives $u_{g2} = \frac{U_{QPC2} - U_{QPC2}'}{V_{g2}} = 0.349845$. Thus, one can apply a gate voltage $V_{g2} = \frac{-U_{QPC2}'}{u_{g2}} + \frac{V_g'}{u_{g2}} = -9.13494 + \frac{V_g'}{0.349845}$, where first term in $V_{g2}$ nullifies the interaction potential, and the second term lifts the threshold energy of QPC2, and the threshold energy of QPC2 is $E_2 + eV_g'$. We tabulate the values of interaction potential and the applied gate voltages in Table \ref{Table4} for different applied voltages such as $V = 1.77 k_B T/e, V = 3.7 k_B T/e $ and $V = 20 k_B T/e$. The voltages $V = 1.14 k_B T/e, 1.77 k_B T/e$ and $3.7 k_B T/e$ are used for studying the setup's performance as QHE. Similarly, $V = 20 k_B T/e$ is used for the setup's performance as QR.} 

\begin{table}[h]
\renewcommand{\arraystretch}{1.5}
\centering
\caption{Estimation of interaction potential and the applied gate voltages for QPC1 and QPC2 in a 3T QSH setup with voltage-temperature probe. Parameters taken are $\tau = 1K, C = 0.01 F, \hbar \omega = 0.1 k_B T$ and $\frac{1}{2}m \omega^2 \lambda^2 = 20 k_B T$.}
\scalebox{0.90}{
\begin{tabular}{|c|c|c|c|c|}
\hline
Applied voltage Bias ($k_B T/e$) & \begin{tabular}[c]{@{}c@{}}Interaction potential\\ for QPC1 ($k_B T/e$)\end{tabular} & \begin{tabular}[c]{@{}c@{}}Applied gate voltage\\ at QPC1 ($k_B T/e$)\end{tabular} & \begin{tabular}[c]{@{}c@{}}Interaction potential\\ at QPC2 ($k_B T/e$)\end{tabular} & \begin{tabular}[c]{@{}c@{}}Applied Gate Voltage\\ at QPC-2 ($k_B T / e$)\end{tabular} \\ \hline
1.14                 &           1.27055                        &    $66.48613 - \frac{V_g'}{0.01911}$                             &        3.195815                          &   $-9.13494 + \frac{V_g'}{0.349845}$                              \\ \hline
1.77                 &       9.59074                            &        $-40.72501 + \frac{V_g'}{0.23550}$                         &          2.550665                        &      $-6.58771 + \frac{V_g'}{0.387185}$                           \\ \hline
3.7                  &              12.78584                    &      $11069.9913 - \frac{V_g'}{0.001155}$                           &         -2.39258                         &         $-10.50344 - \frac{V_g'}{0.22779}$                        \\ \hline
20                   &     10.28782                         &    $514391 + \frac{V_g'}{0.00002}$                              &                2.107945                 &  $-664.96687 + \frac{V_g'}{0.00317}$                               \\ \hline
\end{tabular}}
\label{Table4}
\end{table}

Considering this, one can calculate power and efficiency as a quantum heat engine by changing $V_g^{'}$ at fixed values of $V, \tau$ and we observe that the maximum power approaches the Whitney bound, whereas the maximum efficiency is 0.93 $\eta_c$ as shown in Fig. 3(b) of the main text.

  Similar to the quantum heat engine, we investigate the three-terminal QSH setup as a quantum refrigerator with voltage-temperature probe with our approach and considering identical values of $E_1^{'}$ and $E_2^{'}$ same as used for quantum heat engine. The coefficient of performance $\eta^r$ $= \frac{J}{P}$ and the cooling power $J$ is $-(J_2 + J_3)$ are evaluated with either probe and the parametric plot are shown in Fig. 3(d) of the main text.
 With these conditions, in the three-terminal QSH setup for a voltage-temperature probe, we observe that $J_{max}$ reaches $J^{Wh}_{max}$, while $\eta^r$ reaches about 86$\%$ of $\eta^r_c$ as shown in Fig. 3(d) of the main text. In the main text, we considered $E_1 = E_2 = k_B T$ and achieved the Whitney limits for both the quantum heat engine and refrigerator. For the case, $E_1 \ne E_2$, i.e., if $2E_1 = E_2 = 2k_B T$, the results, see, Fig. \ref{fig:18} are identical to the results of the main text for QSH QHE and QR, see Fig. 3(b) and (d).

\end{document}